\documentclass[aps,prl,reprint,superscriptaddress,nofootinbib,raggedbottom]{revtex4-2}
\usepackage[T1]{fontenc}
\usepackage{amsmath,amssymb,graphicx,booktabs,slashed}
\usepackage{newtxtext,newtxmath,bm,microtype}
\usepackage{tikz}
\usetikzlibrary{arrows.meta,calc,positioning,decorations.pathmorphing,decorations.markings,decorations.pathreplacing,patterns}
\usepackage[colorlinks=true,linkcolor=blue,citecolor=blue,urlcolor=blue]{hyperref}
\newcommand{\dd}{\mathrm{d}}
\newcommand{\ii}{\mathrm{i}}
\newcommand{\tr}{\operatorname{tr}}
\newcommand{\cO}{\mathcal{O}}
\begin{document}
\title{Is Spin Slow in Hot QCD?}
\author{Zi-Hao Liu}
\email{zihaomology@gmail.com}
\affiliation{School of Physical Sciences, University of Chinese Academy of Sciences, Beijing 101408, China}
\author{Mei Huang}
\email{huangmei@ucas.ac.cn}
\affiliation{School of Nuclear Science and Technology, University of Chinese Academy of Sciences, Beijing 101408, China}

\begin{abstract}
Promoting spin to a hydrodynamic variable requires slow relaxation, as mass-suppressed helicity flips in hot QCD suggest. Using the full leading-order collision operator, we show instead that light-quark spin damps at the transport rate $g^4T\ln(1/g)$, nearly as fast as vector currents. Since spatial spin locks to momentum, helicity-conserving soft, small-angle scatterings readily rotate it. Damping involves multiple kinetic modes beyond a single relaxation time, global rotation adds only reversible precession, and stronger coupling further disfavors a slow spin mode in quark-gluon plasma.
\end{abstract}
\maketitle

\section{Introduction}
Relativistic spin hydrodynamics promotes the spin density to an independent dynamical variable alongside the conserved densities \cite{FFJS18,HHHMT19,BFS19,H21,WW22}. Even without an independent spin variable, thermal vorticity and fluid gradients polarize a locally equilibrated plasma \cite{Fu21,Bec21}. Treating spin as a separate hydrodynamic mode is justified only if off-equilibrium polarization relaxes parametrically more slowly than ordinary kinetic excitations \cite{H21,H22}; otherwise spin can enter only as a transient variable, like the dissipative currents of second-order fluid dynamics \cite{DNMR12,WW22}.

\begin{figure*}[t]
 \centering
 \input{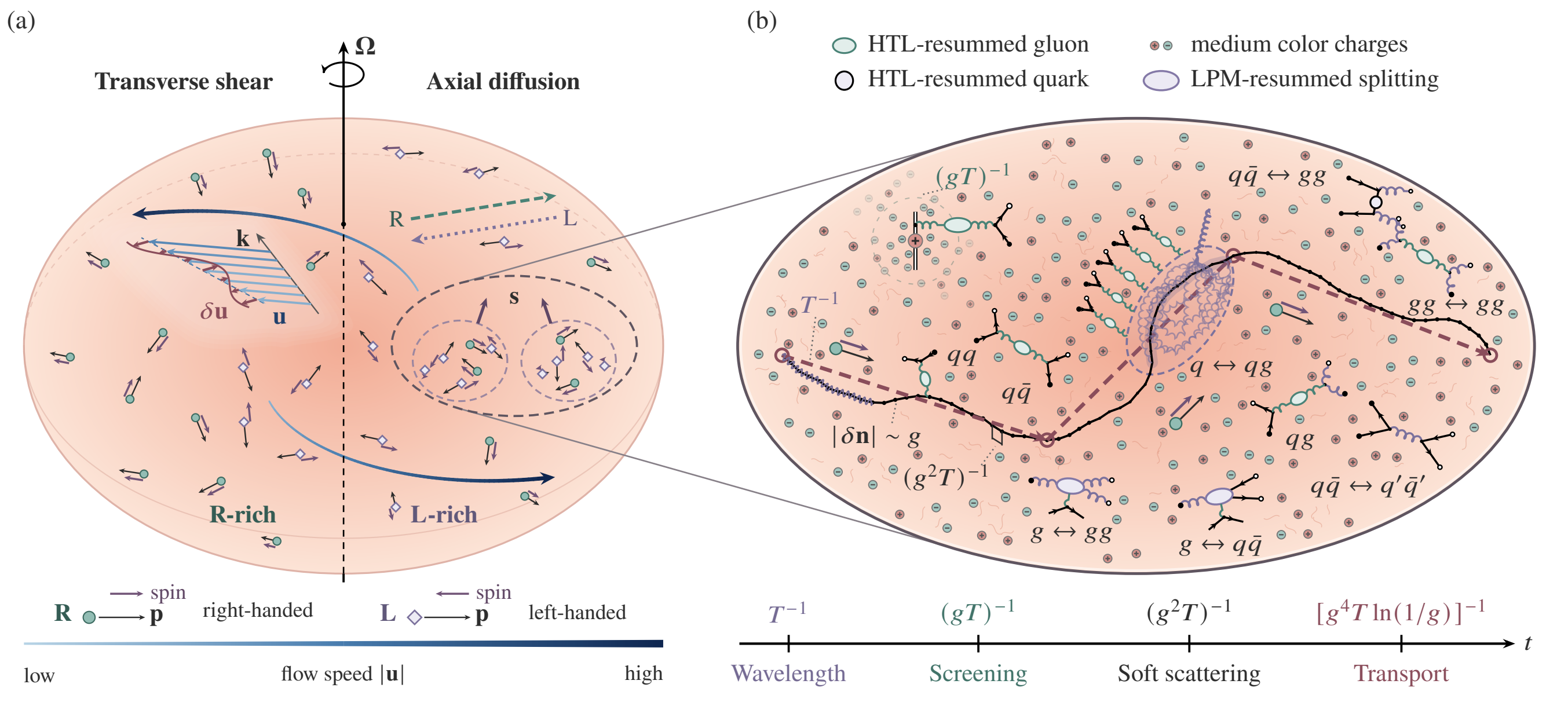}
 \caption{Spin dynamics in a rotating plasma from hydrodynamic to microscopic scales (schematic). (a) Background flow $\mathbf{u}$ (blue) with vorticity $\boldsymbol{\Omega}=\frac12\nabla\times\mathbf{u}$; the velocity profile along the wave vector $\mathbf{k}$ shows a transverse shear perturbation $\delta\mathbf{u}$ as a small modulation of $\mathbf{u}$. Right-handed (R) and left-handed (L) quarks have spin (purple) parallel and antiparallel to their momentum $\mathbf{p}$; dashed and dotted arrows indicate axial-charge diffusion between R-rich and L-rich regions. Dashed circles mark fluid elements, whose spin densities $\mathbf{s}$ sum to a net polarization along $\boldsymbol{\Omega}$. (b) Magnified view of the dashed ellipse in (a), not to scale: a quark with thermal momentum $p\sim T$ and wavelength $T^{-1}$ at weak coupling. Soft scatterings, separated by $(g^2T)^{-1}$, each deflect its direction $\mathbf{n}=\mathbf{p}/p$ by $|\delta\mathbf{n}|\sim g$ and together rotate its momentum and helicity-locked spin by order-unity angles over the transport time $[g^4T\ln(1/g)]^{-1}$ (dashed arrows). Diagrams: collision processes of quarks ($q$, $q'$), antiquarks ($\bar q$), and gluons ($g$), with hard-thermal-loop (HTL) resummed soft exchange and Landau--Pomeranchuk--Migdal (LPM) resummed collinear splitting (violet ellipses). Color charges ($\pm$) screen color fields beyond $(gT)^{-1}$ (dashed circle). Panel (b) is adapted from Refs.~\cite{AMY03a,AMY03b}.}
 \label{fig:transportscales}
\end{figure*}

Approximate chiral symmetry might seem to protect light-quark spin from rapid collisional damping; estimates based on mass-suppressed helicity flips suggest that strange-quark spin barely relaxes before hadronization \cite{KRR20}. For a nearly massless quark, however, spatial spin is locked to the momentum direction, so even helicity-conserving collisions rotate it. Cumulative small-angle scatterings randomize both directions on the transport time scale $\gamma_{\mathrm{tr}}^{-1}\equiv[g^4T\ln(1/g)]^{-1}$ at temperature $T$ and coupling $g\ll1$ (Fig.~\ref{fig:transportscales}). Consistently, leading-logarithmic kinetic analyses for quark masses of order $T$ find transport-scale spin relaxation \cite{LY19,YHH20}, massless-QED calculations resolve gradient-induced polarization and relaxation \cite{FangPuYang2022}, and nonlocal kinetic models yield spin relaxation times comparable to those of dissipative currents \cite{WWSR22}. Meanwhile, axial kinetic theory in the relaxation-time approximation (RTA) exhibits a nonlocal response to spacetime variations of vorticity, suggesting a breakdown of the hydrodynamic gradient expansion \cite{LinTang2024}.

These results motivate determining the kinetic spin response and its coupling to fluid flow beyond leading-logarithmic and single-relaxation-time approximations. Although the quark--gluon plasma (QGP) in heavy-ion collisions is strongly coupled \cite{GM05,S05,KSS05,HS13,BRS18}, weak coupling isolates the microscopic relaxation mechanism. Here we compute the linear response of the light-quark canonical spin density using the complete leading-order QCD collision operator \cite{AMY03a,AMY03b}, both in a static medium and under global rotation with angular velocity $\boldsymbol{\Omega}$ (magnitude $|\boldsymbol{\Omega}|\ll g^2T$), working to leading order in $|\boldsymbol{\Omega}|/T$ and to all orders in $|\boldsymbol{\Omega}|/\gamma_{\mathrm{tr}}$.\footnote{We use natural units ($\hbar=c=k_B=1$) and the Fourier convention $e^{-\ii\omega t+\ii\mathbf{k}\cdot\mathbf{x}}$ with frequency $\omega$, time $t$, wave vector $\mathbf{k}$ ($k\equiv|\mathbf{k}|$), and position $\mathbf{x}$.} We show that small-angle scattering damps thermal spin perturbations on the transport time scale, and that neither fluid rotation nor hydrodynamic coupling produces an independent slow spin mode.

\section{Spin density and kinetic theory}
We consider $\mathrm{SU}(3)$ QCD with three light quark flavors $f\in\{u,d,s\}$ of current mass $m_{0f}=\cO(gT)$ (including the massless limit) at zero chemical potential near thermal equilibrium. In the canonical pseudogauge \cite{H21}, gluons enter total angular momentum through the symmetric energy-momentum tensor and do not contribute to the gauge-invariant spin tensor. The totally antisymmetric Dirac spin tensor reduces the spatial spin density to half the flavor-singlet axial current \cite[Sec.~S1.1]{SM},
\begin{equation}
 s^i=\tfrac12J_5^i,\qquad J_5^\mu=\sum_f\bar q_f\gamma^\mu\gamma_5q_f.
 \label{eq:triplet}
\end{equation}

For a quark with momentum $\mathbf{p}$, direction $\mathbf{n}=\mathbf{p}/p$, and energy $E_p=(p^2+m_{0f}^2)^{1/2}$, the one-particle matrix element of the canonical spin density $j_f^i$ in the helicity basis has a longitudinal part $\frac12n_i\sigma_3$ (with Pauli matrix $\sigma_3$ acting on the two helicity states) and a transverse part proportional to $m_{0f}/(2E_p)$ \cite[Sec.~S1.1]{SM}; charge conjugation gives the same matrix element for antiquarks. At thermal momenta $p\sim T$, transverse spin is suppressed by $m_{0f}/p\sim g$, leaving only the helicity contribution at leading order. The current mass nevertheless enters the leading-order collision operator via chirality-flipping soft-fermion exchange and collinear splitting.

In leading-order effective kinetic theory \cite{AMY03a,AMY03b}, the occupation of each species $a\in\{q_f,\bar q_f,g\}$ is a $2\times2$ matrix in its two physical polarization states:
\begin{equation}
\begin{gathered}
f_a(x,\mathbf{p})
=f_a^{(0)}(p)\mathbf{1}_2+\delta f_a(x,\mathbf{p}),
\\
\delta f_a
=f_a^{(0)}(p)\bar f_a^{(0)}(p)\Phi_a(x,\mathbf{p}),
\end{gathered}
\end{equation}
where $x=(t,\mathbf{x})$, $f_a^{(0)}(p)=(e^{p/T}\pm1)^{-1}$ is the equilibrium Fermi--Dirac ($F$, $+$) or Bose--Einstein ($B$, $-$) distribution, and $\bar f_a^{(0)}=1\mp f_a^{(0)}$. At zeroth order in spatial gradients, the quasiparticle contribution to the spin density is the thermal inner product
\begin{equation}
\begin{gathered}
 \delta s^i=(j^i,\Phi)\\
 \equiv\sum_a d_{c,a}\!\int\!\frac{\dd^3p}{(2\pi)^3}
 f_a^{(0)}\bar f_a^{(0)}\tr_2(j_a^i\Phi_a),
\end{gathered}
 \label{eq:metric}
\end{equation}
with color multiplicities $d_{c,q}=d_{c,\bar q}=3$ and $d_{c,g}=8$, helicity trace $\tr_2$, and $j_g^i=0$. For a fixed Cartesian component $j$, $(j,j)=T^3/4$ defines the leading-order quasiparticle susceptibility $\chi=(j,j)/T=T^2/4$ \cite[Sec.~S1.3]{SM}. The collision operator $C$ comprises \mbox{$2\leftrightarrow2$} scattering with hard-thermal-loop (HTL) resummation of soft exchange and $1\leftrightarrow2$ splitting with Landau--Pomeranchuk--Migdal (LPM) resummation [Fig.~\ref{fig:transportscales}(b)]. It is block diagonal in the multipole angular momentum $J$, acts on $j^i$ in the charge-conjugation-even $J=1$ axial channel, and defines a positive-semidefinite quadratic form $Q[\Phi]\equiv(\Phi,C\Phi)\ge0$ \cite[Sec.~S2]{SM}.

\section{Microscopic spin relaxation}
To test whether collisions permit a slow spin mode, we prepare a static, homogeneous plasma in a weak field $h_0$ conjugate to one spin component $s$ and switch it off at $t=0$. The initial polarization $\delta s(0)=\chi h_0$ then relaxes under $\partial_t\Phi=-C\Phi$ as
\begin{equation}
 \frac{\delta s(t)}{\delta s(0)}=\frac{(j,e^{-Ct}j)}{(j,j)}.
 \label{eq:spinrelease}
\end{equation}

Soft-gluon exchange conserves quark helicity, but a transverse momentum transfer $\mathbf{q}_\perp$ deflects a hard quark by $\delta\mathbf{n}\simeq\mathbf{q}_\perp/p$, rotating its spin. Because the longitudinal spin perturbation has a dipole angular dependence, successive deflections cause angular diffusion with $(\delta\mathbf{n})^2\sim\mathbf{q}_\perp^2/p^2$ and relax the polarization at the transport rate $\gamma_{\mathrm{tr}}$. Unlike momentum transport, where quark and gluon momentum changes cancel by conservation, the four helicity-conserving $qg\to qg$ channels share the same eikonal transition probability; summing over helicities eliminates the interference between quark and gluon polarization perturbations, so gluon recoil cannot cancel the angular diffusion of quark spin \cite[Sec.~S3.2]{SM}.

On any bounded thermal momentum interval $p\sim T$, small-angle quark--gluon scattering alone bounds the damping rate of longitudinal quark polarization from below by a $g$-independent multiple of $\gamma_{\mathrm{tr}}$, and other channels only increase $Q[\Phi]$ \cite[Sec.~S3.2]{SM}. Modes with decay rates $\Gamma\ll\gamma_{\mathrm{tr}}$ thus carry negligible spectral weight at $p\sim T$. Contributions to $\chi$ from other momentum scales are likewise suppressed: hard quarks ($p\gg T$) undergo slower angular deflection ($\propto T^2/p^2$) but have exponentially suppressed thermal weight; soft fermionic modes ($p\sim gT$) contribute only $\cO(g^3T^2)$ due to phase space and Pauli blocking; and the HTL screening cloud of a soft gluon has zero overlap with $j$ \cite[Sec.~S3.3]{SM}. Combined with the initial decay rate $(j,Cj)/(j,j)=\cO(\gamma_{\mathrm{tr}})$, this excludes any finite fraction of spin polarization relaxing at rates $\Gamma\ll\gamma_{\mathrm{tr}}$ or $\Gamma\gg\gamma_{\mathrm{tr}}$.

\begin{figure}[t]
 \centering
 \includegraphics[width=\columnwidth]{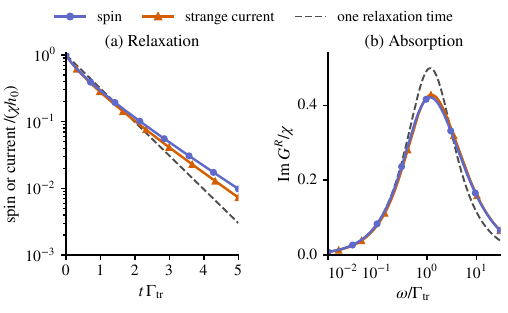}
 \caption{(a) Relaxation of the canonical spin density (circles) and strange-quark vector current (triangles) after switching off a weak conjugate field at $t=0$. (b) Normalized absorptive part of the retarded response function, $\operatorname{Im}G^R(\omega)/\chi$ ($\chi=T^2/4$ for spin; $\chi_J=T^2/3$ for the strange vector current), with rates in units of the current's effective relaxation rate $\Gamma_{\mathrm{tr}}$. Dashed curves show the single-relaxation-time approximation matching $\tau$. Results are computed at leading order using 40 cubic B-splines per channel at $g=10^{-3}$, $m_{0u}=m_{0d}=0$, and $m_{0s}=0.3gT$.}
 \label{fig:finitebasis}
\end{figure}

Figure~\ref{fig:finitebasis} compares the relaxation of canonical spin and the strange-quark vector current at $g=10^{-3}$, where weak-coupling scales are well separated \cite[Sec.~S3.4]{SM}. From each normalized relaxation function $R(t)$ [$R(t)=\delta s(t)/\delta s(0)$ for spin], we define the integrated relaxation time $\tau=\int_0^\infty\dd t\,R(t)=(j,C^{-1}j)/(j,j)$ and effective relaxation rate $\Gamma_{\mathrm{eff}}=\tau/\int_0^\infty\dd t\,tR(t)$. In units of the vector current's effective rate $\Gamma_{\mathrm{tr}}=1.71\times10^{-13}\,T\simeq0.025\,\gamma_{\mathrm{tr}}$, our variational B-spline calculation gives $\Gamma_{\mathrm{eff}}=0.93\,\Gamma_{\mathrm{tr}}$ and $\tau=0.86/\Gamma_{\mathrm{tr}}$ for spin, showing that spin relaxes nearly as fast as the current. Unlike single-exponential decay ($\Gamma_{\mathrm{eff}}\tau=1$), the kinetic spectrum spans multiple decay rates \cite{HT10,Moore18}: the lowest eigenmode carries $36\%$ of $\chi$ and faster modes carry $64\%$, giving $\Gamma_{\mathrm{eff}}\tau=0.80$. Consequently, in the retarded spin response $\delta s(\omega)=G^R(\omega)h(\omega)$ to a driving field $h(\omega)$, the single-relaxation-time approximation $G^R_\tau(\omega)=\chi/(1-\ii\omega\tau)$ deviates from $G^R(\omega)$ by up to $0.11\chi$ near $\omega\simeq2\,\Gamma_{\mathrm{tr}}$ [Fig.~\ref{fig:finitebasis}(b)] \cite[Secs.~S3.1 and S3.4]{SM}.

\section{Vortical polarization and spin lag}
While collisions damp off-equilibrium spin, a rotating fluid sustains polarization in local equilibrium because rotation couples to total angular momentum \cite{BT10}. Writing the fluid three-velocity as a background $\mathbf{u}$ plus an infinitesimal perturbation $\delta\mathbf{u}$, the background vorticity in the local rest frame is $\boldsymbol{\Omega}=\frac12\nabla\times\mathbf{u}$ [Fig.~\ref{fig:transportscales}(a)]. To first order in $|\boldsymbol{\Omega}|/T$, the local-equilibrium distributions read $f_F^{(0)}(p-\frac\lambda2\boldsymbol{\Omega}\cdot\mathbf{n})$ for quarks \cite{CSSYY14} and $f_B^{(0)}(p-\lambda\boldsymbol{\Omega}\cdot\mathbf{n})$ for gluons \cite{Lin26}, where $\lambda=\pm1$ labels the quark helicity sign (physical helicity $\lambda/2$) or gluon circular polarization. Linearizing in $\boldsymbol{\Omega}$ defines $\Phi_{\mathrm{rot}}=\Omega_ij_\Omega^i/T$, with $j_\Omega^i=j^i$ for quarks and eigenvalues $n_i\lambda$ for gluons \cite[Sec.~S4.1]{SM}. The axial current also contains a magnetization term given by the curl of the spin-averaged quark distribution weighted by $\mathbf{n}/p$ \cite{Chen15,Gao19}. A small local boost induces a spin-independent shift $(\mathbf{p}\cdot\mathbf{u}/T)\mathbf{1}_2$, whose magnetization current contributes $2\chi\boldsymbol{\Omega}$ to the spin density. Since $(j,j_\Omega)=(j,j)$, the quasiparticle and magnetization contributions sum to
\begin{equation}
 \mathbf{s}=(\mathbf{j},\Phi_{\mathrm{rot}})+2\chi\boldsymbol{\Omega}=3\chi\boldsymbol{\Omega}=\frac{3T^2}{4}\boldsymbol{\Omega},
 \label{eq:seq}
\end{equation}
reproducing the axial vortical effect $\mathbf{J}_5=\frac32T^2\boldsymbol{\Omega}$ for three colors and three massless flavors \cite{Vilenkin80,Landsteiner14}. Coupling $\boldsymbol{\Omega}$ to spin alone would yield only one-third of \eqref{eq:seq}; the off-equilibrium polarization is therefore $\mathbf{s}-3\chi\boldsymbol{\Omega}$.

Because helicity imbalance can relax at fixed flow velocity, we evolve the spin-averaged flow separately and use $\Phi$ for the polarization-dependent perturbation, including axial charge and higher angular moments generated by free streaming. Free streaming of the spin-averaged local-equilibrium distribution depends only on the symmetric velocity gradient, so vorticity drives polarization through collisions. Indeed, in rigid rotation ($\mathbf{u}=\boldsymbol{\Omega}\times\mathbf{x}$), orbital angular-momentum conservation keeps $\mathbf{p}\cdot\mathbf{u}/T$ constant along free trajectories, while momentum conservation makes the local collision term vanish for this flow perturbation. By contrast, collisions exchange spin and orbital angular momentum, so $C\Phi_{\mathrm{rot}}\ne0$. Stationarity of global rotational equilibrium then requires the first-order nonlocal collision term acting on the spin-averaged distribution to cancel $-C\Phi_{\mathrm{rot}}$, as dictated by angular-momentum conservation \cite{Chen15,HPY17,Wagner22} and verified at leading-logarithmic order in QED \cite{Wang2022QED}. Since vorticity is the sole axial vector among first-order fluid gradients, rotational equilibrium uniquely fixes this source in the $J=1$ axial channel. For wave numbers $k\lesssim\gamma_{\mathrm{tr}}$, the remaining nonlocal terms are parametrically suppressed relative to $C$ by at most $\cO(g^2\ln(1/g))$ \cite[Secs.~S4.2 and S4.3]{SM}, leaving
\begin{equation}
 \partial_t\Phi+\mathbf{n}\cdot\nabla\Phi=-C(\Phi-\Phi_{\mathrm{rot}}).
 \label{eq:kinetic}
\end{equation}

\begin{figure*}[t]
 \centering
 \includegraphics[width=160mm]{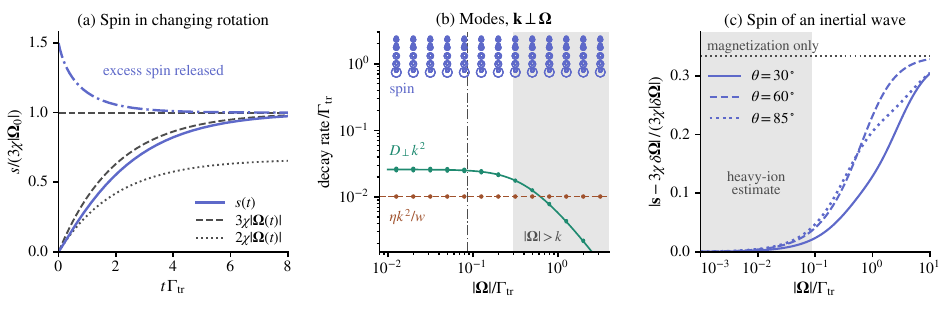}
 \caption{Spin dynamics from \eqref{eq:kinetic} and \eqref{eq:momentum}, retaining $J=1$ polarization, $J=0$ axial charge, and incompressible flow. (a) Total spin density for $\boldsymbol{\Omega}(t)=\boldsymbol{\Omega}_0(1-e^{-t\Gamma_{\mathrm{tr}}/2})$ (solid), local-equilibrium value $3\chi|\boldsymbol{\Omega}(t)|$ (dashed), magnetization part $2\chi|\boldsymbol{\Omega}(t)|$ (dotted), and decay of an initial off-equilibrium spin $1.5\chi|\boldsymbol{\Omega}_0|$ in steady rotation (dash-dotted). (b) Mode decay rates at $k=0.3\,\Gamma_{\mathrm{tr}}$, with $\mathbf{k}\perp\boldsymbol{\Omega}$: nonhydrodynamic spin modes (purple circles, size set by three-component response weights), axial-charge diffusion (green), and shear (brown). Vertical line: $2|\boldsymbol{\Omega}|=k/\sqrt3$, beyond which the incompressible model is extrapolated. Shading ($|\boldsymbol{\Omega}|>k$) marks the additional single-wavelength light-cylinder bound. (c) Fractional off-equilibrium spin carried by inertial waves at propagation angles $\theta$ to $\boldsymbol{\Omega}$ (shaded: heavy-ion regime; dotted: magnetization limit).}
 \label{fig:rotation}
\end{figure*}

Under steady rotation, a uniform off-equilibrium spin parallel to $\boldsymbol{\Omega}$ decays via \eqref{eq:spinrelease} [Fig.~\ref{fig:rotation}(a)]. When $\boldsymbol{\Omega}(t)$ varies, the magnetization part $2\chi\boldsymbol{\Omega}(t)$ responds instantaneously, whereas the helicity distribution relaxes toward $\Phi_{\mathrm{rot}}$. Solving \eqref{eq:kinetic} to first order in a slowly varying uniform vorticity gives \cite[Sec.~S4.2]{SM}
\begin{equation}
 \mathbf{s}(t)=3\chi\boldsymbol{\Omega}(t)-\chi\tau_\Omega\,\partial_t\boldsymbol{\Omega}(t)+\cdots,
 \label{eq:vortical}
\end{equation}
where $\tau_\Omega=(j,C^{-1}j_\Omega)/(j,j)\simeq1.03/\Gamma_{\mathrm{tr}}$. The lag time $\tau_\Omega/3$ enters as a second-order constitutive correction; at $\gamma_{\mathrm{tr}}\ll\omega\ll g^2T$, only the instantaneous magnetization contribution remains.

\section{Hydrodynamic coupling and rotating modes}
To leading order in gradients, relaxing spin couples to conserved momentum and axial charge. Linearizing first about a static background ($\mathbf{u}=0$) with incompressible velocity perturbations ($\nabla\cdot\delta\mathbf{u}=0$), we replace $\Phi_{\mathrm{rot}}$ in \eqref{eq:kinetic} by $\delta\Phi_{\mathrm{rot}}=\delta\Omega_ij_\Omega^i/T$, with $\delta\boldsymbol{\Omega}=\frac12\nabla\times\delta\mathbf{u}$. To retain the slowest possible longitudinal modes, we neglect axial-charge relaxation and couple \eqref{eq:kinetic} to $\partial_tJ_5^0+\nabla\cdot\mathbf{J}_5=0$. Meanwhile, the momentum density $T^{0i}$ includes the spin contribution $\frac12[\nabla\times(\mathbf{j}_\Omega,\Phi)]^i$ \cite{Chen15}; its free-streaming flux in the stress tensor $T^{ij}$ cancels the streaming part of $\partial_t[\frac12\nabla\times(\mathbf{j}_\Omega,\Phi)]$ \cite[Sec.~S4.2]{SM}, so transverse momentum conservation gives
\begin{equation}
 w\,\partial_t\delta\mathbf{u}=\eta\nabla^2\delta\mathbf{u}+\tfrac12\nabla\times\bigl(\mathbf{j}_\Omega,C(\Phi-\delta\Phi_{\mathrm{rot}})\bigr),
 \label{eq:momentum}
\end{equation}
where $w$ is the enthalpy density and $\eta$ the shear viscosity. The collisional torque is the Onsager-reciprocal partner of the vortical source in \eqref{eq:kinetic} \cite{H21}. Eliminating the spin-averaged shear distribution in favor of $\eta$, freely evolving perturbations with periodic or vanishing surface fluxes obey
\begin{equation}
\begin{gathered}
 \frac{\dd}{\dd t}\!\int\!\dd^3x\,\left[\frac{w(\delta\mathbf{u})^2}{2}+\frac{T}{2}(\Phi,\Phi)\right]\\
 =-\!\int\!\dd^3x\,\left[\eta(\partial_i\delta u_j)^2+TQ[\Phi-\delta\Phi_{\mathrm{rot}}]\right]\le0,
\end{gathered}
 \label{eq:dissipation}
\end{equation}
combining viscous shear heating and collisional spin dissipation. Together with the microscopic lower bound, \eqref{eq:dissipation} ensures that any mode decaying at $\Gamma\ll\gamma_{\mathrm{tr}}$ carries only a parametrically small off-equilibrium spin relative to the maximum allowed by its free energy \cite[Sec.~S4.3]{SM}.

After the initial collisional decay, small contributions to the local spin density continue to evolve with momentum and axial-charge transport. For a pure spin perturbation with $k\ll\gamma_{\mathrm{tr}}$ in a nonrotating background, the transverse part transfers angular momentum to the fluid via \eqref{eq:momentum}, generating shear flow. The resulting vorticity perturbation sustains the local-equilibrium spin $3\chi\delta\boldsymbol{\Omega}$, which decays at rate $\eta k^2/w$ and retains only a fraction $3\chi k^2/(4w)\simeq0.009(k/T)^2$ of the initial amplitude. The longitudinal part transports axial charge via $\mathbf{J}_5=2\mathbf{s}$, building a charge gradient. For a flavor-symmetric collision operator, this charge diffuses with constant $D=\tau/3$, carrying a diffusive spin density $\mathbf{s}=-\frac{D}{2}\nabla J_5^0$ opposite to the initial polarization with relative amplitude $-(k\tau)^2/3$ (unequal quark masses replace $D$ by a flavor diffusion matrix \cite[Sec.~S4.3]{SM}). In both hydrodynamic modes, the local spin density is fixed by the vorticity or axial-charge gradient [Fig.~\ref{fig:transportscales}(a)].

For perturbations about a rigidly rotating background $\mathbf{u}=\boldsymbol{\Omega}\times\mathbf{x}$, $\Phi$ denotes the deviation from rotational equilibrium. In the corotating frame, the Coriolis force gives $\dd\mathbf{p}/\dd t=-2\boldsymbol{\Omega}\times\mathbf{p}$, while the local collision operator is evaluated at $\boldsymbol{\Omega}=0$ to leading order for $|\boldsymbol{\Omega}|\ll g^2T$ \cite[Sec.~S5.1]{SM}. Because the Coriolis force rotates momenta and helicity-locked spins together, it commutes with $C$ and preserves the perturbation free energy; the fluid Coriolis force $2\boldsymbol{\Omega}\times\delta\mathbf{u}$ likewise does no work, so \eqref{eq:dissipation} still holds. Uniform spin parallel to $\boldsymbol{\Omega}$ relaxes at the static-medium rates, whereas transverse spin precesses at frequency $2|\boldsymbol{\Omega}|$ with the same damping rates.

For a perturbation with wave vector $\mathbf{k}$ at an angle $\theta$ to $\boldsymbol{\Omega}$, rotation converts shear modes into inertial waves with $\omega=\pm2|\boldsymbol{\Omega}|\cos\theta-\ii\eta k^2/w$ in the incompressible limit $2|\boldsymbol{\Omega}|\ll k/\sqrt3$ \cite{Greenspan68}. Their spin precesses relative to the wave vorticity and departs reversibly from $3\chi\delta\boldsymbol{\Omega}$ by an amount controlled by $|\boldsymbol{\Omega}|\tau_\Omega$ [Fig.~\ref{fig:rotation}(c)]. Axial-charge diffusion perpendicular to $\boldsymbol{\Omega}$ ($D_\perp$) is suppressed once $2|\boldsymbol{\Omega}|\gtrsim\Gamma_{\mathrm{tr}}$ [Fig.~\ref{fig:rotation}(b)]. Within this incompressible moment truncation at $k=0.3\,\Gamma_{\mathrm{tr}}$ and $|\boldsymbol{\Omega}|\le3\,\Gamma_{\mathrm{tr}}$, slow modes carry less than $0.9\%$ of the initial spin amplitude across all three Cartesian components, while all nonhydrodynamic spin relaxation rates exceed $0.73\,\Gamma_{\mathrm{tr}}$ \cite[Sec.~S5.2]{SM}. Causality requires $|\boldsymbol{\Omega}|R<1$ for transverse radius $R$ \cite{AW14,CG17}, and subluminal flow bounds vorticity by the inverse gradient length. An order-unity spin lag ($|\boldsymbol{\Omega}|\gtrsim\gamma_{\mathrm{tr}}$) thus requires a system size or gradient length comparable to the mean free path, outside the hydrodynamic regime.

\section{Discussion and conclusion}
In weakly coupled hot QCD, helicity conservation does not protect spatial spin: small-angle momentum diffusion relaxes light-quark polarization on the transport time scale, nearly as fast as a vector current, and rotation generates no slow spin mode. At $\omega,k\ll\gamma_{\mathrm{tr}}$ and causal rotation rates, spin follows the fluid vorticity and axial-charge gradient through constitutive relations without an independent hydrodynamic degree of freedom, with $\tau_\Omega$ supplying a second-order correction \cite{Li21}. While a nonlocal vortical response also arises in RTA axial kinetic theory \cite{LinTang2024}, \eqref{eq:vortical} gives the leading gradient expansion for the full collision operator. Since no slow mode with finite spin weight separates from the other kinetic modes, nonlocality sets in only at $\omega,k\sim\gamma_{\mathrm{tr}}$ and leaves the ordinary hydrodynamic gradient expansion intact.

For heavy-ion collisions, combining the weak-coupling ratio $\tau w/\eta\simeq7.6$ with $\eta T/w=0.08$--$0.2$ \cite{HS13} and $|\boldsymbol{\Omega}|/T\simeq10^{-3}$--$5\times10^{-2}$ \cite{STAR17,STAR18,Bec15} gives $|\boldsymbol{\Omega}|\tau\simeq10^{-3}$--$0.08$, well within the slow-rotation regime \cite[Sec.~S5.3]{SM}. Microscopic rotation also modifies the self-energies and scattering amplitudes entering $C$, but for ultrarelativistic quarks away from phase transitions or resonances, smooth rotational modifications should affect spin and momentum relaxation comparably because momentum deflections simultaneously rotate the helicity-locked spin \cite[Sec.~S5.4]{SM}.

For heavy quarks ($M\gg T$), chromomagnetic spin flips are suppressed by $(T/M)^2$, making spin relaxation slower than momentum relaxation by $M/(2T)$ \cite{H22}, while spin rotation from stochastic Thomas precession per momentum deflection is suppressed by $T/M$ \cite{KDDJ26}. For light quarks at stronger coupling, frequent large-angle scatterings and broad quasiparticle widths should erase off-equilibrium polarization even faster, further disfavoring an independent spin hydrodynamic mode in the strongly coupled QGP. Transient spin dynamics at kinetic scales nonetheless involve multiple transport-scale modes [Fig.~\ref{fig:finitebasis}(b)], whose description requires going beyond both spin hydrodynamics and a single relaxation time.

\begin{acknowledgments}
This work was supported in part by the National Natural Science Foundation of China (NSFC) Grant Nos. 12235016 and 12221005. Large language models (OpenAI GPT and Anthropic Claude) assisted with symbolic cross-checks, numerical code, figure preparation, and text editing. The authors take full responsibility for the content of this work.
\end{acknowledgments}

\bibliographystyle{apsrev4-2}
\bibliography{references}
\end{document}


\maketitle

This Supplemental Material presents the theoretical derivations, the polarized collision operator, spectral bounds, and numerical analyses supporting the results reported in the Letter. It is organized into five sections:
\begin{itemize}
\item \textbf{Section~\ref{sec:spin-supp}} defines the canonical QCD spin operator in the helicity basis, describes the parallel transport of polarization density matrices via the momentum-space Berry connection, and constructs multipole bases for the parity-even $J=1$ axial-vector and $J=2$ tensor channels.
\item \textbf{Section~\ref{sec:collision-supp}} constructs the complete leading-order polarized collision operator in the effective kinetic theory of Arnold, Moore, and Yaffe (AMY) \cite{AMY03a,AMY03b}, incorporating tree-level hard $2\leftrightarrow2$ scattering, hard-thermal-loop (HTL) resummation of soft exchange, matching across the hard--soft separation scale $q_*$, and Landau--Pomeranchuk--Migdal (LPM) resummation of collinear $1\leftrightarrow2$ splitting. Analysis of the $J=2$ sector shows that gluon linear polarization decouples from the leading-order shear viscosity.
\item \textbf{Section~\ref{sec:spectral-supp}} establishes the spectral representation of spin relaxation. Small-angle quark--gluon scattering excludes finite spectral weight at rates parametrically below $\gamma_{\mathrm{tr}}\equiv g^4T\ln(1/g)$, while on-shell soft collective modes yield a parametrically small contribution to the spin susceptibility. This section also presents the variational B-spline spectrum and convergence tests in a static plasma.
\item \textbf{Section~\ref{sec:rotation-supp}} characterizes local rotational equilibrium, including the $1:2$ partition of the axial vortical effect between polarized quasiparticles and the chiral magnetization current. Angular-momentum conservation in nonlocal collisions yields the vortical source and the Onsager-reciprocal mechanical torque on the fluid, which together bound the off-equilibrium spin carried by hydrodynamic modes.
\item \textbf{Section~\ref{sec:rotframe-supp}} extends the coupled fluid--kinetic equations to a corotating background at arbitrary $|\bm\Omega|/\gamma_{\mathrm{tr}}$, analyzes inertial waves and anisotropic axial-charge diffusion, presents numerical solutions retaining $J=1$ polarization and $J=0$ axial charge, estimates rotation and relaxation scales in relativistic heavy-ion collisions, and bounds rotational corrections to the collision operator.
\end{itemize}
Equation numbers refer to this Supplemental Material unless the Letter is specified. Unless stated otherwise in the tables, relaxation rates are expressed in units of $\Gamma_{\mathrm{tr}}$, the effective relaxation rate of the strange-quark vector current evaluated with $N_b=40$ basis functions per radial channel for the reference parameter set~\eqref{eq:finitebasisconfig} (Tables~\ref{tab:finitebasis} and \ref{tab:masspoints} normalize rates by the strange-current rate at the corresponding basis size or quark mass).

\section{Canonical Spin Density, Helicity Basis, and Multipole Decomposition}
\label{sec:spin-supp}

\subsection{Canonical spin tensor and light-quark regime in hot QCD}
We consider small perturbations about a homogeneous, isotropic, unpolarized thermal equilibrium in $\mathrm{SU}(3)$ QCD at temperature $T$, weak gauge coupling $g\ll1$, and zero chemical potential, with three light quark flavors $f\in\{u,d,s\}$ of real diagonal current masses $m_{0f}$. Throughout, we use natural units ($\hbar=c=k_B=1$) with Minkowski metric $\eta_{ab}=\operatorname{diag}(+,-,-,-)$, Levi-Civita convention $\epsilon^{0123}=+1$, chiral matrix $\gamma_5=\ii\gamma^0\gamma^1\gamma^2\gamma^3$, and Fourier convention $e^{-\ii\omega t+\ii\mathbf k\cdot\mathbf x}$.

In relativistic spin hydrodynamics \cite{FFJS18,HHHMT19,BFS19,H21,Li21,WW22,BH26}, the decomposition of total angular momentum into orbital and intrinsic spin parts depends on the choice of pseudogauge. To define the canonical spin tensor microscopically, we couple the QCD action $I$ to a background tetrad $e_\mu{}^a$ with determinant $e=\det(e_\mu{}^a)$ and spin connection $\omega_{\mu ab}$. Varying $I$ with respect to $\omega_{\mu ab}$ at fixed $e_\mu{}^a$ defines the canonical spin tensor via $\delta I=\tfrac12\int\dd^4x\,e\,S^{\lambda,ab}\delta\omega_{\lambda ab}$ \cite{Hehl76,Weinberg95,H21}. For Dirac quarks with covariant derivative $D_\mu=\partial_\mu-\ii gA_\mu+\tfrac18\omega_{\mu ab}[\gamma^a,\gamma^b]$ (where $A_\mu$ is the gluon gauge field), the anticommutation relation $\{\gamma^\lambda,\tfrac{\ii}{4}[\gamma^a,\gamma^b]\}=-\epsilon^{\lambda ab\rho}\gamma_\rho\gamma_5$ yields a totally antisymmetric canonical quark spin tensor and reproduces (1) of the Letter:
\begin{equation}
 S^{\lambda,ab}=-\epsilon^{\lambda ab\rho}J_{5\rho}/2,
 \qquad s^i\equiv\tfrac12\epsilon^{ijk}S^{0,jk}=J_5^i/2,\qquad S^{0,0i}=0,
 \label{eq:triplet}
\end{equation}
where $J_5^\mu=\sum_f\bar q_f\gamma^\mu\gamma_5q_f$ is the flavor-singlet axial current. By contrast, the Yang--Mills field strength $F_{\mu\nu}=\partial_\mu A_\nu-\partial_\nu A_\mu-\ii g[A_\mu,A_\nu]$ is a spacetime exterior derivative requiring no affine or spin connection, so the Yang--Mills action depends on the tetrad solely through the metric $g_{\mu\nu}=e_\mu{}^a e_\nu{}^b\eta_{ab}$ \cite{Hehl76,LL14,H21}. Consequently, gluons do not contribute to the gauge-invariant canonical spin tensor $S^{\lambda,ab}$. Instead, as shown in Sec.~\ref{sec:rotation-supp}, gluon circular polarization couples dynamically to quark spin through the collision operator and enters total angular-momentum conservation via the symmetric energy-momentum tensor.

To evaluate the one-particle matrix elements of $s^i$ and the polarized scattering amplitudes, we express the Dirac spinors for momentum $\mathbf p$ ($p\equiv|\mathbf p|$), mass $m$, and energy $E=\sqrt{p^2+m^2}$ in a fixed Pauli spin basis \cite{Weinberg95,AMY03a},
\begin{equation}
\begin{aligned}
 U(\mathbf p,m)&=\sqrt{E+m}
 \begin{pmatrix}\one_2\\\bm\sigma\cdot\mathbf p/(E+m)\end{pmatrix},
 \\
 V(\mathbf p,m)&=\sqrt{E+m}
 \begin{pmatrix}(\bm\sigma\cdot\mathbf p)\ii\sigma_2/(E+m)\\\ii\sigma_2\end{pmatrix},
\end{aligned}
\qquad \qquad
 U^\dagger U=V^\dagger V=2E\one_2,
 \label{eq:spinors}
\end{equation}
where $\bm\sigma=(\sigma_1,\sigma_2,\sigma_3)$ are the standard Pauli matrices. For a quark of flavor $f$ with energy $E_p=\sqrt{p^2+m_{0f}^2}$ and momentum direction $\mathbf n=\mathbf p/p$, the one-particle matrix element of the canonical spin density, $j_f^i=U^\dagger\gamma^0\gamma^i\gamma_5U/(4E_p)$, takes the form
\begin{equation}
 j_f^i=\frac12n_i(\mathbf n\cdot\bm\sigma)+\frac{m_{0f}}{2E_p}\bigl[\sigma_i-n_i(\mathbf n\cdot\bm\sigma)\bigr].
 \label{eq:vertex}
\end{equation}
For any local Dirac bilinear $\bar\psi O\psi$, normal ordering of the quantized field yields the one-particle antiquark matrix element $-[V^\dagger\gamma^0 O V]^{\mathsf{T}}/(2E_p)$. Because $O=\tfrac12\gamma^i\gamma_5$ is charge-conjugation even, this operation reproduces \eqref{eq:vertex} in the antiquark spinor basis with the same sign as for quarks. In the helicity basis (where $\mathbf n\cdot\bm\sigma\to\sigma_3$), the first term in \eqref{eq:vertex} reduces to $\tfrac12n_i\sigma_3$ and projects the spin onto $\mathbf n$, whereas the off-diagonal second term represents transverse spin coherence and is suppressed at hard momenta by $m_{0f}/(2E_p)$.

Our weak-coupling analysis is performed in the light-quark mass regime defined by the power counting
\begin{equation}
 m_{0f}=\kappa_f gT,\qquad \kappa_f=\cO(1),\qquad g\to0,
 \label{eq:lightmass}
\end{equation}
which smoothly encompasses the chiral limit ($\kappa_f=0$). At leading order in $g$, hard quasiparticles carry thermal momenta $p\sim T$ with ultrarelativistic dispersion relations and equilibrium Fermi--Dirac or Bose--Einstein distributions
\begin{equation}
 E=p,\qquad f_F^{(0)}(p)=(e^{p/T}+1)^{-1},\quad
 f_B^{(0)}(p)=(e^{p/T}-1)^{-1},
 \label{eq:hardenergy}
\end{equation}
with $\bff_F^{(0)}=1-f_F^{(0)}$ and $\bff_B^{(0)}=1+f_B^{(0)}$ accounting for Pauli blocking and Bose enhancement. Hard $2\leftrightarrow2$ scattering amplitudes can therefore be evaluated in the massless limit $m=0$ in \eqref{eq:spinors} \cite{AMY03a}. Soft momentum transfers $q\sim gT$, by contrast, require HTL resummation \cite{Weldon82a,Weldon82b,BP90,FT90,BI02}. For $\mathrm{SU}(3)$ QCD with $N_f=3$ light flavors, the Debye screening mass $m_D$, the asymptotic hard thermal masses $m_{\infty,g}$ and $m_{\infty,f}$ (governing collinear phase space at $p\sim T$ \cite{AMY01,AMY02}), and the soft-fermion HTL thermal mass $m_{\mathrm{HTL},f}$ are
\begin{equation*}
 m_D^2=\frac{3}{2}g^2T^2,\qquad
 m_{\infty,g}^2=\frac{3}{4}g^2T^2,\qquad
 m_{\infty,f}^2=\frac{1}{3}g^2T^2,\qquad
 m_{\mathrm{HTL},f}^2=\frac{1}{6}g^2T^2.
\end{equation*}
Because gauge interactions preserve chirality, the covariant HTL Dirac self-energy anticommutes with $\gamma_5$ \cite{Weldon82b,LY19,Lin26}. Thermal self-energies thus modify the quasiparticle dispersion relations and soft propagators without introducing chirality-flipping terms in the propagator numerators; the latter arise solely from the current quark mass $m_{0f}$ in \eqref{eq:lightmass}. This distinction allows thermal and current masses to enter at the same soft scale $\cO(gT)$ while playing distinct roles in spin relaxation.

\subsection{Helicity frames, momentum-space Berry connection, and thermal inner product}
For on-shell gluons with four-momentum $P=(p,\mathbf p)$ and $\mathbf p=p\mathbf n$, the two physical circular-polarization states $\lambda=\pm1$ are defined relative to a local right-handed orthonormal triad $(\mathbf e_1(\mathbf n),\mathbf e_2(\mathbf n),\mathbf n)$,
\begin{equation}
 \epsilon_\lambda(P)=\left(0,\frac{\mathbf e_1+\ii\lambda\mathbf e_2}{\sqrt2}\right),
 \quad \mathbf e_2=\mathbf n\times\mathbf e_1,
 \quad \epsilon_\lambda^*\cdot\epsilon_{\lambda'}=-\delta_{\lambda\lambda'},
 \label{eq:polframe}
\end{equation}
with outgoing polarization vectors complex conjugated. In the helicity representation, we order the two quark helicity states as $(+,-)$ corresponding to $\lambda_q=\pm1$ (physical helicity $\lambda_q/2$), and the two gluon circular-polarization states as $(-,+)$ corresponding to $\lambda=\pm1$, consistent with the amplitude convention in which an outgoing gluon of helicity $\lambda$ carries $\epsilon_\lambda^*=\epsilon_{-\lambda}$. A momentum-dependent unitary transformation from the fixed Pauli basis~\eqref{eq:spinors} to the quark helicity basis brings the quark helicity operator $\mathbf n\cdot\bm\sigma$ to $\operatorname{diag}(+1,-1)$, while the gluon helicity operator in the circular-polarization basis reads $\operatorname{diag}(-1,+1)$. Under this transformation of the spinors and polarization density matrices, the momentum-space Berry connection acquires the momentum derivative of the unitary matrix, ensuring covariant differentiation of the polarization matrix and basis-independent physical traces. Below, we express the quark angular matrices in the fixed Pauli basis and the gluon matrices in the circular-polarization basis.

Both the transverse triad $(\mathbf e_1(\mathbf n),\mathbf e_2(\mathbf n))$ and the two-component helicity spinors depend on the momentum direction $\mathbf n$. A local rotation of $(\mathbf e_1,\mathbf e_2)$ about $\mathbf n$ by an angle $\alpha(\mathbf n)$ multiplies the helicity states by phases $e^{-\ii s\alpha(\mathbf n)}$ (with physical helicity $s=\pm\tfrac12$ for quarks and $s=\pm1$ for gluons). When a hard parton undergoes a soft momentum transfer $\mathbf q$ ($q\ll p$) that deflects $\mathbf n$, comparing polarization density matrices at $\mathbf p$ and $\mathbf p+\mathbf q$ requires parallel transport in momentum space via the Berry connection \cite{Berry84,SY12,StY12,SY13,CSSYY14,SDZ15_PRD,SDZ15,Lin26}. As detailed in Sec.~\ref{app:soft}, this parallel transport enters the soft-gluon-exchange collision kernel through the gauge-covariant momentum-space derivative $D_{\mathbf q}\Phi=\mathbf q\cdot\nabla_{\mathbf p}\Phi+\ii[\Phi,B^0]$, where $B^0$ denotes the Berry connection contracted with $\mathbf q$, guaranteeing that all physical collision rates are invariant under local rotations of the transverse reference frame.

In linear response about thermal equilibrium, the $2\times2$ polarization density matrix of parton species $a\in\{q_f,\bar q_f,g\}$ ($f\in\{u,d,s\}$) is parameterized as $f_a(\mathbf p)=f_a^{(0)}(p)\one_2+\delta f_a(\mathbf p)$ with
\begin{equation*}
 \delta f_a(\mathbf p)=f_a^{(0)}(p)\bff_a^{(0)}(p)\Phi_a(\mathbf p),
\end{equation*}
where Hermiticity in the time domain requires $\Phi_a(\omega,\mathbf k)^\dagger=\Phi_a(-\omega,-\mathbf k)$ in Fourier space. The natural thermal inner product on linearized perturbations $\Phi,\Psi$ is induced by the second variation of the equilibrium entropy (the inverse static susceptibility matrix) \cite{AMY03a,AMY03b,HT10},
\begin{equation*}
 (\Phi,\Psi)=\sum_a d_{c,a}\intp f_a^{(0)}(p)\bff_a^{(0)}(p)\tr_2\bigl[\Phi_a^\dagger(\mathbf p)\Psi_a(\mathbf p)\bigr],
\end{equation*}
where $\intp\equiv\int\dd^3p/(2\pi)^3$, $d_{c,q}=d_{c,\bar q}=3$ and $d_{c,g}=8$ are the color multiplicities, and $\tr_2$ denotes the trace over the two physical polarization states. At zeroth order in spatial gradients, (3) of the Letter expresses the quasiparticle contribution to the canonical spin density as the thermal inner product with the spin source $j^i$,
\begin{equation}
 \delta s^i=\sum_a d_{c,a}\intp\tr_2(j_a^i\delta f_a)
 =(j^i,\Phi),
 \label{eq:spinreadout}
\end{equation}
where $j_g^i=0$ because the canonical spin operator~\eqref{eq:triplet} contains no gluon contribution.

\subsection{Multipole decomposition in the \texorpdfstring{$J=1$}{J=1} axial and \texorpdfstring{$J=2$}{J=2} tensor channels}
In an isotropic thermal medium at zero wave number ($\mathbf k=0$), the linearized collision operator $C$ commutes with simultaneous $\mathrm{SO}(3)$ rotations of the particle momentum $\mathbf p$ and spin, as well as with parity and charge conjugation \cite{AMY03a,AMY03b,Lin26}. Linearized perturbations therefore decompose into mutually orthogonal channels labeled by total angular momentum $(J,M)$, parity, and charge conjugation.

The canonical spin density $s^i=\tfrac12J_5^i$ is parity-even, charge-conjugation-even, and transforms as an axial vector ($J=1$). With quark operators written in the fixed Pauli basis and gluon operators in the circular-polarization basis, the parity-even axial $J=1$ sector is spanned by three dipole angular matrices corresponding to longitudinal quark spin (helicity) $B_L^i$, transverse quark spin coherence $B_T^i$, and gluon circular polarization $B_V^i$:
\begin{equation}
 B_L^i=\frac{\sqrt3}{2}n_i(\mathbf n\cdot\bm\sigma),\quad
 B_T^i=\frac{\sqrt{3/2}}2\bigl[\sigma_i-n_i(\mathbf n\cdot\bm\sigma)\bigr],\quad
 B_V^i=\frac{\sqrt3}{2}n_i\operatorname{diag}(-1,+1),
 \label{eq:j1basis}
\end{equation}
where $B_L^i$ and $B_T^i$ act on quark spin indices and $B_V^i$ acts on gluon polarization indices. Acting on the direct sum of quark and gluon polarization spaces, these basis matrices satisfy the trace-orthonormality relation
\begin{equation}
 2\avg{\tr_2(B_A^{i\dagger}B_B^j)}=\delta_{AB}\delta_{ij}\qquad(A,B\in\{L,T,V\})
 \label{eq:j1gram}
\end{equation}
under the normalized spherical average $\avg{\cdots}\equiv\int\frac{\dd\Omega_{\mathbf n}}{4\pi}(\cdots)$. Forming the orthonormal charge-conjugation-even quark combinations $\phi_f^+=(\phi_{q_f}+\phi_{\bar q_f})/\sqrt2$, any charge-conjugation-even axial $J=1$ perturbation is represented by a seven-component vector of radial profiles of $p$, ordered as the three longitudinal quark spin profiles, the gluon circular polarization, and the three transverse quark spin coherences:
\begin{equation}
 \phi(p)=\bigl(\phi_{uL}^+(p),\phi_{dL}^+(p),\phi_{sL}^+(p),\phi_{gV}(p),\phi_{uT}^+(p),\phi_{dT}^+(p),\phi_{sT}^+(p)\bigr)^{\mathsf{T}}.
 \label{eq:j1state}
\end{equation}
In this radial representation, the thermal inner product and the leading-order canonical spin source $j$ (along any fixed Cartesian axis $i$) take the form
\begin{align}
 (\phi,\psi)&=\sum_A\int_0^\infty\!\dd p\,\rho_A^{(1)}(p)\phi_A^*(p)\psi_A(p),
 &\rho_A^{(1)}(p)&=\frac{d_{c,A}p^2}{4\pi^2}f_A^{(0)}(p)\bff_A^{(0)}(p),
 \label{eq:j1radialmetric}\\
 j&=(\sqrt{2/3},\sqrt{2/3},\sqrt{2/3},0,0,0,0)^{\mathsf{T}},
 &(j,j)&=T^3/4.
 \label{eq:sourcenorm}
\end{align}
The normalization $(j,j)=T^3/4$ follows directly from \eqref{eq:j1gram} and the Fermi integral
\begin{equation}
 \int_0^\infty\!\dd p\,p^2\frac{e^{p/T}}{(e^{p/T}+1)^2}=\frac{\pi^2T^3}{6}.
 \label{eq:fermionintegral}
\end{equation}
Each quark flavor enters $j$ with coefficient $\sqrt{2/3}$ (incorporating the factor $\sqrt2$ from combining quarks and antiquarks into $\phi_{fL}^+$) and contributes $T^3/12$ to $(j,j)$, yielding the quasiparticle spin susceptibility $\chi\equiv(j,j)/T=T^2/4$. The transverse-spin piece of the vertex~\eqref{eq:vertex} is suppressed at hard momenta by $m_{0f}/(2E_p)=\cO(g)$ and contributes to $(j,j)$ only at $\cO(g^2T^3)$. For comparison, the spatial vector current $J^i=\bar q_s\gamma^i q_s$ of a single quark flavor resides in the charge-conjugation-odd $J=1$ vector channel with source coefficient $2\sqrt{2/3}$ and normalization $(j_J,j_J)=T^3/3$ ($\chi_J=T^2/3$).

When evaluating explicit collision integrals, it is convenient to absorb the angular normalization factors into unnormalized radial coefficients:
\begin{equation*}
\begin{aligned}
 \Phi_{q_f}^i(\mathbf p)=\Phi_{\bar q_f}^i(\mathbf p)
 &=\ell_f(p)n_i(\mathbf n\cdot\bm\sigma)
  +\ell_{fT}(p)\bigl[\sigma_i-n_i(\mathbf n\cdot\bm\sigma)\bigr],\\
 \Phi_g^i(\mathbf p)&=\ell_g(p)n_i\operatorname{diag}(-1,+1),
\end{aligned}
\end{equation*}
where $\ell_f=\ell_q=\ell_{\bar q}$ and $\ell_{fT}=\ell_{qT}=\ell_{\bar qT}$ for flavor $f$ in the charge-conjugation-even sector. These longitudinal quark, transverse quark, and circular gluon polarization coefficients are related to the normalized radial profiles~\eqref{eq:j1state} via
\begin{equation}
 \ell_q=\ell_{\bar q}=\sqrt{3/8}\,\phi_{fL}^+,\qquad
 \ell_{qT}=\ell_{\bar qT}=\frac{\sqrt3}{4}\,\phi_{fT}^+,\qquad
 \ell_g=\frac{\sqrt3}{2}\,\phi_{gV}.
 \label{eq:unnormalized}
\end{equation}
By rotational invariance, the radial $J=1$ collision matrix elements follow from averaging the full collision matrix elements $Q[\Phi,\Psi]\equiv(\Phi,C\Psi)$ over the three Cartesian directions,
\begin{equation}
 (\phi,C_{J=1}\psi)=\frac13\sum_{i=1}^3Q\bigl[\Phi^i[\phi],\Phi^i[\psi]\bigr],
 \label{eq:j1projection}
\end{equation}
where $\Phi^i[\phi]$ denotes the polarization density matrices reconstructed above using \eqref{eq:unnormalized}. As shown in Sec.~\ref{sec:collision-supp}, the four longitudinal quark and circular gluon polarization profiles $(\phi_{uL}^+,\phi_{dL}^+,\phi_{sL}^+,\phi_{gV})$ decouple completely at leading order from the three transverse quark spin profiles $(\phi_{uT}^+,\phi_{dT}^+,\phi_{sT}^+)$.

Beyond circular polarization in the $J=1$ channel, on-shell gluons also admit linear polarization---the coherence between opposite circular-polarization states $\lambda=+1$ and $\lambda=-1$ ($\Delta\lambda=\pm2$)---which transforms as a traceless symmetric transverse tensor and first enters in the parity-even $J=2$ channel \cite{Lin26}. Because the spin-averaged quadrupole deformation of the momentum distribution that governs the shear viscosity $\eta$ also resides in the parity-even $J=2$ sector \cite{AMY03a,AMY03b,GMT18}, we examine whether collisions mix these two subsectors and modify the leading-order shear viscosity of hot QCD. To treat both subsectors on an equal footing, let $H_A$ ($A=1,\dots,5$) be an orthonormal basis of real symmetric traceless $3\times3$ matrices satisfying $\tr(H_AH_B)=\delta_{AB}$ and the completeness relation $\sum_{A=1}^5H_A^{ij}H_A^{kl}=\tfrac12(\delta^{ik}\delta^{jl}+\delta^{il}\delta^{jk})-\tfrac13\delta^{ij}\delta^{kl}$. The spin-independent shear ($S$) and traceless gluon linear-polarization ($L$) angular basis matrices are defined as
\begin{equation}
 Y_A^S(\mathbf n)=\frac{\sqrt{15}}2(n_i H_A^{ij} n_j)\one_2,\qquad
 Y_A^L(\mathbf n)=\sqrt{5/2}\left[\epsilon^\dagger H_A\epsilon-\tfrac12\tr_2(\epsilon^\dagger H_A\epsilon)\one_2\right],
 \label{eq:j2basis}
\end{equation}
where $\epsilon=(\epsilon_-,\epsilon_+)$ is the $3\times2$ matrix formed by the spatial parts of the gluon polarization vectors~\eqref{eq:polframe}. These basis matrices satisfy $\avg{\tr_2[Y_A^{X\dagger}Y_B^Y]}=\delta_{AB}\delta_{XY}$ for $X,Y\in\{S,L\}$. Working in the charge-conjugation-even basis as in the $J=1$ channel, we reconstruct the perturbation $\Phi_A[\phi]$ for each tensor component $A$ from the radial profiles as $\Phi_{A,q_f}=\Phi_{A,\bar q_f}=\phi_{S,f}Y_A^S/\sqrt2$ and $\Phi_{A,g}=\phi_{S,g}Y_A^S+\phi_{L,g}Y_A^L$. The five independent radial profiles comprise the three charge-conjugation-even quark combinations $\phi_{S,f}$ and the two gluon profiles $(\phi_{S,g},\phi_{L,g})$, with radial weights and shear sources
\begin{equation}
\begin{gathered}
 \rho_a^{(2)}(p)=\frac{d_{c,a}p^2f_a^{(0)}(p)\bff_a^{(0)}(p)}{2\pi^2},
 \quad a\in\{u,d,s,g\},\\
 \pi_{S,f}(p)=\frac{2\sqrt2p}{\sqrt{15}},\qquad
 \pi_{S,g}(p)=\frac{2p}{\sqrt{15}},\qquad \pi_L(p)=0.
\end{gathered}
 \label{eq:j2source}
\end{equation}
Here both gluon profiles carry the radial weight $\rho_g^{(2)}$, while the factor $1/\sqrt2$ in the quark and antiquark perturbations preserves their combined radial weight $\rho_f^{(2)}$ and rescales the quark shear source by $\sqrt2$. In Sec.~\ref{app:j2}, we construct the collision operator in the $J=2$ sector and prove that the off-diagonal blocks $C_{SL}^{J=2}$ and $C_{LS}^{J=2}$ vanish identically across all four leading-order collision processes.

\section{Polarized Collision Operator in Effective Kinetic Theory}
\label{sec:collision-supp}

The leading-order effective kinetic theory combines hard $2\leftrightarrow2$ scattering, screened soft exchange, and collinear $1\leftrightarrow2$ splitting \cite{AMY03a,AMY03b}. Its polarized collision quadratic form decomposes as
\begin{equation*}
 Q[\Phi,\Psi]=Q_H[\Phi,\Psi]+Q_g[\Phi,\Psi]+Q_f[\Phi,\Psi]+Q_{12}[\Phi,\Psi],
\end{equation*}
where $Q_H$ describes hard $2\leftrightarrow2$ scattering, $Q_g$ and $Q_f$ describe HTL-resummed soft-gluon and soft-fermion exchange, and $Q_{12}$ describes LPM-resummed $1\leftrightarrow2$ splitting and merging. Each functional is real, symmetric, and positive semidefinite on Hermitian perturbations (and sesquilinear in complex Fourier amplitudes). We separate hard and soft three-momentum transfers at an intermediate cutoff scale $gT\ll q_*\ll T$ \cite{BMPR90,BT91a,BT91b}, following (4.2)--(4.5) of Ref.~\cite{AMY03b}. Below we retain the full polarization structure and the current quark masses of \eqref{eq:lightmass}, establish cutoff independence across $q_*$, and evaluate the projections onto the $J=1$ and $J=2$ channels.

\subsection{Hard \texorpdfstring{$2\leftrightarrow2$}{2 <-> 2} scattering amplitudes and collision functional}
\label{app:hard}
The 34 distinct flavor-resolved $2\leftrightarrow2$ reaction channels, each counted once, comprise elastic $qq$, $\bar q\bar q$, $q\bar q$, $qg$, $\bar qg$, and $gg$ scattering, as well as annihilation $q\bar q\leftrightarrow gg$ and $q_f\bar q_f\leftrightarrow q_h\bar q_h$ ($f\ne h$). Detailed balance implies $F_{1234}\equiv f_1^{(0)}f_2^{(0)}\bff_3^{(0)}\bff_4^{(0)}=F_{3412}$. We treat the tree-level amplitudes $A_r$ (stripped of $g^2$ and color factors) as operators mapping incoming to outgoing two-particle helicity states, weighted by the color factor matrix $K_{rs}=\sum_{\mathrm{colors}}c_r^*c_s$. The linearized collision variation, identical-particle symmetry factor, and Lorentz-invariant phase-space measure are
\begin{align}
 U_\Phi[A]&=(\Phi_3\otimes\one+\one\otimes\Phi_4)A-A(\Phi_1\otimes\one+\one\otimes\Phi_2),\nonumber\\
 \nu_{ab;cd}&=[(1+\delta_{ab})(1+\delta_{cd})(1+\delta_{\{a,b\},\{c,d\}})]^{-1},\nonumber\\
 \dd\Pi_{1234}&=\prod_{r=1}^4\frac{\dd^3p_r}{(2\pi)^3\,2p_r}(2\pi)^4\delta^{(4)}(P_1+P_2-P_3-P_4).
 \label{eq:hardmeasure}
\end{align}
For unpolarized perturbations $\Phi_r\propto\one_2$, $U_\Phi[A]$ reduces to $(\Phi_3+\Phi_4-\Phi_1-\Phi_2)A$, as in (2.22) of Ref.~\cite{AMY03b}. Retaining open polarization indices yields the hard collision functional
\begin{equation}
 Q_H[\Phi,\Psi]=g^4\sum_R\nu_R\int\dd\Pi_{1234}F_{1234}\Theta_H
           \operatorname{Re}\sum_{rs}K_{rs}\tr\bigl(U_\Phi[A_r]^\dagger U_\Psi[A_s]\bigr),
 \label{eq:hardform}
\end{equation}
where the phase-space step function $\Theta_H$ restricts all $t$- and $u$-channel spatial momentum transfers to exceed $q_*$.

We use massless hard kinematics with Mandelstam variables $s=(P_1+P_2)^2$, $t=(P_1-P_3)^2$, and $u=(P_1-P_4)^2$, and generator normalization $\tr(t^At^B)=\delta^{AB}/2$. The $q_fq_h\to q_fq_h$ helicity amplitudes read
\begin{equation}
 (A_t)_{cd;ab}=\frac{(\bar u_{3c}\gamma^\mu u_{1a})(\bar u_{4d}\gamma_\mu u_{2b})}{t},
 \qquad
 (A_u)_{cd;ab}=-\frac{(\bar u_{4d}\gamma^\mu u_{1a})(\bar u_{3c}\gamma_\mu u_{2b})}{u},
 \label{eq:qqamplitude}
\end{equation}
where the $u$-channel exchange term applies for identical flavors ($f=h$). Antiquark--antiquark scattering is obtained by replacing the spinor currents with $(\bar v_1\gamma^\mu v_3)(\bar v_2\gamma_\mu v_4)$ and $-(\bar v_1\gamma^\mu v_4)(\bar v_2\gamma_\mu v_3)$ under the same $(cd;ab)$ index ordering. For $q_f\bar q_h\to q_f\bar q_h$,
\begin{equation}
 (A_t)_{cd;ab}=\frac{(\bar u_{3c}\gamma^\mu u_{1a})(\bar v_{2b}\gamma_\mu v_{4d})}{t},
 \quad
 (A_s)_{cd;ab}=-\frac{(\bar v_{2b}\gamma^\mu u_{1a})(\bar u_{3c}\gamma_\mu v_{4d})}{s}
 \quad(f=h),
 \label{eq:qqbaramplitude}
\end{equation}
while flavor-changing annihilation ($f\ne h$) proceeds solely via the $s$ channel. A single four-fermion diagram carries color weight $C_Fd_{c,q}/2=2$, whereas two interfering channels have the color matrix
\begin{equation}
 K_{ff}=\begin{pmatrix}2&-2/3\\-2/3&2\end{pmatrix},
 \label{eq:ffcolor}
\end{equation}
since the fermionic exchange sign is already absorbed into $A_u$ and $A_s$.

With all momenta defined as incoming and $k_o=-k_i-k_j$, we write the three-gluon vertex factor as
\begin{equation}
 \mathcal J^\mu(k_i,k_j;e_i,e_j)
 =(e_i\cdot e_j)(k_i-k_j)^\mu
 +e_j^\mu(k_j-k_o)\cdot e_i+e_i^\mu(k_o-k_i)\cdot e_j.
 \label{eq:threecurrent}
\end{equation}
For quark--gluon scattering $q(P_1)g_a(P_2)\to q(P_3)g_b(P_4)$, the channel amplitudes are
\begin{align}
 A_s=\bar u_3\slashed\epsilon_4^*\frac{\slashed P_1+\slashed P_2}{s}\slashed\epsilon_2u_1,
 \quad
 A_u=\bar u_3\slashed\epsilon_2\frac{\slashed P_1-\slashed P_4}{u}\slashed\epsilon_4^*u_1,
 \quad
 A_t=\bar u_3\slashed{\mathcal J}(P_2,-P_4;\epsilon_2,\epsilon_4^*)u_1/t,
 \label{eq:qgamplitude}
\end{align}
and the associated color structures $(t^bt^a,t^at^b,[t^a,t^b])$ yield
\begin{equation}
 K_{qg}=\begin{pmatrix}
 16/3&-2/3&-6\\-2/3&16/3&6\\-6&6&12
 \end{pmatrix}.
 \label{eq:qgcolor}
\end{equation}
For antiquark--gluon scattering, the corresponding helicity-space amplitudes read
\begin{align}
 A_1&=\left[\bar v_1\slashed\epsilon_4^*
          \frac{-\slashed P_3+\slashed P_2}{(-P_3+P_2)^2}\slashed\epsilon_2v_3\right]^{\mathsf{T}},
\quad
 A_2&=\left[\bar v_1\slashed\epsilon_2
          \frac{-\slashed P_3-\slashed P_4}{(-P_3-P_4)^2}\slashed\epsilon_4^*v_3\right]^{\mathsf{T}},
 \quad
 A_3=\left[\bar v_1\slashed{\mathcal J}(P_2,-P_4;\epsilon_2,\epsilon_4^*)v_3/t\right]^{\mathsf{T}},
 \label{eq:antigamplitude}
\end{align}
where the matrix transpose restores the incoming-to-outgoing antiquark spin-index ordering, and the color matrix remains $K_{qg}$. Note that the exchanged momentum in $A_1$ is $P_2-P_3=-(P_1-P_4)$, matching the convention used in the soft-fermion exchange amplitude below. For pair annihilation $q(P_1)\bar q(P_2)\to g(P_3)g(P_4)$,
\begin{align}
 (A_t)_{\lambda_3\lambda_4;ab}
 =\bar v_{2b}\slashed\epsilon_4^*\frac{\slashed P_1-\slashed P_3}{t}\slashed\epsilon_3^*u_{1a},
 \quad
 (A_u)_{\lambda_3\lambda_4;ab}
 =\bar v_{2b}\slashed\epsilon_3^*\frac{\slashed P_1-\slashed P_4}{u}\slashed\epsilon_4^*u_{1a},
 \nonumber\\
 (A_s)_{\lambda_3\lambda_4;ab}
 =\bar v_{2b}\slashed{\mathcal J}(-P_3,-P_4;\epsilon_3^*,\epsilon_4^*)u_{1a}/s,
 \label{eq:annamplitude}
\end{align}
with color matrix \eqref{eq:qgcolor} and the identical-gluon symmetry factor included in $\nu_R$. For $gg\to gg$, setting $k=(P_1,P_2,-P_3,-P_4)$ and $e=(\epsilon_1,\epsilon_2,\epsilon_3^*,\epsilon_4^*)$, the three channel pairings $(12;34)$, $(13;24)$, and $(14;23)$ take the form
\begin{equation}
 A_{ij;kl}=\frac{\mathcal J(k_i,k_j;e_i,e_j)\cdot \mathcal J(k_k,k_l;e_k,e_l)}{(k_i+k_j)^2}
                   +V_{ij;kl},
 \label{eq:ggamplitude}
\end{equation}
where the channel-decomposed four-gluon contact terms and color matrix are
\begin{equation}
\begin{aligned}
 V_s&=(e_1\cdot e_3)(e_2\cdot e_4)-(e_1\cdot e_4)(e_2\cdot e_3),\\
 V_t&=(e_1\cdot e_2)(e_3\cdot e_4)-(e_1\cdot e_4)(e_2\cdot e_3),\\
 V_u&=(e_1\cdot e_2)(e_3\cdot e_4)-(e_1\cdot e_3)(e_2\cdot e_4),
\end{aligned}
 \quad
 K_{gg}=\begin{pmatrix}72&36&-36\\36&72&36\\-36&36&72\end{pmatrix}. \label{eq:ggcontact}
\end{equation}
By virtue of the color Jacobi identity, gauge-dependent pieces cancel between the exchange and contact terms in on-shell Ward identities, and the helicity selection rules used in Sec.~\ref{app:j2} apply directly to the full amplitude.

To evaluate the phase-space integrals, we adopt the momentum-transfer parametrization of (A15)--(A21) in Ref.~\cite{AMY03b}. In those formulas, our assignment $Q=P_1-P_4=P_3-P_2$, $q=|\mathbf Q|$, $\hat{\mathbf q}=\mathbf Q/q$, $Q^0=qy$, $P_1^0=p$, and $P_2^0=k$ corresponds to $Q_{\mathrm{AMY}}=-Q$, $\omega_{\mathrm{AMY}}=-qy$, and $(P',K')=(P_4,P_3)$, which fixes the integration limits and polar angles via (A16)--(A18) and (A21) therein. Denoting the relative azimuthal angle about $\hat{\mathbf q}$ by $\varphi$ and averaging the integrand over global $\mathrm{SO}(3)$ rotations reduces the phase-space measure to
\begin{equation}
 \dd\Pi_{1234}=\frac{q\,\dd p\,\dd k\,\dd q\,\dd y\,\dd\varphi}{512\pi^6}.
 \label{eq:hardjacobian}
\end{equation}
Projecting onto the $J=1$ and $J=2$ channels introduces the multipole averaging factors $1/3$ and $1/5$, respectively. The corresponding $t$-channel parametrization follows by interchanging labels $3$ and $4$.

\subsection{HTL-resummed soft gluon and soft fermion exchanges}
\label{app:soft}
Soft momentum transfers ($q\sim gT$) are screened by HTL self-energies. For soft-gluon exchange between hard partons with $\mathbf p=p\mathbf n$, $\mathbf k=k\hat{\mathbf k}$, $P_3=P+Q$, and $P_4=K-Q$, expanding the external currents to first order in $q/p$ generates both a momentum shift and a Berry rotation of the local helicity frame, which combine into the momentum-space covariant derivative \cite{SY12,SY13,CSSYY14,SDZ15,Lin26}
\begin{equation}
 D_{\mathbf q}\Phi=\mathbf q\cdot\nabla_p\Phi+\ii[\Phi,B^0].
 \label{eq:covderivative}
\end{equation}
Expanding the temporal and spatial Dirac currents with the spinors~\eqref{eq:spinors} yields the quark Berry connection components
\begin{equation}
 B_q^0=-\frac{(\mathbf n\times\mathbf q)\cdot\bm\sigma}{2p},\qquad
 \mathbf B_q=\mathbf nB_q^0+\frac{\mathbf n\times\mathbf q}{2p}(\mathbf n\cdot\bm\sigma).
 \label{eq:spinconnection}
\end{equation}
For gluons, defining $\mathbf e_1(\mathbf n)$ by projecting a fixed reference vector $\mathbf e_{\mathrm{ref}}$ transverse to $\mathbf n$ (supplemented by a second coordinate patch near $\mathbf n\parallel\mathbf e_{\mathrm{ref}}$) and expanding \eqref{eq:threecurrent} gives
\begin{equation}
 B_g^0=-\frac{\mathbf e_{\mathrm{ref}}\cdot\mathbf n}
 {\sqrt{1-(\mathbf e_{\mathrm{ref}}\cdot\mathbf n)^2}}
 \frac{\mathbf e_2\cdot\mathbf q}{p}\operatorname{diag}(-1,+1),\qquad
 \mathbf B_g=\mathbf nB_g^0+\frac{\mathbf n\times\mathbf q}{p}\operatorname{diag}(-1,+1),
 \label{eq:gluonconnection}
\end{equation}
where all physical traces are independent of $\mathbf e_{\mathrm{ref}}$. Writing $\mathbf n_\perp=\mathbf n-(\mathbf n\cdot\hat{\mathbf q})\hat{\mathbf q}$, $\hat{\mathbf k}_\perp=\hat{\mathbf k}-(\hat{\mathbf k}\cdot\hat{\mathbf q})\hat{\mathbf q}$, and $t_\perp=\mathbf n_\perp\cdot\hat{\mathbf k}_\perp$, the longitudinal (electric) and transverse (magnetic) scattering variations read
\begin{align}
 X_E&=(D_{\mathbf q}\Phi_a)\otimes\one-\one\otimes(D_{\mathbf q}\Phi_b),
 \nonumber\\
 X_T&=t_\perp\bigl[(\mathbf q\cdot\nabla_p\Phi_a)\otimes\one
                   -\one\otimes(\mathbf q\cdot\nabla_k\Phi_b)\bigr]
 \nonumber\\
 &\quad+\ii[\Phi_a,\mathbf B_{a\perp}\cdot\hat{\mathbf k}_\perp]\otimes\one
         -\ii\one\otimes[\Phi_b,\mathbf B_{b\perp}\cdot\mathbf n_\perp].
 \label{eq:softgains}
\end{align}
Both the connection commutators and the momentum-gradient terms scale as $\cO(q/p)$, whereas subleading spin-independent recoil corrections multiply the vanishing zeroth-order variation and drop out. Setting $y=q^0/q=\mathbf n\cdot\hat{\mathbf q}=\hat{\mathbf k}\cdot\hat{\mathbf q}$ and $L_R=\ln[(1+y)/(1-y)]-\ii\pi$ for $|y|<1$, we employ the HTL self-energies $\Pi_E=\Pi_{00}^{\mathrm{AMY}}$ and $\Pi_T=\Pi_T^{\mathrm{AMY}}$ from (A11) and (A12) of Ref.~\cite{AMY03b}, evaluated at $\omega=q^0=qy$ and spatial momentum $\mathbf q$. In our $(+---)$ metric convention, $D_E=[q^2+\Pi_E]^{-1}$ and $D_T=[(q^0)^2-q^2-\Pi_T]^{-1}$, so that (A8)--(A10) of Ref.~\cite{AMY03b} correspond to $D_{\mathrm{AMY}}^{00}=-D_E$ and $D_{\mathrm{AMY}}^{ij}=-(\delta^{ij}-\hat q^i\hat q^j)D_T$. With $X[\Phi]=D_EX_E[\Phi]+D_TX_T[\Phi]$, the soft-gluon collision functional is
\begin{align}
 Q_g[\Phi,\Psi]=\frac{g^4}{4}\sum_{a,b}\frac{(d_{c,a}C_a)(d_{c,b}C_b)}{d_A}
 \int\!\frac{\dd^3p\,\dd^3k\,\dd^3q}{(2\pi)^9}
 &\;2\pi\delta[(\mathbf n-\hat{\mathbf k})\cdot\mathbf q]\nonumber\\
 {}\times\Theta(q_*-q)\Theta^g_{ab}f_a^{(0)}\bff_a^{(0)}f_b^{(0)}\bff_b^{(0)}
 &\;\operatorname{Re}\tr\bigl(X[\Phi]^\dagger X[\Psi]\bigr).
 \label{eq:softgluonform}
\end{align}
Here $(C_F,C_A,d_A)=(4/3,3,8)$ yields color weights $(2,12,72)$ for $(qq,qg,gg)$. The secondary cutoff factor $\Theta^g_{ab}=\Theta(|p\mathbf n-k\hat{\mathbf k}|-q_*)$ is present whenever the crossed channel also admits gluon exchange (i.e., for identical species or pairs involving a gluon) and equals unity otherwise. In the coordinates $(p,k,q,y,\varphi)$, the momentum integration measure in \eqref{eq:softgluonform} becomes
\begin{equation}
 \frac{p^2k^2q}{32\pi^6}\,\dd p\,\dd k\,\dd q\,\dd y\,\dd\varphi,
 \qquad 0<q<q_*,\quad -1<y<1.
 \label{eq:softmeasure}
\end{equation}
The trace in \eqref{eq:softgluonform} retains longitudinal--transverse interference. Target-recoil cross terms between $\Phi_a$ and $\Phi_b$ vanish for the traceless $J=1$ and gluon linear-polarization ($L$) matrices, but survive for the spin-independent shear matrices $Y_A^S\propto\one_2$ in \eqref{eq:j2softss}.

Soft-fermion exchange mediates parton conversion in $qg\leftrightarrow gq$, $\bar qg\leftrightarrow g\bar q$, and $q\bar q\leftrightarrow gg$. We use the retarded HTL fermion self-energy components $\Sigma_R^{0,\mathrm{AMY}}$ and $\bm\Sigma_R^{\mathrm{AMY}}$ from (A1) and (A2) of Ref.~\cite{AMY03b} with $m_F^2=m_{\mathrm{HTL},f}^2$, evaluated at $\omega=q^0=qy$ and spatial momentum $\mathbf q$. Retaining the current quark mass $m_{0f}$ gives the dressed propagator
\begin{equation}
 S_R(Q)=\frac{\slashed R+m_{0f}}{R^2-m_{0f}^2},\qquad R=Q-\Sigma_R,\qquad
 S_A(Q)=\gamma^0S_R(Q)^\dagger\gamma^0.
 \label{eq:softfermionpropagator}
\end{equation}
For hard external momenta $P=(p,p\mathbf n)$ and $K=(k,k\hat{\mathbf k})$ satisfying $\mathbf n\cdot\hat{\mathbf q}=\hat{\mathbf k}\cdot\hat{\mathbf q}=y$, the four stripped conversion amplitudes are
\begin{equation}
\begin{gathered}
 A_{qg}^{\zeta}=\bar U(K)\slashed\epsilon_2S_\zeta(Q)\slashed\epsilon_4^*U(P),
 \qquad
 A_{\bar qg}^{\zeta}=
         [\bar V(P)\slashed\epsilon_4^*S_\zeta(-Q)\slashed\epsilon_2V(K)]^{\mathsf{T}},
 \\
 A_{{\mathrm{ann}},t}^{\zeta}=\bar V(K)\slashed\epsilon_4^*S_\zeta(Q)\slashed\epsilon_3^*U(P),
 \qquad
 A_{{\mathrm{ann}},u}^{\zeta}=\bar V(K)\slashed\epsilon_3^*S_\zeta(Q)\slashed\epsilon_4^*U(P),
 \label{eq:softamplitudes}
\end{gathered}
 \qquad \zeta=R,A,
\end{equation}
where the external momentum assignments $(1,2;3,4)$ are $(P,K;K,P)$ for the two Compton channels, and $(P,K;P,K)$ and $(P,K;K,P)$ for the $t$- and $u$-channel annihilation diagrams, respectively. Using Hermitian conjugation,
\begin{equation}
 [\bar U(K)\slashed\epsilon_2S_R(Q)\slashed\epsilon_4^*U(P)]^\dagger
 =\bar U(P)\slashed\epsilon_4S_A(Q)\slashed\epsilon_2^*U(K),
 \label{eq:cutreverse}
\end{equation}
together with charge conjugation $C_DS_R(Q)^{\mathsf T}C_D^{-1}=S_A(-Q)$ ($C_D=\ii\gamma^2\gamma^0$) and the thermal identity $f_F^{(0)}\bff_B^{(0)}=[2\sinh(p/T)]^{-1}$, we obtain the soft-fermion collision functional
\begin{align}
 Q_f[\Phi,\Psi]=\frac{g^4}{2}\sum_f\sum_{c=1}^{4}
 \int\!\frac{\dd^3p\,\dd^3k\,\dd^3q}{(2\pi)^9}
 \frac{2\pi\delta[(\mathbf n-\hat{\mathbf k})\cdot\mathbf q]}{16p^2k^2}
 &\;\Theta(q_*-q)\Theta(|p\mathbf n-k\hat{\mathbf k}|-q_*)
 \nonumber\\[-2pt]
 {}\times\frac{1}{4\sinh(p/T)\sinh(k/T)}\,\frac{16}{3}
 &\;\frac{1}{2}\sum_{\zeta=R,A}\operatorname{Re}\tr
 U_\Phi[A_c^\zeta]^\dagger U_\Psi[A_c^\zeta],
 \label{eq:softfermionform}
\end{align}
where the color factor is $16/3=C_F^2d_{c,q}$ and the kinematic factor $1/(16p^2k^2)$ arises from the spinor normalization~\eqref{eq:spinors}. Projecting onto the $J=1$ channel via the collinear endpoint identity
\begin{equation}
 \slashed\epsilon_\lambda^*(P)U(P)(\mathbf n\cdot\bm\sigma)
         =\lambda\slashed\epsilon_\lambda^*(P)U(P)
 \label{eq:endpointhelicity}
\end{equation}
shows that $A_{\lambda_{q3}\lambda_4;\lambda_{q1}\lambda_2}=\delta_{\lambda_4,\lambda_{q1}}\delta_{\lambda_2,\lambda_{q3}}a_{\lambda_{q3}\lambda_{q1}}$: the chirality-conserving numerator $\slashed R$ and the mass insertion $m_{0f}$ connect equal and opposite quark helicities, respectively, without interfering. Moreover, because any transverse spin insertion $X_\perp$ anticommutes with $\mathbf n\cdot\bm\sigma$,
\begin{equation}
 \sum_{\lambda=\pm1}\slashed\epsilon_\lambda^*(P)U(P)X_\perp\bar U(P)
                           \slashed\epsilon_\lambda(P)=0,
 \label{eq:transversezero}
\end{equation}
quark helicity and transverse spin remain completely decoupled even at finite $m_{0f}$. With $\Delta\ell_p=\ell_f(p)-\ell_g(p)$, each of the four channels in \eqref{eq:softfermionform} contributes to the longitudinal helicity block as
\begin{equation}
 \frac13\bigl[(N_V+N_m)(\Delta\ell_p^2+\Delta\ell_k^2)-2(\mathbf n\cdot\hat{\mathbf k})(N_V-N_m)\Delta\ell_p\Delta\ell_k\bigr],
 \label{eq:conversiongain}
\end{equation}
where $N_V$ and $N_m$ denote the squared-amplitude weights from $\slashed R$ and $m_{0f}$, respectively:
\begin{equation}
 N_V=\frac{16\bigl\{2\operatorname{Re}[(P\cdot R)(K\cdot R^*)]-(P\cdot K)(R\cdot R^*)\bigr\}}{|R^2-m_{0f}^2|^2},
 \qquad N_m=\frac{16m_{0f}^2P\cdot K}{|R^2-m_{0f}^2|^2}.
 \label{eq:conversionnorms}
\end{equation}
In the $J=1$ transverse quark spin block, the gain terms vanish by \eqref{eq:transversezero}, leaving only the diagonal loss terms $\tfrac23(N_V+N_m)[\ell_{fT}(p)^2+\ell_{fT}(k)^2]$.

\subsection{Hard--soft separation scale \texorpdfstring{$q_*$}{q*} and eikonal matching}
\label{app:matching}
To verify cutoff independence at complete leading order, we examine the intermediate regime $gT\ll q\ll\Lambda\sim T$ for smooth radial profiles supported at $p,k\gtrsim\Lambda$. In this overlap region, the hard and soft integrands can be expanded in powers of $q/\Lambda$ and $m/q$, respectively, where $m\in\{m_D,m_{\mathrm{HTL},f},m_{0f}\}=\cO(gT)$. Generalizing the spin-averaged matching of (4.2)--(4.5) in Ref.~\cite{AMY03b}, we establish the polarized matching and bound the subleading remainders below.

In the eikonal limit $q\ll p,k$, soft-gluon exchange preserves the helicities of both hard partons. Up to Berry phases, the stripped $t$-channel amplitude approaches $(2p)(2k)(1-\mathbf n\cdot\hat{\mathbf k})/(-t)$. The external-current factors $2p$ and $2k$ contribute $16p^2k^2$ to the squared amplitude, which combines with the hard phase-space measure $q/(512\pi^6)$ in \eqref{eq:hardjacobian} to reproduce the soft measure $p^2k^2q/(32\pi^6)$ in \eqref{eq:softmeasure}. Small-angle deflection tilts each parton's transverse polarization plane by $\cO[(q/\Lambda)\sqrt{1-y^2}]$, while the thermal weight approaches $F_{1234}\to f_a^{(0)}\bff_a^{(0)}f_b^{(0)}\bff_b^{(0)}$. Using $-t=q^2(1-y^2)$, the bare propagator factor decomposes as
\begin{equation}
 \frac{1-\mathbf n\cdot\hat{\mathbf k}}{-t}
 =\left.D_E\right|_{\Pi_E=0}+t_\perp\left.D_T\right|_{\Pi_T=0}
 =\frac1{q^2}-\frac{t_\perp}{q^2(1-y^2)}.
 \label{eq:eikonalmatch}
\end{equation}
Expanding $U_\Phi[A_t]$ to first order in $q/p$ recovers both the momentum derivatives and the Berry connection commutators in $X_E$ and $X_T$. The small-$q$ asymptote of the hard gluon-exchange integrand thus matches the bare ($m_D\to0$) limit of \eqref{eq:softgluonform} with all polarization indices retained, and their common $\int\!\dd q/q$ scaling generates the leading Coulomb logarithm. Likewise, for fermion exchange, both the small-$q$ hard integrand and the bare limit of \eqref{eq:conversiongain} yield $N_V\to16pk(1-\cos\varphi)/q^2$ and $N_m\to0$, producing the same logarithmic integral $\int\!\dd q/q$ once multiplied by the phase-space factor $q\,\dd q$.

We now bound the subleading matching remainders across $-1<y<1$. For soft-gluon exchange at $q\gg m_D$, $D_E=(q^2+\Pi_E)^{-1}$ differs from $\left.D_E\right|_{\Pi_E=0}=q^{-2}$ by a relative correction $\cO[(m_D^2/q^2)(1+|\ln(1-y^2)|)]$, which is integrable over $y\in(-1,1)$. Although $\operatorname{Re}D_E^{-1}$ vanishes in an exponentially narrow layer $1-|y|\sim e^{-cq^2/m_D^2}$ near the light cone, Landau damping bounds $|D_E|\le|\operatorname{Im}\Pi_E|^{-1}=\cO(m_D^{-2})$, restricting this layer's contribution to relative order $(q^4/m_D^4)e^{-cq^2/m_D^2}$. In the transverse (magnetic) channel, the variations in \eqref{eq:softgains} satisfy
\begin{equation}
 \tr(X_E^\dagger X_E)=\cO(q^2/\Lambda^2),\qquad
 \tr(X_T^\dagger X_T)=\cO\bigl(q^2(1-y^2)^2/\Lambda^2\bigr).
 \label{eq:gainbound}
\end{equation}
The transverse velocity projections $\mathbf n_\perp,\hat{\mathbf k}_\perp\propto(1-y^2)^{1/2}$ in $X_T$ cancel the $(1-y^2)^{-1}$ light-cone singularity of $\left.D_T\right|_{\Pi_T=0}$, rendering $\bigl|\left.D_T\right|_{\Pi_T=0}\bigr|^2\tr(X_T^\dagger X_T)=\cO(1/q^2\Lambda^2)$ uniformly bounded in $y$. Because $\operatorname{Re}\Pi_T\ge0$ for $|y|<1$, we have $|D_T|\le\bigl|\left.D_T\right|_{\Pi_T=0}\bigr|$, and the relative HTL correction away from the light cone scales as $m_D^2/[q^2(1-y^2)]$. Splitting the $y$ integration at the boundary layer $1-y^2\sim m_D^2/q^2$ yields a net HTL remainder of relative order $(m_D^2/q^2)\ln(q/m_D)$. At the opposite infrared end ($q\ll m_D$), static magnetic gluons are dynamically screened by Landau damping, $\operatorname{Im}\Pi_T\simeq-(\pi/4)m_D^2y$, giving
\begin{equation}
 \int_{-1}^1\!\dd y\,\frac{q^2}{|q^2-\ii(\pi/4)m_D^2y|^2}
                         =\cO(m_D^{-2}),
 \label{eq:magneticintegrable}
\end{equation}
which combines with the phase-space factor $q\,\dd q$ to render the ultrasoft magnetic region $q\sim g^2T$ infrared finite and parametrically subleading.

For soft-fermion exchange, decomposing the dressed inverse propagator along $\hat{\mathbf q}$ via $R^0-R^z=-C_-$ and $R^0+R^z=C_+$ gives
\begin{equation}
 C_-=q(1-y)+\frac{m_{{\mathrm{HTL}},f}^2}{q}[1+(1-y)L_R/2],\qquad
 C_+=q(1+y)+\frac{m_{{\mathrm{HTL}},f}^2}{q}[1-(1+y)L_R/2],
 \label{eq:fermionfactors}
\end{equation}
so that
\begin{equation}
 R^2-m_{0f}^2=-C_-C_+-m_{0f}^2,
 \qquad R^0-yR^z=-\frac{m_{{\mathrm{HTL}},f}^2}{q}[y+(1-y^2)L_R/2].
 \label{eq:fermiondenom}
\end{equation}
Outside the light-cone boundary layers $1\mp y\lesssim m^2/q^2$ (with $m^2=m_{{\mathrm{HTL}},f}^2+m_{0f}^2$), the HTL and current-mass corrections to $N_V$ and $N_m$ are of relative order $m^2/[q^2(1\mp y)]$ up to logarithms. Inside these layers, the numerator of \eqref{eq:conversionnorms} is bounded by $pkq^2[m_{{\mathrm{HTL}},f}^4/q^4+(1-y^2)^2+(m_{0f}^2/q^2)(1-y^2)]$, so integrating over $y$ again produces a net difference of relative order $(m^2/q^2)\ln(q/m)$ between the dressed and bare integrands.

Finally, phase-space regions where two channels become simultaneously soft force the incoming velocities $\mathbf n$ and $\hat{\mathbf k}$ to be nearly collinear:
\begin{equation}
 |\mathbf n-\hat{\mathbf k}|^2=2(1-y^2)(1-\cos\varphi),\qquad
 \int_{|\mathbf n-\hat{\mathbf k}|<\epsilon}\!\dd y\,\dd\varphi
                                   \lesssim\epsilon.
 \label{eq:intersection}
\end{equation}
Consequently, simultaneous $t$- and $u$-channel soft overlaps, secondary cutoff step functions $\Theta(|p\mathbf n-k\hat{\mathbf k}|-q_*)$, and regular interference terms in a soft region contribute at most at relative order $(q_*/\Lambda)[1+\ln(q_*/gT)]$. Similarly, for hard momentum transfers $|\mathbf Q|>q_*$ at small $|u|\gtrsim sq_*^2/T^2$, mass-induced helicity flips are bounded by
\begin{equation}
 \Gamma_{\mathrm{flip}}^{|\mathbf Q|>q_*}
 \lesssim\frac{g^4m_{0f}^2}{T}\int^{\cO(T^2)}\!\dd s
         \int_{csq_*^2/T^2}^{\cO(s)}
         \frac{\dd|u|}{(|u|+M_f^2)^2}
 \lesssim g^4T\frac{m_{0f}^2}{q_*^2}
                     [1+\ln(q_*/M_f)],
 \label{eq:hardflipbound}
\end{equation}
confirming that all leading-order $\cO(g^4T)$ chirality flips reside in the soft-fermion and collinear sectors.

Denoting the overall coefficient of $\dd q/q$ in the overlap region $gT\ll q\ll T$ by $\mathcal B[\Phi,\Psi]$ (obtained by substituting the bare propagators $\left.D_E\right|_{\Pi_E=0}$, $\left.D_T\right|_{\Pi_T=0}$, and $\slashed Q/Q^2$ into \eqref{eq:softgluonform} and \eqref{eq:softfermionform} and integrating over the remaining coordinates) and suppressing the fixed arguments $\Phi,\Psi$, the hard ($q>q_*$) and soft ($q<q_*$) contributions take the asymptotic forms
\begin{equation}
\begin{aligned}
 Q_H&=\mathcal B\ln\frac{T}{q_*}+A_H+\mathcal B\,\cO\Bigl(\frac{q_*}{T}\Bigl[1+\ln\frac{q_*}{gT}\Bigr]\Bigr),\\
 Q_g+Q_f&=\mathcal B\ln\frac{q_*}{gT}+A_S+\mathcal B\,\cO\Bigl(\frac{g^2T^2}{q_*^2}\ln\frac{q_*}{gT}\Bigr),
\end{aligned}
 \label{eq:matchingexpansion}
\end{equation}
where $A_H$ and $A_S$ are independent of $q_*$. Differentiating with respect to $\ln q_*$ gives
\begin{equation}
\begin{aligned}
 \frac{\dd Q_H}{\dd\ln q_*}&=-\mathcal B\Bigl[1+\cO\Bigl(\frac{q_*}{T}\Bigl[1+\ln\frac{q_*}{gT}\Bigr]\Bigr)\Bigr],\\
 \frac{\dd(Q_g+Q_f)}{\dd\ln q_*}&=\mathcal B\Bigl[1+\cO\Bigl(\frac{g^2T^2}{q_*^2}\ln\frac{q_*}{gT}\Bigr)\Bigr].
\end{aligned}
 \label{eq:matching}
\end{equation}
The $\ln q_*$ dependence cancels in the sum $Q_H+Q_g+Q_f$. For our reference choice $q_*=\sqrt g\,T$, the hard and soft matching remainders scale as $\cO(\sqrt g[1+\ln(1/g)])$ and $\cO(g\ln(1/g))$, respectively, both vanishing as $g\to0$ (and more generally for any $q_*=g^\alpha T$ with $0<\alpha<1$).

\subsection{Collinear \texorpdfstring{$1\leftrightarrow2$}{1 <-> 2} splitting and LPM resummation}
\label{app:lpm}
Near-collinear $1\leftrightarrow2$ processes ($q\leftrightarrow qg$, $\bar q\leftrightarrow\bar qg$, $g\leftrightarrow q\bar q$, and $g\leftrightarrow gg$) with transverse momentum $k_\perp\sim gT$ have a quantum formation time $t_{\mathrm{form}}\sim(g^2T)^{-1}$ comparable to the soft-scattering mean free path, requiring LPM resummation \cite{LP53,Migdal56,BDMPS96,Zakharov96} as formulated in (B1)--(B6) of Ref.~\cite{AMY03b}. To determine the polarized response, we retain the full helicity structure of the splitting vertices, including the chirality-flipping terms proportional to the current quark mass.

Consider a hard parent parton $a$ of momentum $P$ splitting into daughters $b$ and $c$ with longitudinal momentum fractions $z=1-x$ and $x$, and transverse momenta $-\mathbf k_\perp$ and $+\mathbf k_\perp$. The collinear energy mismatch (inverse formation time) is
\begin{equation}
 \delta E(\mathbf k_\perp)=\frac{k_\perp^2+M_b^2}{2zP}+\frac{k_\perp^2+M_c^2}{2xP}-\frac{M_a^2}{2P},
 \qquad M_f^2=m_{0f}^2+m_{\infty,f}^2,\quad M_g^2=m_{\infty,g}^2.
 \label{eq:formation}
\end{equation}
Because soft scatterings during formation probe transverse impact parameters $b\sim(gT)^{-1}\gg P^{-1}$ and preserve helicity, the $4\times2$ collinear splitting vertex factorizes from transverse momentum broadening. Defining $k_{\perp\lambda}=(k_{\perp1}-\ii\lambda k_{\perp2})/\sqrt2$ and expanding the Dirac current to first order in $k_\perp/P$ and $m_{0f}/P$ yields the quark bremsstrahlung vertices $V_{\lambda_c}^{qg}$ for emitted-gluon helicity $\lambda_c=\pm1$. Stripped of coupling and color factors, and with daughter-quark rows and parent-quark columns ordered as $(+,-)$, these matrices read \cite{AP77,AMY02,Lin26}
\begin{equation}
 V_-^{qg}=\frac1{\sqrt z}
 \begin{pmatrix}2zk_{\perp-}/x&\sqrt2m_{0f}x\\0&2k_{\perp-}/x\end{pmatrix},
 \quad
 V_+^{qg}=\frac1{\sqrt z}
 \begin{pmatrix}2k_{\perp+}/x&0\\-\sqrt2m_{0f}x&2zk_{\perp+}/x\end{pmatrix}.
 \label{eq:lpmqvertices}
\end{equation}
For antiquark bremsstrahlung, the anti-fundamental generator $-t^{A*}$ absorbs the charge-conjugation sign, leaving the helicity vertices identical to \eqref{eq:lpmqvertices}. For pair creation $g\to q\bar q$ (with quark rows and antiquark columns),
\begin{equation}
 V_-^{q\bar q}=\frac1{\sqrt{xz}}
 \begin{pmatrix}0&2xk_{\perp+}\\-2zk_{\perp+}&-\sqrt2m_{0f}\end{pmatrix},
 \quad
 V_+^{q\bar q}=\frac1{\sqrt{xz}}
 \begin{pmatrix}\sqrt2m_{0f}&-2zk_{\perp-}\\2xk_{\perp-}&0\end{pmatrix},
 \label{eq:lpmpairvertices}
\end{equation}
while the three-gluon splitting vertex is
\begin{equation}
 V_{\lambda_b\lambda_c;\lambda_a}
 =-2\left[\delta_{\lambda_a\lambda_b}\frac{k_{\perp\lambda_c}}x
           +\delta_{\lambda_a\lambda_c}\frac{k_{\perp\lambda_b}}z
           -\delta_{\lambda_b,-\lambda_c}k_{\perp,-\lambda_a}\right].
 \label{eq:lpmggvertices}
\end{equation}
Exchanging the two daughters ($x\leftrightarrow z$, $\mathbf k_\perp\to-\mathbf k_\perp$) reverses the signs of both the stripped gluon vertex and its color factor. The chirality-conserving terms carry azimuthal harmonic number $m=1$ via $k_{\perp\pm}$, whereas the mass-dependent terms carry $m=0$ and vanish in the chiral limit $m_{0f}\to0$. Specifically, the $m_{0f}$ terms flip the quark helicity in $q\to qg$ and produce a quark--antiquark pair of equal helicity in $g\to q\bar q$ (corresponding to $\Delta\lambda_q=\pm2$, or a physical helicity change of $\pm1$).

The collinear gain-minus-loss variation $U_\Phi=(\Phi_b\otimes\one+\one\otimes\Phi_c)V-V\Phi_a$ acts as the inhomogeneity in the polarized LPM integral equation,
\begin{equation*}
 (\ii\delta E+\mathcal C_\perp)F_\Phi(\mathbf k_\perp)=U_\Phi(\mathbf k_\perp).
\end{equation*}
The transverse-broadening collision kernel $\mathcal C_\perp$ follows from (B3) and (B6) of Ref.~\cite{AMY03b} under the identification $(p',p,k,\mathbf h)=(P,zP,xP,P\mathbf k_\perp)$ for $q\to qg$ and $g\to gg$, and via the crossing relation (B1b) for $g\to q\bar q$:
\begin{equation}
 (\mathcal C_\perp F)(\mathbf k_\perp)=\int\!\frac{\dd^2q}{(2\pi)^2}g^2T
 \left(\frac1{q^2}-\frac1{q^2+m_D^2}\right)\sum_{r\in\{a,b,c\}}C_r\bigl[F(\mathbf k_\perp)-F(\mathbf k_\perp-s_r\mathbf q)\bigr],
 \label{eq:lpmkernel}
\end{equation}
with longitudinal scaling factors $s=(1,x,z)$ and three-body color factors $C_r=(C_A/2,C_F-C_A/2,C_A/2)$ for $q\to qg$, $(C_F-C_A/2,C_A/2,C_A/2)$ for $g\to q\bar q$, and $(C_A/2,C_A/2,C_A/2)$ for $g\to gg$. Color neutrality of the three-parton state ($\mathbf T_a+\mathbf T_b+\mathbf T_c=0$) guarantees that the dipole potential is positive semidefinite, $\sum_{r<s}(-\mathbf T_r\cdot\mathbf T_s)[1-\cos(\mathbf q\cdot\mathbf r_{rs})]=\tfrac12\sum_{r,s}\mathbf T_r\cdot\mathbf T_s\,e^{\ii\mathbf q\cdot(\mathbf r_s-\mathbf r_r)}\ge0$, which in turn ensures $\mathcal W[\Phi,\Phi]\ge0$ for $\mathcal W[\Phi,\Psi]\equiv\inth\tr(U_\Phi^\dagger F_\Psi+F_\Phi^\dagger U_\Psi)=2\inth\tr(F_\Phi^\dagger\mathcal C_\perp F_\Psi)$. Summing over all splitting channels and incorporating inverse $2\to1$ merging via detailed balance yields the collinear collision functional
\begin{equation}
 Q_{12}[\Phi,\Psi]=\sum_{a;bc}\frac{g^2\mathcal G_{a;bc}}{16\pi S_{bc}}
 \int\!\frac{\dd^3P}{(2\pi)^3}\int_0^1\frac{\dd x}{P^2xz}
 f_a^{(0)}(P)\bff_b^{(0)}(zP)\bff_c^{(0)}(xP)\mathcal W[\Phi,\Psi],
 \label{eq:lpmform}
\end{equation}
with overall color factors $\mathcal G_{a;bc}=d_{c,q}C_F=4$ ($q\to qg$, $\bar q\to\bar qg$, $g\to q\bar q$) and $d_A C_A=24$ ($g\to gg$), and identical-daughter symmetry factor $S_{gg}=2$ ($S_{bc}=1$ otherwise).

Decomposing the splitting vertex as $V=k_{\perp1}V_1+k_{\perp2}V_2+V_0$ and defining $U_\Phi[V_r]=(\Phi_b\otimes\one+\one\otimes\Phi_c)V_r-V_r\Phi_a$, azimuthal orthogonality separates the vector ($m=1$) and scalar ($m=0$) contributions:
\begin{equation}
 \mathcal W[\Phi,\Psi]=W_1\sum_{r=1,2}\tr\bigl(U_\Phi[V_r]^\dagger U_\Psi[V_r]\bigr)
                 +W_0\tr\bigl(U_\Phi[V_0]^\dagger U_\Psi[V_0]\bigr),
 \label{eq:lpmscalarvector}
\end{equation}
where the formation weights $W_1$ and $W_0$ (of dimensions $T^3$ and $T$, respectively) are
\begin{equation}
 W_1=2\operatorname{Re}\inth k_{\perp1}(\ii\delta E+\mathcal C_\perp)^{-1}k_{\perp1},\qquad
 W_0=2\operatorname{Re}\inth(\ii\delta E+\mathcal C_\perp)^{-1}1.
 \label{eq:formationintegrals}
\end{equation}
Varying \eqref{eq:lpmscalarvector} with respect to each leg's perturbation gives the individual collision operators
\begin{align}
 \mathcal C_a[V_r]&=\tfrac12\{V_r^\dagger V_r,\Phi_a\}
         -V_r^\dagger(\Phi_b\otimes\one+\one\otimes\Phi_c)V_r,\nonumber\\
 \mathcal C_b[V_r]&=\tr_c\bigl[\tfrac12\{V_rV_r^\dagger,\Phi_b\otimes\one+\one\otimes\Phi_c\}-V_r\Phi_aV_r^\dagger\bigr],\nonumber\\
 \mathcal C_c[V_r]&=\tr_b\bigl[\tfrac12\{V_rV_r^\dagger,\Phi_b\otimes\one+\one\otimes\Phi_c\}-V_r\Phi_aV_r^\dagger\bigr],
 \label{eq:lpmlineactions}
\end{align}
which manifestly conserve energy and net quark flavor.

Evaluating the helicity traces in \eqref{eq:lpmscalarvector} with the $J=1$ profiles \eqref{eq:unnormalized} decouples longitudinal and transverse polarizations by azimuthal orthogonality. For real profiles and a fixed Cartesian index $i$, we define the quadratic weights $N_{1,0}^{L,T}$ via
\begin{equation*}
\begin{aligned}
 \sum_{r=1,2}\tr\bigl(U_\Phi[V_r]^\dagger U_\Phi[V_r]\bigr)
 &=n_i^2N_1^L+(1-n_i^2)N_1^T,\\
 \tr\bigl(U_\Phi[V_0]^\dagger U_\Phi[V_0]\bigr)
 &=n_i^2N_0^L+(1-n_i^2)N_0^T,
\end{aligned}
\end{equation*}
where $L$ encompasses quark helicity and gluon circular polarization, $T$ denotes transverse quark spin, and the subscripts $1,0$ label the vertex components $V_{1,2}$ and $V_0$. Writing $(\ell_a,\ell_b,\ell_c)$ for the radial polarization profiles evaluated at $(P,zP,xP)$, quark bremsstrahlung with $(\ell_a,\ell_b,\ell_c)=(\ell_f(P),\ell_f(zP),\ell_g(xP))$ gives
\begin{equation}
 N_{1,qg}^L=\frac8{x^2z}\bigl[(\ell_b-\ell_a+\ell_c)^2+z^2(\ell_b-\ell_a-\ell_c)^2\bigr],\qquad
 N_{0,qg}^L=\frac{4m_{0f}^2x^2}{z}(\ell_c-\ell_a-\ell_b)^2.
 \label{eq:lpmtransfer}
\end{equation}
Similarly, pair creation with $(\ell_a,\ell_b,\ell_c)=(\ell_g(P),\ell_q(zP),\ell_{\bar q}(xP))$ and gluon splitting with $(\ell_g(P),\ell_g(zP),\ell_g(xP))$ yield
\begin{align}
 N_{1,{\mathrm{pair}}}^L&=\frac8{xz}
         \bigl[z^2(\ell_b-\ell_c-\ell_a)^2+x^2(\ell_b-\ell_c+\ell_a)^2\bigr],
 &N_{0,{\mathrm{pair}}}^L&=\frac{4m_{0f}^2}{xz}(\ell_b+\ell_c-\ell_a)^2,
 \label{eq:lpmpairnorm}\\
 N_{1,gg}^L&=\frac8{x^2z^2}
         \bigl[(\ell_b+\ell_c-\ell_a)^2+z^4(\ell_b-\ell_c-\ell_a)^2+x^4(-\ell_b+\ell_c-\ell_a)^2\bigr],
 &N_{0,gg}^L&=0.
 \label{eq:lpmggnorm}
\end{align}
For the transverse quark spin profiles, setting $(\ell_a,\ell_b)=(\ell_{fT}(P),\ell_{fT}(zP))$ in bremsstrahlung and $(\ell_b,\ell_c)=(\ell_{qT}(zP),\ell_{\bar qT}(xP))$ in pair creation gives
\begin{align}
 N_{1,qg}^T&=\frac8{x^2z}\bigl[(1+z^2)(\ell_a^2+\ell_b^2)-4z\ell_a\ell_b\bigr],
 &N_{0,qg}^T&=\frac{4m_{0f}^2x^2}{z}(\ell_a^2+\ell_b^2),
 \nonumber\\
 N_{1,{\mathrm{pair}}}^T&=\frac8{xz}\bigl[(x^2+z^2)(\ell_b^2+\ell_c^2)-4xz\ell_b\ell_c\bigr],
 &N_{0,{\mathrm{pair}}}^T&=\frac{4m_{0f}^2}{xz}(\ell_b^2+\ell_c^2),
 \label{eq:lpmtransversenorm}
\end{align}
and $N_{1,gg}^T=N_{0,gg}^T=0$. Both $N_{1,qg}^T$ and $N_{1,\mathrm{pair}}^T$ are manifestly positive semidefinite [e.g., $(1+z^2)(\ell_a^2+\ell_b^2)-4z\ell_a\ell_b=2z(\ell_a-\ell_b)^2+(1-z)^2(\ell_a^2+\ell_b^2)\ge0$]. Averaging over directions using $\avg{n_i^2}=1/3$ and $\avg{1-n_i^2}=2/3$, the complete $J=1$ collinear quadratic form per Cartesian component is
\begin{equation}
 Q_{12,J=1}[\phi,\phi]=\sum_{a;bc}\frac{g^2\mathcal G_{a;bc}}{32\pi^3S_{bc}}
 \int_0^\infty\!\dd P\int_0^1\!\frac{\dd x}{xz}
 f_a^{(0)}\bff_b^{(0)}\bff_c^{(0)}
 \left[W_1\left(\frac{N_1^L}{3}+\frac{2N_1^T}{3}\right)
         +W_0\left(\frac{N_0^L}{3}+\frac{2N_0^T}{3}\right)\right].
 \label{eq:lpmradial}
\end{equation}

To evaluate $W_0$ and $W_1$ numerically, we Fourier transform to impact-parameter space $\mathbf b$ conjugate to $\mathbf k_\perp$, following the scalar and vector boundary-value method in (A.6)--(A.16) of Ref.~\cite{ABB11} and the three-body QCD formulation in (E.1)--(E.4) of Ref.~\cite{GMT16}. The three-body dipole potential reads \cite{AGZ02}
\begin{equation}
 D(b)=\sum_{j\in\{a,b,c\}}C_j\gamma(s_jb),\qquad
 \gamma(b)=\frac{g^2T}{2\pi}\bigl[K_0(m_Db)+\ln(m_Db/2)+\gamma_E\bigr].
 \label{eq:lpmpositionpotential}
\end{equation}
Introducing the dimensionless variables $r=gTb$, $d=2Pxz/T$, $\widehat D=D/(g^2T)$, and the effective squared masses
\begin{equation}
 \mu_{qg}^2=x^2(\kappa_f^2+1/3)+3z/4,
 \quad \mu_{\mathrm{pair}}^2=\kappa_f^2+1/3-3xz/4,
 \quad \mu_{gg}^2=3(1-xz)/4,
 \label{eq:lpmamass}
\end{equation}
the radial wave functions $u_m(r)$ in the azimuthal sectors $m=0,1$ satisfy
\begin{equation}
 u_m''(r)+\frac{u_m'(r)}r
 -\left(\frac{m^2}{r^2}+\mu^2-\ii d\widehat D(r)\right)u_m(r)=0,
 \qquad m=0,1,
 \label{eq:lpmradialode}
\end{equation}
subject to decay at infinity and the short-distance vacuum asymptotics $u_0\sim K_0(\mu r)$ and $u_1\sim\mu K_1(\mu r)$ as $r\to0$. Under the Fourier convention $e^{\ii\mathbf k_\perp\cdot\mathbf b}$, these boundary conditions correspond to
\begin{align*}
 \int\!\frac{\dd^2k_\perp}{(2\pi)^2}e^{\ii\mathbf k_\perp\cdot\mathbf b}
 (\ii\delta E+\mathcal C_\perp)^{-1}1
 &=-\frac{\ii dT}{2\pi}u_0(r),\\
 \int\!\frac{\dd^2k_\perp}{(2\pi)^2}e^{\ii\mathbf k_\perp\cdot\mathbf b}
 (\ii\delta E+\mathcal C_\perp)^{-1}k_{\perp1}
 &=\frac{dgT^2}{2\pi}\frac{b_1}{b}u_1(r).
\end{align*}
Since the angular integrals of $1$ and $(b_1/b)^2$ equal $2\pi$ and $\pi$, respectively, Parseval's theorem yields the formation weights~\eqref{eq:formationintegrals} as
\begin{equation}
 \frac{W_0}{T}=\frac{d^2}{\pi}\int_0^\infty\!r\,\dd r\,\widehat D(r)|u_0(r)|^2,
 \qquad
 \frac{W_1}{g^2T^3}=\frac{d^2}{2\pi}\int_0^\infty\!r\,\dd r\,\widehat D(r)|u_1(r)|^2.
 \label{eq:lpmradialweights}
\end{equation}
Equation~\eqref{eq:lpmradialode} also demonstrates that coherent multiple scattering suppresses $W_0$ and $W_1$ relative to their single-scattering (Bethe--Heitler) limits. Writing $u_m=u_m^{(0)}+\delta u_m$, where $u_m^{(0)}$ denotes the real vacuum Bessel solution $K_0(\mu r)$ or $\mu K_1(\mu r)$ sharing the same $r\to0$ singularity, the remainder $\delta u_m$ is regular at the origin, vanishes at infinity, and satisfies $(-\Delta_m+\mu^2)\delta u_m=\ii d\widehat Du_m$ with $\Delta_m=\partial_r^2+r^{-1}\partial_r-m^2/r^2$. Multiplying by $r\,\delta u_m^*$ and integrating by parts gives $\operatorname{Re}\int_0^\infty r\,\dd r\,\widehat D\,\delta u_m^*u_m=0$, which implies
\begin{equation}
 \int_0^\infty\!r\,\dd r\,\widehat D|u_m|^2=\int_0^\infty\!r\,\dd r\,\widehat D|u_m^{(0)}|^2-\int_0^\infty\!r\,\dd r\,\widehat D|\delta u_m|^2\le\int_0^\infty\!r\,\dd r\,\widehat D|u_m^{(0)}|^2<\infty.
 \label{eq:lpmsuppression}
\end{equation}
Thus $W_0$ and $W_1$ are strictly nonnegative and uniformly bounded from above by their finite Bethe--Heitler values.

In the soft-daughter limit where a longitudinal momentum fraction $\epsilon\in\{x,z\}$ approaches zero, $\mu^2$ remains strictly positive while $d=\cO(\epsilon)$, so $u_m(r)\to u_m^{(0)}(r)$ and both $W_0$ and $W_1$ scale as $\cO(\epsilon^2)$. For soft-gluon emission ($x\to0$), combinations involving only the parent and hard-daughter profiles vanish as $\cO(x^2)$; for example, \eqref{eq:lpmtransversenorm} contains
\begin{equation}
 (1+z^2)(\ell_a^2+\ell_b^2)-4z\ell_a\ell_b=2z(\ell_a-\ell_b)^2+x^2(\ell_a^2+\ell_b^2)=\cO(x^2)
 \label{eq:transverseendpoint}
\end{equation}
for smooth profiles satisfying $\ell_b-\ell_a=\cO(x)$, and the chirality-flipping weight $N_{0,qg}^{L,T}$ likewise carries a factor $m_{0f}^2x^2$. Similarly, at soft-fermion endpoints, the $\cO(\epsilon^2)$ scaling of $W_{0,1}$ cancels the $\epsilon^{-2}$ kinematic factors from phase space and the splitting vertex, ensuring that excising an interval of width $\epsilon$ around an endpoint alters the collinear integral by at most $\cO(g^4T\epsilon)$. By contrast, for circularly polarized soft gluons [$\ell_c=\ell_g(xP)$ in \eqref{eq:lpmtransfer} and \eqref{eq:lpmggnorm}], the formation weight scales as $W_1\sim x^2$ while phase space and Bose enhancement give $(xz)^{-1}\bff_B^{(0)}(xP)\sim x^{-2}$; combined with the vertex factor $x^{-2}\ell_g(xP)^2$, the integrand scales as $x^{-2}|\ell_g(xP)|^2$ as $x\to0$. Infrared convergence of $Q_{12}$ therefore requires the gluon circular-polarization profile to satisfy the form-domain condition
\begin{equation}
 \int_0^T|\ell_g(p)|^2p^{-2}\dd p<\infty,
 \label{eq:gluonformdomain}
\end{equation}
which holds for all trial functions vanishing at least linearly as $p\to0$. Once \eqref{eq:gluonformdomain} is imposed, the cross term between $\Delta\ell=\ell_f(zP)-\ell_f(P)=\cO(x)$ and the soft-gluon profile $\ell_g(xP)$ in \eqref{eq:lpmtransfer},
\begin{equation}
 (1+z^2)(\Delta\ell^2+\ell_g^2)+2(1-z^2)\Delta\ell\ell_g,
 \qquad \Delta\ell=\ell_f(zP)-\ell_f(P)=\cO(x),
 \label{eq:circularendpoint}
\end{equation}
is integrable as well. Finally, for semi-collinear emissions with $k_\perp^2\sim gT^2$ and hard parent and daughter momenta, $\delta E\sim gT$ and $\mathcal C_\perp\sim g^3T\ln(1/g)$, so expanding the resolvent in \eqref{eq:formationintegrals} gives a dissipative part of order $\delta E^{-1}\mathcal C_\perp\delta E^{-1}\sim g\ln(1/g)/T$. Combined with the transverse phase space and the vector splitting vertex, the semi-collinear region contributes at order
\begin{equation*}
 \frac{g^2}{T^2}\int_{k_\perp^2\sim gT^2}\!\dd^2k_\perp\,k_\perp^2\,
 \delta E^{-1}\mathcal C_\perp\delta E^{-1}
 =\cO\bigl(g^5T\ln(1/g)\bigr),
\end{equation*}
which is parametrically subleading to the leading-order transport rate \cite{GMT16,GMT18}.

\subsection{\texorpdfstring{$J=2$}{J=2} collision operator and decoupling of gluon linear polarization from shear viscosity}
\label{app:j2}
The $J=2$ spin-averaged shear ($S$) and gluon linear-polarization ($L$) profiles are expanded in the tensor basis~\eqref{eq:j2basis}, where individual quark and antiquark legs carry $\phi_{S,q_f}=\phi_{S,\bar q_f}=\phi_{S,f}/\sqrt2$. At $\mathbf k=0$, $\mathrm{SO}(3)$ rotational invariance implies
\begin{equation}
 C_{JM\alpha;J'M'\beta}=\delta_{JJ'}\delta_{MM'}C^{(J)}_{\alpha\beta},
 \label{eq:rotblock}
\end{equation}
which decouples the $J=1$ and $J=2$ channels. Within the $J=2$ sector, the absence of shear--linear-polarization ($SL$) mixing follows from the microscopic structure of each collision process:
\begin{enumerate}
\item \emph{Hard $2\leftrightarrow2$ scattering.} Inserting $Y_A^L$ on a gluon leg flips its helicity by $\Delta\lambda=\pm2$. In massless tree-level QCD with all momenta defined as outgoing, four-gluon amplitudes vanish unless two gluons carry each helicity sign, and $q\bar qgg$ amplitudes with a chirality-conserving quark line require opposite gluon helicities \cite{PT86,MP91}. Consequently, interference between shear ($\Delta\lambda=0$) and linear-polarization ($\Delta\lambda=\pm2$) insertions vanishes identically in \eqref{eq:hardform}.
\item \emph{Soft-gluon exchange.} Because $Y_A^S\propto\one_2$ whereas both $D_{\mathbf q}Y_A^L$ and its Berry connection commutators are traceless in helicity space, all $SL$ cross traces vanish regardless of whether the two insertions act on the same parton or on different scattering partons.
\item \emph{Soft-fermion exchange.} For $q\ll p,k$, the hard quark and gluon at each conversion vertex share the collinear four-momentum $P=(p,p\mathbf n)$. For any symmetric traceless tensor $L_{ij}$ transverse to $\mathbf n$, the Clifford algebra gives
\begin{equation}
 \slashed e_i\slashed P\slashed e_j+\slashed e_j\slashed P\slashed e_i=2\delta_{ij}\slashed P,
 \qquad L_{ij}\slashed e_i\slashed P\slashed e_j=0,
 \label{eq:cliffordj2}
\end{equation}
so the mixed $SL$ vertex contraction vanishes prior to contracting with the soft propagator $S_\zeta(Q)$.
\item \emph{Collinear splitting.} With all collinear partons aligned along $\mathbf n$ and $\epsilon_\lambda\cdot\mathbf n=0$, completeness of the traceless symmetric tensor basis $H_A$ yields
\begin{equation}
 \sum_{A=1}^5(n_i H_A^{ij} n_j)
 \left[\epsilon^\dagger H_A\epsilon-\tfrac12\tr(\epsilon^\dagger H_A\epsilon)\one_2\right]=0,
 \label{eq:lpmj2zero}
\end{equation}
which eliminates the $SL$ off-diagonal block even in the presence of the current-mass terms in \eqref{eq:lpmqvertices} and \eqref{eq:lpmpairvertices}.
\end{enumerate}

The hard-scattering contribution to the radial $J=2$ collision operator is obtained by substituting the reconstructed perturbations $\Phi_A[\phi]$ [defined below \eqref{eq:j2basis}] into \eqref{eq:hardform}. As in \eqref{eq:j1projection}, rotational invariance gives, for real radial profiles,
\begin{equation}
 (\phi,C_H^{J=2}\psi)=\frac15\sum_{A=1}^5
 Q_H\bigl[\Phi_A[\phi],\Phi_A[\psi]\bigr],
 \label{eq:j2projection}
\end{equation}
where the radial inner product uses the weights in \eqref{eq:j2source} and the factor $1/5$ averages over the five tensor components $A=1,\ldots,5$ (with color and helicity sums, coupling factors, thermal distributions, and phase-space integrations already incorporated in $Q_H$, and complex Fourier amplitudes extended by sesquilinearity).

For soft-gluon exchange, we set $q_\parallel=\mathbf n\cdot\mathbf q$, $\mathbf q_\perp=\mathbf q-q_\parallel\mathbf n$, and $Z=D_E+t_\perp D_T$. Expanding the shear perturbation to first order in $\mathbf q$ yields the traceless symmetric tensor
\begin{equation}
 \mathcal T_a^{ij}[\phi_S]=\frac{\sqrt{15}}2\left[
         q_\parallel\phi_S'\bigl(n_i n_j-\tfrac13\delta_{ij}\bigr)
               +\frac{\phi_S}{p}(n_i q_{\perp j}+q_{\perp i} n_j)\right].
 \label{eq:j2scalarderivative}
\end{equation}
Averaging the helicity trace in \eqref{eq:softgluonform} over the five tensor components gives the shear-channel integrand, including two-particle recoil cross terms:
\begin{equation}
 \mathcal Q^g_{SS}[\phi_S,\psi_S]=\frac45|Z|^2
 \bigl(\mathcal T_a[\phi_{S,a}]-\mathcal T_b[\phi_{S,b}]\bigr):\bigl(\mathcal T_a[\psi_{S,a}]-\mathcal T_b[\psi_{S,b}]\bigr).
 \label{eq:j2softss}
\end{equation}
For gluon linear polarization, writing $\delta\mathbf n=\mathbf q_\perp/p$, $\xi=(\mathbf n\times\mathbf q)\cdot\hat{\mathbf k}/p$, and $\epsilon^i=(\epsilon_-^i,\epsilon_+^i)$ for the spatial rows of the polarization matrix in \eqref{eq:j2basis}, we have
\begin{align}
 D_{\mathbf q}Y_A^L&=\sqrt{5/2}
 \left[-(\epsilon^i)^\dagger\bigl(\delta n_i n_k H_A^{kj}+H_A^{ik}n_k\delta n_j\bigr)\epsilon^j
                   +(n_i H_A^{ij}\delta n_j)\one_2\right],\nonumber\\
 \mathcal U_A&=q_\parallel\phi_L'Y_A^L+\phi_L D_{\mathbf q}Y_A^L,
 \qquad \mathcal R_A=\ii\xi[\phi_L Y_A^L,\operatorname{diag}(-1,+1)].
 \label{eq:j2lineartransport}
\end{align}
With $X_E^L=\mathcal U_a\otimes\one-\one\otimes \mathcal U_b$ and $X_T^L=t_\perp X_E^L+\mathcal R_a\otimes\one-\one\otimes \mathcal R_b$, the helicity traces eliminate all cross terms between particles $a$ and $b$. Averaging over the five tensor components leaves the single-particle contribution to the integrand of \eqref{eq:softgluonform} in the $LL$ block:
\begin{equation}
 \mathcal Q^g_{LL,a}[\phi_L,\psi_L]
 =2\left\{|Z|^2\left[q_\parallel^2\phi_L'\psi_L'+\frac{q_\perp^2}{p^2}\phi_L\psi_L\right]
                           +4|D_T|^2\xi^2\phi_L\psi_L\right\},
 \label{eq:j2softll}
\end{equation}
where the $q_\parallel^2$ and $q_\perp^2$ terms describe radial and angular momentum diffusion, respectively, and the $\xi^2$ term arises from the magnetic Berry phase rotation. In the leading-logarithmic approximation, the angular integral of either diagonal block yields $4\pi q^{-2}(\phi'^2+6\phi^2/p^2)$; for the $L$ channel, the covariant derivative $D_{\mathbf q}Y_A^L$ and the magnetic commutator $\mathcal R_A$ contribute $2$ and $4$, respectively, to the $J=2$ Casimir $J(J+1)=6$.

For soft-fermion exchange, we define $N_{\mathrm{conv}}=(16/3)\tfrac12\sum_{\zeta=R,A}\tr[(A_c^\zeta)^\dagger A_c^\zeta]$. The endpoint identity~\eqref{eq:cliffordj2} determines the angular weights multiplying the radial products $\phi_i\psi_j$ in \eqref{eq:softfermionform} after averaging over the five tensor components. Labeling the incoming ($1,2$) and outgoing ($3,4$) external legs by $i,j\in\{1,2,3,4\}$, the $SS$, $SL$, and $LL$ bilinear coefficients read
\begin{equation}
 \mathcal Q^f_{iS,jS}=\frac{N_{\mathrm{conv}}}{2}\varsigma_i\varsigma_jP_2(\mathbf n_i\cdot\mathbf n_j),
 \quad \mathcal Q^f_{iS,jL}=\mathcal Q^f_{iL,jS}=0,
 \quad \mathcal Q^f_{iL,jL}=\frac{N_{\mathrm{conv}}}{2}\delta_{ij},
 \quad \varsigma=(-1,-1,+1,+1),
 \label{eq:j2fermiongram}
\end{equation}
where $L$ is restricted to gluon legs. The resulting leading-logarithmic single-particle conversion damping rates match (4.5) and (2.18) of Ref.~\cite{AMY03b}:
\begin{equation}
 \gamma_{q\to g}^{\log}(p)=\frac{g^4T^2}{18\pi p}\coth\left(\frac{p}{2T}\right)
                     \ln\frac{q_*}{gT},
 \qquad
 \gamma_{g,f}^{\log}(p)=\frac{g^4T^2}{24\pi p}\tanh\left(\frac{p}{2T}\right)
                     \ln\frac{q_*}{gT},
 \label{eq:j2conversioncheck}
\end{equation}
where $\gamma_{g,f}^{\log}(p)$ sums over quarks and antiquarks of a single flavor and coincides with the leading-logarithmic loss rate of gluon linear polarization.

For collinear $1\leftrightarrow2$ splitting with legs ordered as $(a,b,c)$ and $\varsigma=(-1,1,1)^{\mathsf T}$, the quadratic form in the radial profiles within the $SS$ block of the collinear integrand is governed by the rank-one matrices $d_1\varsigma\varsigma^{\mathsf T}$ and $d_0\varsigma\varsigma^{\mathsf T}$ multiplying $W_1$ and $W_0$, respectively, with coefficients
\begin{equation}
\begin{aligned}
 d_1^{qg}&=\frac{4(1+z^2)}{x^2z}, & d_1^{q\bar q}&=\frac{4(x^2+z^2)}{xz}, & d_1^{gg}&=\frac{4(1+x^4+z^4)}{x^2z^2},\\
 d_0^{qg}&=\frac{2m_{0f}^2x^2}{z}, & d_0^{q\bar q}&=\frac{2m_{0f}^2}{xz}, & d_0^{gg}&=0,
\end{aligned}
 \label{eq:j2lpmcoeff}
\end{equation}
where $d_{1,0}^{\bar qg}=d_{1,0}^{qg}$. In the $LL$ block, quark bremsstrahlung and pair production contribute only diagonal gluon loss terms at entries $(c,c)$ and $(a,a)$, respectively, with the same coefficients $d_{1,0}$. Pure gluon splitting $g\to gg$ has $\mathcal Q_{LL}^{(0)}=0$ and a $W_1$ coefficient matrix
\begin{equation}
 R_g(x)=\begin{pmatrix}
 d_g&-8/x^2&-8/z^2\\-8/x^2&d_g&8\\-8/z^2&8&d_g
 \end{pmatrix},\qquad d_g=8(1+x^{-2}+z^{-2}).
 \label{eq:j2lpmgram}
\end{equation}
For any nonzero profile vector $\boldsymbol\ell=(\ell_a,\ell_b,\ell_c)^{\mathsf T}$,
\begin{equation}
 \boldsymbol\ell^{\mathsf{T}}R_g\boldsymbol\ell=8\bigl[x^{-2}(\ell_a-\ell_b)^2+z^{-2}(\ell_a-\ell_c)^2
 +(\ell_b+\ell_c)^2+\ell_a^2+z^{-2}\ell_b^2+x^{-2}\ell_c^2\bigr]>0\quad(\boldsymbol\ell\ne0),
 \label{eq:j2positive}
\end{equation}
confirming that the $L$ sector possesses no collisional zero mode. Combined with $\pi_L=0$ and $C_{SL}^{J=2}=C_{LS}^{J=2}=0$, we obtain $C_{J=2}^{-1}\pi=(C_{SS}^{-1}\pi_S,0)^{\mathsf T}$, proving that gluon linear polarization decouples completely from the leading-order shear viscosity of hot QCD.

\section{Spectral Theory, Transport-Scale Bounds, and Homogeneous Spin Relaxation}
\label{sec:spectral-supp}

\subsection{Spectral decomposition, Kubo response, and relaxation times}
Homogeneous spin relaxation is governed by the spectral decomposition of the collision operator in the thermal Hilbert space. We define $C$ as the nonnegative self-adjoint operator associated with the closure of the collision quadratic form $Q$ (the Friedrichs extension \cite{Kato95}), whose form domain is restricted by the infrared condition~\eqref{eq:gluonformdomain} and whose collinear sector is controlled by \eqref{eq:lpmsuppression}. For each relaxation-rate threshold $\Gamma\ge0$, let $P_\Gamma$ denote the orthogonal spectral projector of $C$ onto $[0,\Gamma]$ \cite{HT10,Moore18}. Its spectral increments $\dd P_\Gamma$ define the normalized, nonnegative spectral measure of the spin susceptibility,
\begin{equation*}
 \dd\mu_s(\Gamma)=\frac{(j,\dd P_\Gamma\,j)}{(j,j)},\qquad
 \int_0^\infty\!\dd\mu_s(\Gamma)=1.
\end{equation*}

A weak static field $h_0$ conjugate to $s$, such as axial-vector torsion or a spin chemical potential \cite{H21}, prepares the equilibrium polarization $\Phi(0)=h_0j/T$. Once the field is switched off at $t=0$, the homogeneous kinetic equation $\partial_t\Phi=-C\Phi$ yields (4) of the Letter, with the spectral representation
\begin{equation}
 \frac{\delta s(t)}{\delta s(0)}
 =\frac{(j,e^{-Ct}j)}{(j,j)}
 =\int_0^\infty\!\dd\mu_s(\Gamma)\,e^{-\Gamma t}.
 \label{eq:continuumdecay}
\end{equation}
For a time-dependent driving field $h(t)$ conjugate to $s^i$, detailed balance requires the collision integral to vanish at the instantaneous equilibrium distribution $\Phi_{\mathrm{eq}}(t)=h(t)j/T$, introducing the source term $+h(t)Cj/T$ into the linearized kinetic equation \cite{H21,Landsteiner14}. For harmonic time dependence $e^{-\ii\omega t}$, we obtain
\begin{equation}
\begin{gathered}
 \partial_t\Phi=-C(\Phi-hj/T),\qquad
 (C-\ii\omega)\Phi(\omega)=h(\omega)Cj/T,\\
 \frac{G^R(\omega)}{\chi}=\frac{\bigl(j,C(C-\ii\omega)^{-1}j\bigr)}{(j,j)}
 =\int_0^\infty\!\dd\mu_s(\Gamma)\,\frac{\Gamma}{\Gamma-\ii\omega},
 \label{eq:forcedresponse}
\end{gathered}
\end{equation}
with quasiparticle spin susceptibility $\chi=(j,j)/T=T^2/4$. Here $G^R(\omega)$ is the quasiparticle contribution to the spatially integrated retarded spin correlator $G^R_{\mathrm{full}}(t)=+\ii\theta(t)\langle[s^i(t),s^i(0)]\rangle$. Ultraviolet contact terms analytic at kinetic frequencies $\omega\sim\gamma_{\mathrm{tr}}$ can shift the full static susceptibility at $\omega=0$ without affecting transport-scale relaxation (see Sec.~\ref{sec:rotation-supp}).

Because $C$ is an integrodifferential operator on momentum space rather than a scalar multiple of the identity, multiple eigenmodes contribute to $\dd\mu_s(\Gamma)$ \cite{HT10,Moore18}. The integrated relaxation time,
\begin{equation*}
 \tau\equiv\int_0^\infty\!\dd t\,\frac{\delta s(t)}{\delta s(0)}
 =\int_0^\infty\!\frac{\dd\mu_s(\Gamma)}{\Gamma}
 =\frac{(j,C^{-1}j)}{(j,j)},
\end{equation*}
governs the low-frequency expansion $G^R(\omega)/\chi=1+\ii\omega\tau+\cO(\omega^2)$ and, for a flavor-symmetric collision operator, determines the axial-charge diffusion constant $D=\tau/3$. It differs from the mean decay time
\begin{equation*}
 \Gamma_{\mathrm{eff}}^{-1}\equiv\frac{\int_0^\infty\dd t\,t\,\delta s(t)}{\int_0^\infty\dd t\,\delta s(t)}
 =\frac{1}{\tau}\int_0^\infty\!\frac{\dd\mu_s(\Gamma)}{\Gamma^2}
 =\frac{(j,C^{-2}j)}{(j,C^{-1}j)},
\end{equation*}
which weights slower spectral components more heavily and whose inverse defines the effective relaxation rate $\Gamma_{\mathrm{eff}}$ reported in the Letter. Applying the Cauchy--Schwarz inequality to the normalized measure $\dd\mu_s(\Gamma)$ gives
\begin{equation*}
 \tau^2=\left(\int_0^\infty\!\frac{\dd\mu_s(\Gamma)}{\Gamma}\right)^2\le\int_0^\infty\!\dd\mu_s(\Gamma)\int_0^\infty\!\frac{\dd\mu_s(\Gamma)}{\Gamma^2}=\frac{\tau}{\Gamma_{\mathrm{eff}}},
\end{equation*}
so that $\Gamma_{\mathrm{eff}}\le 1/\tau$, with equality if and only if $\dd\mu_s(\Gamma)$ is concentrated at a single relaxation rate. The strange-quark vector current $J^i=\bar q_s\gamma^iq_s$ has a $C$-odd source $j_J=(0,0,2\sqrt{2/3})^{\mathsf{T}}$ in the flavor basis $(u,d,s)$ and susceptibility $\chi_J=(j_J,j_J)/T=T^2/3$; its effective relaxation rate defines the reference unit $\Gamma_{\mathrm{tr}}\equiv(j_J,C^{-1}j_J)/(j_J,C^{-2}j_J)$.

To compute the relaxation spectrum numerically, we expand the seven radial profiles~\eqref{eq:j1state} in a finite variational basis \cite{AMY03a,AMY03b,HT10}, $\phi(p)\simeq\sum_A c_A b_A(p)$, where each basis function $b_A(p)$ is supported in a single radial channel and constructed from cubic B-splines in $\ln(p/T)$ \cite{deBoor78,Johnson1988,Bachau2001}. The basis functions for gluon circular polarization include an overall factor $p/T$ to enforce the infrared regularity condition~\eqref{eq:gluonformdomain}, and the two boundary splines are omitted so that trial functions vanish smoothly at the endpoints of the momentum grid. Projecting the linearized Boltzmann equation onto this basis yields the decay eigenvalues $\lambda_n$ and orthonormal polarization eigenmodes $\Phi_n$ via the generalized eigenvalue problem
\begin{equation}
\begin{gathered}
 \Phi_n(p)=\sum_A(c_n)_A b_A(p),\qquad
 (\Phi_n,\Phi_m)=\delta_{nm},\\
 Q[b_A,\Phi_n]=\lambda_n(b_A,\Phi_n).
\end{gathered}
 \label{eq:galerkin}
\end{equation}
The positive-definite Gram matrix has mass dimension $[(b_A,b_B)]=T^3$, while the symmetric collision matrix has $[Q[b_A,b_B]]=T^4$, giving $[\lambda_n]=T$. The normalized spectral weights $w_n$ and the susceptibility fraction $r_N$ omitted by the finite basis are
\begin{equation}
 w_n=\frac{|(j,\Phi_n)|^2}{(j,j)},\qquad
 r_N=1-\sum_{n=0}^{N-1}w_n\ge0.
 \label{eq:finiteweights}
\end{equation}
Projecting $j$ onto the $N$-dimensional variational subspace gives $\sum_{n=0}^{N-1}\Phi_n(\Phi_n,j)$, which approximates the spectral measure by $\dd\mu_s(\Gamma)\simeq\sum_{n=0}^{N-1}w_n\delta(\Gamma-\lambda_n)\,\dd\Gamma$ and captures a fraction $1-r_N$ of the continuum susceptibility $\chi=(j,j)/T$. Normalizing by $\chi$ yields
\begin{equation}
 \frac{G^R(\omega)}{\chi}\simeq\sum_{n=0}^{N-1}w_n\frac{\lambda_n}{\lambda_n-\ii\omega},
 \qquad
 \frac{\delta s(t)}{\chi h_0}\simeq\sum_{n=0}^{N-1}w_ne^{-\lambda_nt},
 \label{eq:spectralquantities}
\end{equation}
with static limit $G^R(0)/\chi=1-r_N$. The single-relaxation-time approximation $G^R_\tau(\omega)$ shown in Fig.~2 of the Letter matches both the initial polarization and the integrated relaxation time $\tau$,
\begin{equation}
 \frac{G^R_\tau(\omega)}{\chi}
 =\frac{\sum_nw_n}{1-\ii\omega\tau},\qquad
 \tau=\frac{m_1}{\sum_nw_n},\qquad
 m_k=\sum_{n=0}^{N-1}\frac{w_n}{\lambda_n^k},
 \label{eq:benchmarks}
\end{equation}
reproducing the static limit and low-frequency slope of $G^R(\omega)$ without free parameters, while the effective relaxation rate is $\Gamma_{\mathrm{eff}}=m_1/m_2$.

\subsection{Lower bound on angular diffusion and exclusion of slow spectral weight}
\label{sec:slowweight-supp}
Positive semidefiniteness of the collision operator ($Q[\Phi,\Phi]\ge0$) and a transport-scale initial decay rate ($(j,Cj)/(j,j)=\cO(\gamma_{\mathrm{tr}})$ with $\gamma_{\mathrm{tr}}\equiv g^4T\ln(1/g)$) do not by themselves preclude a finite fraction of the spectral measure $\mu_s$ from residing at anomalously slow rates $\Gamma\ll\gamma_{\mathrm{tr}}$. In particular, the angular deflection rate of a hard quark falls as $g^4T^3\ln(1/g)/p^2$ at high momentum \cite{HT10,Moore18}, whereas the radial thermal weight $\rho_f^{(1)}(p)\propto p^2f_F^{(0)}(p)[1-f_F^{(0)}(p)]$ in \eqref{eq:j1radialmetric} is concentrated at thermal momenta $p\sim T$. Establishing a lower bound on the damping rate of longitudinal quark polarization within any bounded thermal momentum interval $[aT,bT]$ ($0<a<b<\infty$) is therefore sufficient to rule out any parametrically slow component of the spin response. We now derive this bound using small-angle quark--gluon scattering alone.

Consider $t$-channel quark--gluon and antiquark--gluon scattering ($q_fg\to q_fg$ and $\bar q_fg\to\bar q_fg$) with external momenta restricted to a thermal interval containing $[aT,bT]$, relative scattering angles confined to a fixed noncollinear region of nonzero measure (where $s$ is bounded below by a $g$-independent multiple of $T^2$ and $y=q^0/q$ is bounded away from $\pm1$ so that $q_\perp^2\sim q^2$), and spatial momentum transfers restricted to $q_*<q<g^{1/4}T$ with $q_*=\sqrt g\,T$. This momentum-transfer window lies entirely within the hard $2\leftrightarrow2$ domain ($q>q_*$), where tree-level amplitudes apply, and captures a fixed fraction $\ln(g^{1/4}T/q_*)=\tfrac14\ln(1/g)$ of the full Coulomb logarithm. Because these kinematic cuts respect global $\mathrm{SO}(3)$ rotational and parity invariance, the retained collisions do not mix the $J=1$ axial channel with other $J$ sectors. Moreover, massless tree-level amplitudes conserve quark chirality, so dropping the nonnegative helicity-off-diagonal entries of $\tr(U_\Phi^\dagger U_\Phi)$ eliminates the transverse quark spin coherences $\phi_{fT}^+$.

In the small-angle regime ($q\ll p,k$), all four helicity-conserving $qg\to qg$ transition probabilities $(\lambda_q=\pm1,\lambda_g=\pm1)$ approach a common eikonal limit and remain above half of that value throughout the retained interval. Writing the helicity-diagonal gain-minus-loss variation for quark helicity $\lambda=\pm1$ and gluon helicity $\lambda'=\pm1$ as $\Delta_0+\lambda\Delta_q+\lambda'\Delta_g$---where $\Delta_0$ is the variation of the spin-averaged distribution and $\Delta_q,\Delta_g$ are the variations of the quark and gluon helicity polarizations---and summing over all four helicity combinations gives
\begin{equation}
 \sum_{\lambda,\lambda'=\pm1}\bigl|\Delta_0+\lambda\Delta_q+\lambda'\Delta_g\bigr|^2=4\bigl(|\Delta_0|^2+|\Delta_q|^2+|\Delta_g|^2\bigr)\ge4|\Delta_q|^2 .
 \label{eq:helicitysumsupp}
\end{equation}
Equation~\eqref{eq:helicitysumsupp} explains why momentum conservation does not protect spin: in spin-averaged momentum transport ($\Delta_0$), target recoil cancels forward momentum loss to conserve total momentum, whereas for quark helicity ($\Delta_q$), scattering off the two gluon helicities $\lambda'=\pm1$ with equal eikonal strength eliminates the cross term $\Delta_q^*\Delta_g$, preventing gluon recoil from canceling the angular diffusion of quark helicity.

For a real longitudinal quark polarization profile $\ell(p)n_i$ deflected by a small angle $\theta\sim q/p$ from $\mathbf p_1$ to $\mathbf p_3$, averaging $[\ell(p_3)n_{3i}-\ell(p_1)n_{1i}]^2$ over the three Cartesian components $i=1,2,3$ gives
\begin{equation*}
 \frac13\sum_i[\ell(p_3)n_{3i}-\ell(p_1)n_{1i}]^2
 =\frac13\bigl[\cos\theta\,[\ell(p_3)-\ell(p_1)]^2+(1-\cos\theta)\bigl(\ell(p_1)^2+\ell(p_3)^2\bigr)\bigr],
\end{equation*}
with $\cos\theta>0$ at small angles. Retaining the nonnegative angular-deflection term $(1-\cos\theta)\ell(p_1)^2\simeq\tfrac12\theta^2\ell(p_1)^2$ along with the squared $t$-channel propagator yields the momentum-transfer integral
\begin{equation*}
 \int_{q_*}^{g^{1/4}T}\!\dd q\,q\,\frac{1}{q^4}\frac{q^2}{p^2}
 =\frac{1}{p^2}\int_{q_*}^{g^{1/4}T}\!\frac{\dd q}{q}
 =\frac{\ln(1/g)}{4p^2}.
\end{equation*}
Multiplying by the coupling factor $g^4$ and the thermal target phase space $\propto T^3$ gives an angular damping rate of order $g^4T^3\ln(1/g)/p^2$. Because all omitted terms and all other collision channels ($qq$, $gg$, soft exchange, and LPM splitting) give nonnegative contributions to $Q[\phi,\phi]$, the complete leading-order collision quadratic form satisfies the lower bound
\begin{equation}
 {\cal N}_{ab}^{L}[\phi]
 \equiv\sum_f\int_{aT}^{bT}\!\dd p\,\rho_f^{(1)}(p)\phi_{fL}^{+*}(p)\phi_{fL}^+(p),
 \qquad
 Q[\phi,\phi]\ge c_{ab}\gamma_{\mathrm{tr}}{\cal N}_{ab}^{L}[\phi],
 \label{eq:lowerbound}
\end{equation}
where $\mathcal N_{ab}^L[\phi]$ is $2/T$ times the quadratic free energy of longitudinal quark polarization on $[aT,bT]$, and $c_{ab}>0$ is a $g$-independent dimensionless constant determined by $a$ and $b$. By contrast, for a spherically symmetric ($\ell=0$, $J=0$) axial-charge perturbation $\Phi_{q_f}\propto(\mathbf n\cdot\bm\sigma)$ lacking the directional factor $n_i$, the Legendre factor $1-P_0(\cos\theta)$ vanishes, leaving the massless axial charge undamped by elastic scattering.

We now convert the lower bound~\eqref{eq:lowerbound} into an upper bound on the slow spectral weight of the canonical spin susceptibility. Let
\begin{equation}
 W_s(\Gamma)=\int_0^{\Gamma}\!\dd\mu_s(\Gamma')
 \label{eq:slowfraction}
\end{equation}
denote the fraction of the spin susceptibility carried by modes with relaxation rates at most $\Gamma$, and let $\psi=P_\Gamma j$ be the corresponding slow projection of the spin source. Orthogonality of $P_\Gamma$ implies $(j,\psi)=(\psi,\psi)=(j,j)W_s(\Gamma)$, while its spectral support gives $Q[\psi,\psi]\le\Gamma(\psi,\psi)$. Applying \eqref{eq:lowerbound} then yields
\begin{equation*}
 \mathcal N_{ab}^L[\psi]\le\frac{Q[\psi,\psi]}{c_{ab}\gamma_{\mathrm{tr}}}\le\frac{\Gamma(\psi,\psi)}{c_{ab}\gamma_{\mathrm{tr}}}.
\end{equation*}
Because this inequality constrains longitudinal quark polarization only within $[aT,bT]$, we also bound the susceptibility fraction $r_{ab}$ arising from momenta outside this interval. Since the leading-order spin source $j$ has only longitudinal quark components, the susceptibility contributed by $[aT,bT]$ is $\mathcal N_{ab}^{L}[j]/T$, whereas the total susceptibility is $(j,j)/T$. From the thermal inner product~\eqref{eq:j1radialmetric}, source normalization~\eqref{eq:sourcenorm}, and Fermi integral~\eqref{eq:fermionintegral}, we have
\begin{equation*}
 \mathcal N_{ab}^{L}[j]
 =\frac{3}{2\pi^2}\int_{aT}^{bT}\!\dd p\,
 p^2\frac{e^{p/T}}{(e^{p/T}+1)^2},
 \qquad
 (j,j)=\frac{3}{2\pi^2}\frac{\pi^2T^3}{6}=\frac{T^3}{4},
\end{equation*}
so the susceptibility fraction outside $[aT,bT]$ is
\begin{equation}
\begin{aligned}
 r_{ab}&\equiv1-\frac{\mathcal N_{ab}^{L}[j]}{(j,j)}
 =\frac{6}{\pi^2T^3}
 \left(\int_0^{aT}+\int_{bT}^{\infty}\right)\!\dd p\,
 p^2\frac{e^{p/T}}{(e^{p/T}+1)^2}\\
 &=1-\frac{6}{\pi^2T^3}\int_{aT}^{bT}\!\dd p\,
 p^2\frac{e^{p/T}}{(e^{p/T}+1)^2}.
\end{aligned}
 \label{eq:outsidesource}
\end{equation}
Splitting $(j,\psi)$ into contributions from inside and outside $[aT,bT]$, applying the Cauchy--Schwarz inequality to each piece, and using $\mathcal N_{ab}^{L}[j]\le(j,j)$ gives
\begin{equation*}
 |(j,\psi)|\le
 \sqrt{(j,j)\mathcal N_{ab}^{L}[\psi]}
 +\sqrt{r_{ab}(j,j)(\psi,\psi)}.
\end{equation*}
For $\psi\ne0$, dividing by $\sqrt{(j,j)(\psi,\psi)}=(j,j)\sqrt{W_s(\Gamma)}$ and substituting the upper bound on $\mathcal N_{ab}^{L}[\psi]$ yields
\begin{equation}
 \sqrt{W_s(\Gamma)}\le\sqrt{r_{ab}}+\sqrt{\frac{\Gamma}{c_{ab}\gamma_{\mathrm{tr}}}},
 \label{eq:weightbound}
\end{equation}
which also holds trivially when $\psi=0$. In \eqref{eq:outsidesource}, the low-momentum tail ($p<aT$) is suppressed by phase space ($\propto p^2\dd p$) and the high-momentum tail ($p>bT$) is exponentially suppressed by the Fermi distribution; specifically, $f_F^{(0)}(p)[1-f_F^{(0)}(p)]\le\min(1/4,\,e^{-p/T})$ implies
\begin{equation*}
 r_{ab}\le\frac{a^3}{2\pi^2}
 +\frac{6}{\pi^2}(b^2+2b+2)e^{-b}.
\end{equation*}
Choosing $a\ll1$ and $b\gg1$ renders $r_{ab}$ arbitrarily small. Fixing $a$ and $b$, taking $g\to0$ at fixed $\Gamma/\gamma_{\mathrm{tr}}$, and then sending $\Gamma/\gamma_{\mathrm{tr}}\to0$ gives
\begin{equation}
 \Gamma\ll\gamma_{\mathrm{tr}}
 \quad\Longrightarrow\quad W_s(\Gamma)\ll1.
 \label{eq:noslowweight}
\end{equation}
Thus, no finite fraction of the canonical light-quark spin polarization carried by hard quasiparticles survives for times parametrically longer than $\gamma_{\mathrm{tr}}^{-1}$.

Conversely, because the initial decay rate of the spin source is of order the transport scale,
\begin{equation}
 \frac{(j,Cj)}{(j,j)}=\int_0^\infty\!\dd\mu_s(\Gamma)\,\Gamma=\cO(\gamma_{\mathrm{tr}}),
 \label{eq:meandecay}
\end{equation}
Markov's inequality bounds the spectral weight above any $\Gamma_{\mathrm{fast}}$ by $(j,Cj)/[(j,j)\Gamma_{\mathrm{fast}}]$, excluding any finite fraction of spin polarization from relaxing at rates $\Gamma\gg\gamma_{\mathrm{tr}}$. Together, \eqref{eq:noslowweight} and \eqref{eq:meandecay} confine all finite spectral weight of the leading-order spin susceptibility to transport-scale rates $\Gamma\sim\gamma_{\mathrm{tr}}$. Splitting the frequency integral in \eqref{eq:forcedresponse} at an intermediate threshold $\Gamma>0$ similarly yields
\begin{equation}
 \left|\frac{G^R(\omega)}{\chi}-1\right|\le W_s(\Gamma)+\frac{|\omega|}{\Gamma},
 \label{eq:diagonalbound}
\end{equation}
demonstrating that spin polarization closely tracks any conjugate driving field varying slowly relative to $\gamma_{\mathrm{tr}}$ ($|\omega|\ll\gamma_{\mathrm{tr}}$).

\subsection{Spin susceptibility of on-shell soft collective excitations}
\label{app:softsource}
In effective kinetic theory, virtual soft collective modes with momenta $q\sim gT$ enter through the HTL-resummed propagators in $Q_g, Q_f$ and the transverse broadening kernel $\mathcal C_\perp$ in $Q_{12}$. To verify that on-shell soft collective excitations---namely normal quark and plasmino branches and gluonic plasmons at $p\sim gT$ \cite{Weldon82a,Weldon82b,BP90,Nair94,FRS95,FR96,BI02}---cannot evade the hard-quasiparticle bound~\eqref{eq:noslowweight}, we bound their direct contribution to the canonical spin susceptibility. At $p\sim gT$, the phase-space volume scales as $\int_{p\sim gT}\dd^3p\sim g^3T^3$. While Bose enhancement $f_B^{(0)}(1+f_B^{(0)})\simeq T^2/\omega^2\sim g^{-2}$ partially compensates this phase-space suppression for soft bosons at $\omega\sim gT$, Pauli blocking restricts the fermionic statistical weight to $f_F^{(0)}(1-f_F^{(0)})\le\tfrac14$.

For a timelike on-shell HTL quark or plasmino mode of flavor $f$ at soft four-momentum $K=(E,\mathbf p)$ ($p\sim gT$), the inverse retarded propagator reads \cite{Weldon82b,BP90,BI02}
\begin{equation}
 S_R^{-1}(K)=\slashed K-m_{0f}-m_{{\mathrm{HTL}},f}^2
          \left\langle\frac{\slashed v}{v\cdot K+\ii0}\right\rangle_v,
 \qquad v=(1,\mathbf v),\quad \mathbf v^2=1,
 \label{eq:htlpole}
\end{equation}
where $\langle\cdots\rangle_v\equiv\int\frac{\dd\Omega_v}{4\pi}(\cdots)$ averages over the velocity directions of hard thermal particles. With $\alpha_i=\gamma^0\gamma^i$, the wave-function renormalization matrix $\mathcal Z$ and the HTL-dressed canonical spin vertex $\mathcal V_i$ (obtained via the axial Ward identity at fixed hard-quasiparticle distributions) take the forms
\begin{align}
 \mathcal Z&=\partial_E\bigl[\gamma^0S_R^{-1}(K)\bigr]
 =\one_4+m_{{\mathrm{HTL}},f}^2\left\langle
          \frac{\one_4-\bm\alpha\cdot\mathbf v}{(E-\mathbf v\cdot\mathbf p)^2}\right\rangle_v,
 \nonumber\\
 \mathcal V_i&=\frac12\left[\alpha_i+m_{{\mathrm{HTL}},f}^2
 \left\langle\frac{v_i(\one_4-\bm\alpha\cdot\mathbf v)}{(E-\mathbf v\cdot\mathbf p)^2}\right\rangle_v\right]\gamma_5.
 \label{eq:softvertexnorm}
\end{align}
Canonically normalized on-shell soft-quasiparticle spinors $u_r(\mathbf p)$ satisfy $u_r^\dagger\mathcal Zu_s=\delta_{rs}$. Because $\one_4-\bm\alpha\cdot\mathbf v$ is positive semidefinite for any unit vector $\mathbf v$, commutes with $\gamma_5$, and satisfies $|v_i|\le1$, the matrix element of $\mathcal V_i$ in any normalized pole state is bounded by
\begin{equation}
 |u_r^\dagger\mathcal V_i u_s|\le\frac12 u_r^\dagger\mathcal Z u_r=\frac12.
 \label{eq:softvertexbound}
\end{equation}
Combining \eqref{eq:softvertexbound} with Pauli blocking $f_F^{(0)}(1-f_F^{(0)})\le\tfrac14$ and the soft phase-space volume $\cO(g^3T^3)$ shows that on-shell soft fermionic modes contribute at most $\chi_{\mathrm{soft},F}=\cO(g^3T^2)$, a parametrically small fraction $\cO(g^3)$ of the hard-quark spin susceptibility $\chi=T^2/4$.

On-shell soft gluonic modes (plasmons), by contrast, do not couple directly to the canonical spin operator~\eqref{eq:triplet}, which contains no gluon fields. Although a single soft gluon modulates the hard-quark distributions through its HTL screening cloud, this polarization cloud carries adjoint color and is proportional to the spin identity $\one_2$, so its linear overlap with the color-singlet, spin-traceless source $j$ vanishes identically. Coupling soft gluons to $j$ therefore requires either spin-dependent vertex corrections suppressed by $q/p\sim g$ or color-singlet two-plasmon composites. Integrating the Bose-enhanced statistical weight up to the Landau-damping cutoff $p\lesssim gT\sqrt{\ln(1/g)}$ \cite{FRS95,FR96} yields at most $\chi_{\mathrm{soft},B}=\cO(gT^2)$ up to logarithms, even for an order-unity effective vertex. The total soft collective contribution to $\chi$ is therefore parametrically suppressed relative to $\chi=T^2/4$. Moreover, because the lower bound~\eqref{eq:lowerbound} on hard-quark angular damping relies solely on $2\leftrightarrow2$ scattering of hard external particles and is independent of soft-mode distributions, on-shell soft modes modify the bounds~\eqref{eq:weightbound}, \eqref{eq:adiabatic-supp}, and \eqref{eq:slowstate-supp} by at most terms vanishing as $g\to0$. Finally, soft transverse gluons contribute only $\cO(gT^3)$ to the rotational susceptibility norm $(j_\Omega,j_\Omega)$.

\subsection{Numerical spectrum, process breakdown, and basis convergence}
\label{sec:finitebasis-supp}
We evaluate the charge-conjugation-even $J=1$ axial channel and the $C$-odd strange-quark vector current using the complete leading-order collision operator discretized in cubic B-spline bases with $N_b$ basis functions per radial channel (giving total matrix dimensions $7N_b$ for spin and $3N_b$ for the vector current) at the benchmark weak-coupling parameter set
\begin{equation}
 g=10^{-3},\qquad (\kappa_u,\kappa_d,\kappa_s)=(0,0,0.3),\qquad
 q_*=\sqrt g\,T,\qquad 0.01<p/T<24.
 \label{eq:finitebasisconfig}
\end{equation}
The choice $g=10^{-3}$ cleanly separates the asymptotic scales $gT\ll q_*\ll T$ ($gT/q_*=q_*/T=\sqrt g\simeq0.0316$) with current quark masses $m_{0f}=\kappa_fgT$. All responses are normalized by the exact continuum susceptibilities $(j,j)=T^3/4$ and $(j_J,j_J)=T^3/3$, and relaxation rates are expressed in units of the strange-quark vector-current effective relaxation rate $\Gamma_{\mathrm{tr}}=1.713\times10^{-13}\,T$ at $N_b=40$. The time-domain relaxation curve $\delta s(t)/(\chi h_0)$ and absorptive frequency response $\operatorname{Im}G^R(\omega)/\chi$ in Fig.~2 of the Letter are evaluated from \eqref{eq:spectralquantities} at $N_b=40$, alongside the single-relaxation-time approximation $G^R_\tau(\omega)$ from \eqref{eq:benchmarks}. All assembled collision matrices are numerically symmetric and positive definite with strictly positive eigenvalues $\lambda_n>0$.

\begin{table}[t]
 \centering
 \begingroup\renewcommand{\arraystretch}{1.25}
 \begin{tabular}{@{}lrr@{}}\toprule
 Quantity & $N_b=24$ & $N_b=40$\\\midrule
 Lowest axial eigenvalue $\lambda_0/T$ & $1.277\times10^{-13}$ & $1.279\times10^{-13}$\\
 Effective rate of the strange vector current $\Gamma_{\mathrm{tr}}/T$ & $1.710\times10^{-13}$ & $1.713\times10^{-13}$\\
 Ratio of spin and current effective rates $\Gamma_{\mathrm{eff}}/\Gamma_{\mathrm{tr}}$ & $0.9277$ & $0.9278$\\
 Lowest axial eigenvalue in rate units $\lambda_0/\Gamma_{\mathrm{tr}}$ & $0.7471$ & $0.7467$\\
 Susceptibility weight of the lowest axial mode $w_0$ & $0.3595$ & $0.3569$\\
 Fraction of spin susceptibility omitted, $r_N$ & $1.2\times10^{-7}$ & $8.2\times10^{-8}$\\\bottomrule
 \end{tabular}
 \endgroup
 \caption{Spectral quantities of the complete leading-order collision operator for the parameter set~\eqref{eq:finitebasisconfig} with $N_b=24$ and $40$ B-spline basis functions per radial channel (axial matrix dimensions $168$ and $280$; vector-current dimensions $72$ and $120$). Here $\Gamma_{\mathrm{tr}}=m_1/m_2$ is the effective relaxation rate of the strange-quark vector current at each $N_b$, and $\lambda_0$ is the lowest axial eigenvalue.}
 \label{tab:finitebasis}
\end{table}

\begin{figure}[t]
 \centering
 \includegraphics[width=160mm]{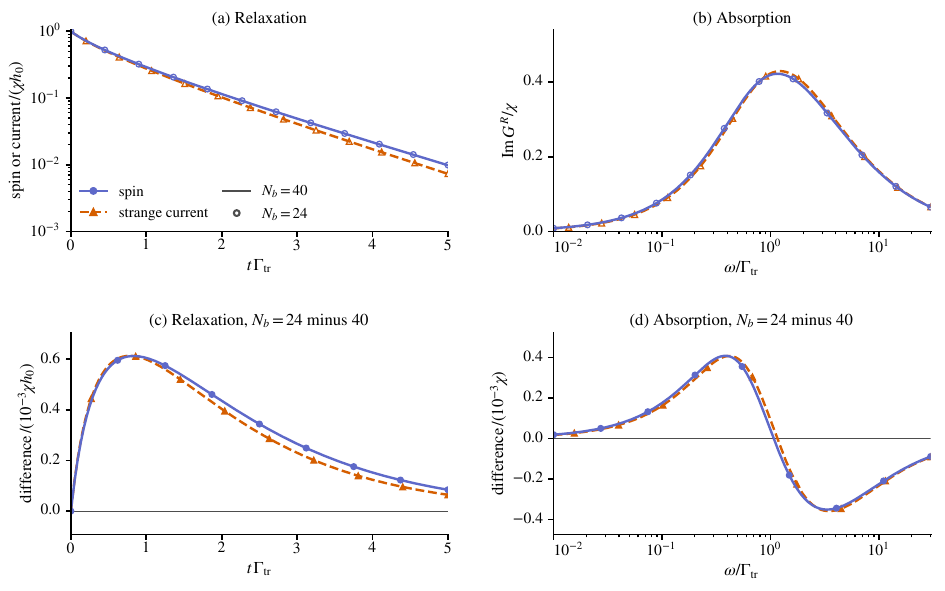}
 \caption{Basis and quadrature convergence of the canonical spin density (purple solid lines and circles) and strange-quark vector current (orange dashed lines and triangles) for the parameter set~\eqref{eq:finitebasisconfig}. Panels (a) and (b) display the time-domain relaxation and absorptive frequency response for $N_b=40$ (curves) and $N_b=24$ (open markers). Panels (c) and (d) show the corresponding point-by-point differences ($N_b=24$ minus $N_b=40$) in units of $10^{-3}$. Responses are normalized by the exact continuum susceptibilities $\chi$, and rates by $\Gamma_{\mathrm{tr}}$ at $N_b=40$.}
 \label{fig:discretization}
\end{figure}

Table~\ref{tab:finitebasis} compares the principal spectral quantities for $N_b=24$ and $40$. The cubic B-spline basis on $0.01<p/T<24$ omits a negligible fraction $r_N<1.2\times10^{-7}$ of the spin susceptibility. The lowest resolved axial eigenmode at $\lambda_0\simeq0.747\,\Gamma_{\mathrm{tr}}$ carries $w_0\simeq36\%$ of the spin susceptibility, while the remaining $64\%$ is distributed across faster transport-scale eigenmodes. The ratio of the spin and vector-current effective relaxation rates is $\Gamma_{\mathrm{eff}}/\Gamma_{\mathrm{tr}}=0.9278$, shifting by only $9\times10^{-5}$ between $N_b=24$ and $40$. Across the entire frequency range $10^{-3}\le\omega/\Gamma_{\mathrm{tr}}\le100$ and time interval $0\le t\Gamma_{\mathrm{tr}}\le5$, the $N_b=24$ and $40$ results differ by at most $7.9\times10^{-4}$ in $G^R(\omega)/\chi$ and $6.1\times10^{-4}$ in $\delta s(t)/(\chi h_0)$ (Fig.~\ref{fig:discretization}), reflecting the combined convergence of the radial B-spline basis, five-dimensional angular quadratures, and impact-parameter LPM meshes.

\begin{table}[t]
 \centering
 \begingroup\renewcommand{\arraystretch}{1.2}
 \begin{tabular}{@{}lcccc@{}}\toprule
 $\omega/\Gamma_{\mathrm{tr}}$ & $0.5$ & $1$ & $2$ & $5$\\\midrule
 $|G^R(\omega)-G^R_\tau(\omega)|/\chi$ & $0.035$ & $0.080$ & $0.110$ & $0.093$\\\bottomrule
 \end{tabular}
 \endgroup
 \caption{Deviation of the single-relaxation-time approximation $G^R_\tau(\omega)$ in \eqref{eq:benchmarks} from the full multi-mode response $G^R(\omega)$ at representative frequencies for $N_b=40$. Values for $N_b=24$ differ by less than $10^{-4}$.}
 \label{tab:errorunits}
\end{table}

Because $64\%$ of the spin susceptibility resides in modes with $\lambda_n>\lambda_0$, the single-relaxation-time approximation $G^R_\tau(\omega)$ deviates from the full multi-mode response by up to $11\%$ of $\chi$ near $\omega\simeq2\,\Gamma_{\mathrm{tr}}$ (Table~\ref{tab:errorunits}). In the time domain, at $t\Gamma_{\mathrm{tr}}=1$, the canonical spin density and strange-quark vector current retain comparable fractions ($0.290$ and $0.275$) of their initial amplitudes.

\begin{table}[t]
 \centering
 \begingroup\renewcommand{\arraystretch}{1.2}
 \begin{tabular}{@{}lrrrr@{}}\toprule
 & \multicolumn{2}{c}{Lowest axial mode $\Phi_0$} & \multicolumn{2}{c}{Projected spin source}\\
 \cmidrule(lr){2-3}\cmidrule(lr){4-5}
 Collision process & $N_b=24$ & $N_b=40$ & $N_b=24$ & $N_b=40$\\\midrule
 Hard $2\leftrightarrow2$ scattering ($Q_H$) & $51.6$ & $51.7$ & $52.1$ & $51.8$\\
 Soft-gluon exchange ($Q_g$) & $36.0$ & $35.9$ & $36.4$ & $36.8$\\
 Soft-fermion conversion ($Q_f$) & $3.9$ & $3.8$ & $11.3$ & $11.3$\\
 LPM collinear splitting ($Q_{12}$) & $8.5$ & $8.6$ & $0.14$ & $0.14$\\\bottomrule
 \end{tabular}
 \endgroup
 \caption{Percentage contributions of the four microscopic collision processes to the axial damping quadratic form $Q[\phi,\phi]$ at $q_*=\sqrt g\,T$, evaluated on the normalized lowest axial eigenmode $\Phi_0$ ($(\Phi_0,\Phi_0)=1$) and on the spin source projected onto the variational subspace.}
 \label{tab:fractions}
\end{table}

Table~\ref{tab:fractions} lists the percentage contributions of the four microscopic collision channels to the damping quadratic form for the lowest axial mode $\Phi_0$ and for the projected spin source. Hard $2\leftrightarrow2$ scattering and soft-gluon exchange both conserve quark chirality and relax spin via momentum deflections, together accounting for $87.6\%$--$88.6\%$ of the total damping in both cases. The remaining $\sim12\%$ is partitioned differently between the two states: for the projected spin source, $11.3\%$ arises from soft-fermion conversion between quarks and gluons and only $0.14\%$ from collinear splitting, because $j$ is momentum independent for quarks and vanishes for gluons (so $\Delta\ell=\ell_f(zP)-\ell_f(P)=0$ in \eqref{eq:circularendpoint}, leaving only small current-mass corrections). By contrast, the lowest eigenmode $\Phi_0$ admixes gluon circular polarization to minimize its decay rate, receiving $8.6\%$ of its damping from LPM collinear splitting and $3.8\%$ from soft-fermion conversion.

\begin{table}[t]
 \centering
 \begingroup\renewcommand{\arraystretch}{1.2}
 \begin{tabular}{@{}lrrrrr@{}}\toprule
 $\kappa=(\kappa_u,\kappa_d,\kappa_s)$ & $\Gamma_{\mathrm{tr}}/T$ & Spin $m_1/m_2$ $[T]$ & Spin/current & $\lambda_0/\Gamma_{\mathrm{tr}}$ & $w_0$\\\midrule
 $(0,0,0)$ & $1.716\times10^{-13}$ & $1.587\times10^{-13}$ & $0.9249$ & $0.7449$ & $0.3597$\\
 $(0,0,0.3)$ & $1.710\times10^{-13}$ & $1.586\times10^{-13}$ & $0.9277$ & $0.7471$ & $0.3595$\\
 $(0,0,1)$ & $1.673\times10^{-13}$ & $1.580\times10^{-13}$ & $0.9445$ & $0.7597$ & $0.3510$\\
 $(0,0,2)$ & $1.628\times10^{-13}$ & $1.570\times10^{-13}$ & $0.9643$ & $0.7708$ & $0.3083$\\
 $(0.3,0.3,0.3)$ & $1.710\times10^{-13}$ & $1.585\times10^{-13}$ & $0.9269$ & $0.7463$ & $0.3592$\\\bottomrule
 \end{tabular}
 \endgroup
 \caption{Dependence of the relaxation spectrum on the dimensionless current-mass parameters $\kappa=(\kappa_u,\kappa_d,\kappa_s)$ at $N_b=24$, with $g$, $q_*$, and the momentum grid of \eqref{eq:finitebasisconfig}. Because the Gram matrix $(b_A,b_B)$, the leading-order source $j$, and the hard-scattering and soft-gluon quadratic forms $Q_H, Q_g$ are independent of $\kappa$ at leading order, variations across rows stem entirely from soft-fermion exchange ($Q_f$) and collinear splitting ($Q_{12}$). Each row is normalized by its corresponding strange-current rate $\Gamma_{\mathrm{tr}}$.}
 \label{tab:masspoints}
\end{table}

Table~\ref{tab:masspoints} examines the sensitivity of the relaxation spectrum to the dimensionless current-quark mass parameters $\kappa=(\kappa_u,\kappa_d,\kappa_s)$ at $N_b=24$. Varying $\kappa_s$ from $0$ (the chiral limit) to $2$ shifts the effective spin relaxation rate $m_1/m_2$ by only $1.1\%$ (from $1.587\times10^{-13}\,T$ to $1.570\times10^{-13}\,T$), confirming that angular diffusion of helicity-locked spins dominates over mass-induced chirality flips throughout the light-quark regime. Meanwhile, the ratio $\Gamma_{\mathrm{eff}}/\Gamma_{\mathrm{tr}}$ increases slightly from $0.925$ to $0.964$ because a larger strange-quark mass modifies the exchanged-fermion propagator in \eqref{eq:softfermionform} and increases the collinear formation energy via $M_s^2=m_{0s}^2+m_{\infty,s}^2$ in \eqref{eq:lpmamass}, which suppresses soft-fermion conversion and collinear emission for strange quarks and reduces $\Gamma_{\mathrm{tr}}/T$ from $1.716\times10^{-13}$ to $1.628\times10^{-13}$.

\FloatBarrier
\section{Rotational Equilibrium, Vortical Kinetic Theory, and Hydrodynamic Modes}
\label{sec:rotation-supp}

\subsection{Rotational equilibrium Wigner function and the axial vortical effect}
We now analyze how fluid rotation polarizes the plasma in local equilibrium and how rapidly the spin density responds to a time-dependent or spatially varying vorticity. We reserve $\mathbf u$ for the background three-velocity and $\delta\mathbf u$ for an infinitesimal velocity perturbation, so that the total velocity is $\mathbf u+\delta\mathbf u$, with corresponding vorticities $\bm\Omega=\tfrac12\nabla\times\mathbf u$ and $\delta\bm\Omega=\tfrac12\nabla\times\delta\mathbf u$.

For a relativistic fluid in global rotational equilibrium at angular velocity $\bm\Omega$ (with magnitude $|\bm\Omega|$), the grand-canonical density operator is $\hat\rho_{\mathrm{rot}}\propto\exp[-(H-\bm\Omega\cdot\mathbf J)/T]$, where $\mathbf J=\mathbf L+\mathbf S$ is the conserved total angular momentum \cite{Zubarev79,Bec12,BT10,Bec13,AW14,CG17,BH26}. In the quantum kinetic theory of QCD \cite{Lin26} [see (142) and (145) therein], expanding the quark and gluon Wigner functions to first order in spatial gradients in the local rest frame of the fluid yields two distinct contributions to the phase-space structure:
\begin{enumerate}
\item \emph{Helicity-polarized local-equilibrium distributions.} The spin--vorticity coupling $-\bm\Omega\cdot\mathbf S$ shifts the local-equilibrium phase-space distributions. To first order in $|\bm\Omega|/T$, these take the form $f_F^{(0)}(p-\tfrac\lambda2\bm\Omega\cdot\mathbf n)$ for quarks \cite{CSSYY14} and $f_B^{(0)}(p-\lambda\bm\Omega\cdot\mathbf n)$ for gluons \cite{Lin26}, where $\lambda=\pm1$ denotes the quark helicity sign (physical helicity $\lambda/2$) or gluon circular polarization. Expanding these distributions yields the rotational perturbation $\Phi_{\mathrm{rot}}=\Omega_ij_\Omega^i/T$, where in the seven-component $J=1$ radial basis~\eqref{eq:j1state} the rotational source vector $j_\Omega$ reads
\begin{equation}
 j_\Omega=\bigl(\sqrt{2/3},\sqrt{2/3},\sqrt{2/3},2/\sqrt3,0,0,0\bigr)^{\mathsf T},\qquad
 (j,j_\Omega)=\frac{T^3}4,\qquad (j_\Omega,j_\Omega)=\frac{41T^3}{36}.
 \label{eq:jomega}
\end{equation}
Here the gluon entry corresponds to $j_{\Omega,g}^i=n_i\operatorname{diag}(-1,+1)$ in the gluon ordering $(-,+)$. The normalization $(j_\Omega,j_\Omega)=\tfrac{41}{36}T^3$ follows from combining the quark contribution $\tfrac14T^3$ with the gluon Bose integral $\int_0^\infty\dd p\,p^2f_B^{(0)}(p)[1+f_B^{(0)}(p)]=\pi^2T^3/3$. For any perturbation $\Phi$, the inner product $(\mathbf j_\Omega,\Phi)$ measures the total helicity angular momentum density carried by the quasiparticles, weighting quarks by $\pm\tfrac12$ and gluons by $\pm1$. Projecting $\Phi_{\mathrm{rot}}$ onto the canonical quark spin source $j$ gives the quasiparticle contribution to the equilibrium spin density, $(\mathbf j,\Phi_{\mathrm{rot}})=\chi\bm\Omega$.

\item \emph{Chiral magnetization current.}
In chiral kinetic theory, Lorentz covariance requires a magnetization
contribution to the particle-number current of each helicity $\lambda/2$,
weighted by $\lambda\mathbf n/(2p)$ \cite{CSSYY14}.
Because the axial current supplies an additional helicity factor $\lambda$
(with $\lambda^2=1$), the resulting magnetization term depends on the spin-summed distribution $\tr_2\delta f_q$
\cite{Gao19,Lin26}.
Summing over flavors, antiquarks, and colors and using
$\mathbf s=\mathbf J_5/2$ gives
\begin{equation}
 \mathbf s_{\mathrm{mag}}=\frac14\nabla\times\sum_{q,\bar q}d_{c,q}\intp\frac{\mathbf n}{p}\tr_2\delta f_q(\mathbf x,\mathbf p) .
 \label{eq:magnetization}
\end{equation}
\begin{samepage}
Unlike the quasiparticle term $(\mathbf j,\Phi)$, $\mathbf s_{\mathrm{mag}}$ depends solely on the spin-averaged quark distribution $\tr_2\delta f_q$. As a spatial curl, it integrates to zero for a localized fluid and contributes no net global angular momentum. A small local boost $f_F^{(0)}(p-\mathbf p\cdot\mathbf u)$ yields
$\delta f_q=f_F^{(0)}(p)[1-f_F^{(0)}(p)]
(\mathbf p\cdot\mathbf u/T)\one_2$ to first order in $\mathbf u$.
Substituting this shift into \eqref{eq:magnetization} and using
$\intp n_i n_j f_F^{(0)}(1-f_F^{(0)})
=\frac{T^3}{36}\delta_{ij}$ per color and spin state gives,
for three massless flavors,
\begin{equation*}
 \mathbf s_{\mathrm{mag}}=\frac14T^2\nabla\times\mathbf u=\frac12T^2\bm\Omega=2\chi\bm\Omega.
\end{equation*}
\end{samepage}
\end{enumerate}

Because collisional corrections to the Wigner function vanish in global rotational equilibrium, the total equilibrium canonical spin density in (5) of the Letter is the sum of the quasiparticle and magnetization contributions:
\begin{equation*}
 \mathbf s_{\mathrm{eq}}=(\mathbf j,\Phi_{\mathrm{rot}})+\mathbf s_{\mathrm{mag}}=\chi\bm\Omega+2\chi\bm\Omega=3\chi\bm\Omega=\frac{3T^2}{4}\bm\Omega.
\end{equation*}
This reproduces the leading-order axial vortical effect for $N_c=3, N_f=3$ massless quarks, $\mathbf J_5=2\mathbf s_{\mathrm{eq}}=\frac{N_cN_f}{6}T^2\bm\Omega=\frac32T^2\bm\Omega$ \cite{Vilenkin79,Vilenkin80,SonSurowka09,Landsteiner11,Landsteiner14}, as well as its $1:2$ partition between polarized quasiparticle distributions and magnetization currents in the comoving fluid frame \cite{CSSYY14,Gao19}.

It is instructive to contrast the rotational coefficient $3\chi$ with the linear response to an external axial-vector field $h_i$ conjugate to $s^i$ \cite{Landsteiner14,H21}. At finite kinetic frequencies $\omega\sim\gamma_{\mathrm{tr}}$ and $\mathbf k=0$, driving the system with $h_i(\omega)$ perturbs the quasiparticle distributions and induces $(j,\Phi)$ with susceptibility $\chi=(j,j)/T=T^2/4$ [\eqref{eq:forcedresponse}]. For free massless quarks in the static limit ($\omega=0, \mathbf k\to0$), however, a constant uniform axial gauge field $A_\mu^5=(0,\mathbf h/2)$ in the massless Dirac Lagrangian $\bar q\gamma^\mu(\ii\partial_\mu+A_\mu^5\gamma_5)q$ merely shifts the canonical spatial momenta of right- and left-handed quarks by $\mathbf p\to\mathbf p\mp\mathbf h/2$. In dimensional or gauge-invariant ultraviolet regularization, a contact term analytic at $\omega\sim\gamma_{\mathrm{tr}}$ cancels the response to this constant shift, so the full static susceptibility conjugate to $\mathbf h$ vanishes at $\omega=0$ \cite{Landsteiner14}. Fluid rotation $\bm\Omega$, by contrast, couples to the conserved total angular momentum generator $\mathbf J=\mathbf L+\mathbf S$ in $\exp[-(H-\bm\Omega\cdot\mathbf J)/T]$, which cannot be gauged away by a momentum shift and fixes the equilibrium spin density to $3\chi\bm\Omega$ \cite{Bec12,AW14,Landsteiner14}.

A symmetric shear velocity gradient $\sigma_{ij}=\tfrac12(\partial_i u_j+\partial_j u_i-\tfrac23\delta_{ij}\nabla\cdot\mathbf u)$ also polarizes quarks at individual momenta $\mathbf p$ in local equilibrium via the Wigner function \cite{Fu21,Bec21}. Because $\sigma_{ij}$ is symmetric and traceless, any axial vector constructed from $\sigma_{ij}$, the momentum direction $n_k$, and $\epsilon_{ijk}$ averages to zero over the momentum sphere $\avg{\cdots}$. Shear-induced polarization therefore yields no momentum-integrated spin density $\mathbf s(\mathbf x)$ at first order in gradients.

\subsection{Nonlocal collisions, angular-momentum conservation, and coupled fluid--spin equations}
We now derive the kinetic equation governing spin in a vortical flow and the reciprocal collisional torque exerted on the fluid velocity.

Because the fluid velocity fixes the spin-independent shift of the phase-space distributions while the helicity imbalance relaxes independently, we separate the local-equilibrium flow perturbation $\Phi_{\mathrm{flow}}$ from the polarized perturbation $\Phi$. Retaining $\Phi_{\mathrm{flow}}$ and accounting for the spin-averaged shear correction through the viscosity $\eta$ below, we write
\begin{equation*}
 f_a=f_a^{(0)}\one_2
 +f_a^{(0)}\bff_a^{(0)}\Phi_{\mathrm{full},a},
 \qquad
 \Phi_{\mathrm{full}}=\Phi_{\mathrm{flow}}+\Phi,
\end{equation*}
where $\Phi$ is traceless in helicity space and incorporates the axial-charge and higher angular moments generated by free streaming. A small local boost shifts the rest-frame energy from $p$ to $p-\mathbf p\cdot\mathbf u$; using $\partial_p f_a^{(0)}=-f_a^{(0)}\bff_a^{(0)}/T$, a Taylor expansion gives
\begin{equation*}
\begin{aligned}
 f_a^{(0)}(p-\mathbf p\cdot\mathbf u)
 &=f_a^{(0)}(p)-(\mathbf p\cdot\mathbf u)\partial_p f_a^{(0)}(p)
 +\cO(u^2)=f_a^{(0)}(p)+f_a^{(0)}(p)\bff_a^{(0)}(p)
 \frac{\mathbf p\cdot\mathbf u}{T}+\cO(u^2),\\
 \Phi_{\mathrm{flow},a}
 &=\frac{\mathbf p\cdot\mathbf u}{T}\one_2.
\end{aligned}
\end{equation*}

Consider first a rigidly rotating plasma, $\mathbf u=\bm\Omega\times\mathbf x$, confined within the light cylinder $|\mathbf u|=|\bm\Omega|\rho<1$ (where $\rho$ is the transverse distance from the rotation axis) and described by the stationary global-equilibrium density operator $\exp[-(H-\bm\Omega\cdot\mathbf J)/T]$ \cite{Bec12,AW14,CG17}. Near the axis ($|\bm\Omega|\rho\ll1$), where the linear boost expansion applies, the equilibrium polarization to first order in $\bm\Omega$ is $\Phi=\Phi_{\mathrm{rot}}=\Omega_ij_\Omega^i/T$. The flow perturbation $\Phi_{\mathrm{flow}}$ is constant along free-particle trajectories because $\mathbf n\cdot\nabla[\mathbf p\cdot(\bm\Omega\times\mathbf x)]=\mathbf n\cdot(\mathbf p\times\bm\Omega)=0$. At a fixed collision point, its gain-minus-loss variation in \eqref{eq:hardmeasure} likewise vanishes by momentum conservation:
\begin{equation*}
\begin{aligned}
 U_{\Phi_{\mathrm{flow}}}[A]
 =(\Phi_{\mathrm{flow},3}\otimes\one+\one\otimes\Phi_{\mathrm{flow},4})A
 -A(\Phi_{\mathrm{flow},1}\otimes\one+\one\otimes\Phi_{\mathrm{flow},2})=\frac{\mathbf u\cdot
 (\mathbf p_3+\mathbf p_4-\mathbf p_1-\mathbf p_2)}{T}A=0,
\end{aligned}
\end{equation*}
with an identical cancellation for collinear splitting and merging, so that $C\Phi_{\mathrm{flow}}=0$. By contrast, $C\Phi_{\mathrm{rot}}\ne0$ because local scattering and splitting processes exchange helicity and orbital angular momentum.

Stationarity of global rotational equilibrium requires the nonlocal collision term to cancel $-C\Phi_{\mathrm{rot}}$ \cite{Chen15,HPY17,WS21,WW22,WWSR22,Wagner22,Lin26}. For fixed incoming and outgoing helicities, evaluating $\Phi_{\mathrm{flow}}$ at the displaced wave-packet centers $\mathbf x+\Delta\mathbf x_r$ adds
\begin{equation*}
 \sum_{\mathrm{out}}
 \frac{\mathbf p_r\cdot(\bm\Omega\times\Delta\mathbf x_r)}{T}
 -\sum_{\mathrm{in}}
 \frac{\mathbf p_r\cdot(\bm\Omega\times\Delta\mathbf x_r)}{T}
 =\frac{\bm\Omega\cdot\Delta\mathbf L}{T}
 =-\frac{\bm\Omega\cdot\Delta\mathbf S}{T}
\end{equation*}
to the collisional variation, where $\Delta$ denotes outgoing minus incoming angular momentum and the last step uses total angular-momentum conservation, $\Delta\mathbf L+\Delta\mathbf S=0$. Because this exactly cancels the local variation $\bm\Omega\cdot\Delta\mathbf S/T$ arising from $\Phi_{\mathrm{rot}}$, the nonlocal collision term acting on the spin-averaged flow supplies $+C\Phi_{\mathrm{rot}}$ when the local collision operator is normalized as $-C\Phi$. Combining the local and nonlocal contributions prior to momentum integration shows that $C$ acts on the departure $\Phi-\Phi_{\mathrm{rot}}$, to which the infrared regularity condition~\eqref{eq:gluonformdomain} applies. Since vorticity is the sole axial vector constructible from first-order fluid gradients in the charge-conjugation-even $J=1$ channel, rotational equilibrium uniquely fixes this source term for a general vorticity field $\bm\Omega(\mathbf x,t)=\tfrac12\nabla\times\mathbf u$. Including a spatially uniform conjugate field $h_i(t)$, the linearized kinetic equation reads
\begin{equation}
 \partial_t\Phi+\mathbf n\cdot\nabla\Phi=-C\Bigl(\Phi-\Phi_{\mathrm{rot}}-\frac{h^ij^i}{T}\Bigr),
 \label{eq:kinetic-supp}
\end{equation}
which reproduces (6) of the Letter at $h=0$ and reduces to \eqref{eq:forcedresponse} in a static medium ($\Phi_{\mathrm{rot}}=0$).

To analyze coupled hydrodynamic and kinetic modes, we linearize about a static unpolarized background ($\mathbf u=0$) with an infinitesimal incompressible velocity perturbation $\delta\mathbf u$ ($|\delta\mathbf u|\ll1$, $\nabla\cdot\delta\mathbf u=0$) and vortical source $\delta\Phi_{\mathrm{rot}}=\delta\Omega_i j_\Omega^i/T$. The total perturbation is $\Phi_{\mathrm{full}}=\delta\Phi_{\mathrm{flow}}+\Phi$, with $\delta\Phi_{\mathrm{flow},a}=(\mathbf p\cdot\delta\mathbf u/T)\one_2$, and \eqref{eq:kinetic-supp} applies with $\Phi_{\mathrm{rot}}\to\delta\Phi_{\mathrm{rot}}$.

Microscopic conservation of energy, linear momentum, and total angular momentum ensures that the symmetric (Belinfante) energy-momentum tensor $T^{\mu\nu}$ satisfies $\partial_\mu T^{\mu\nu}=0$ \cite{Chen15,H21,Li21}, where angular-momentum conservation ($\partial_\lambda J_{\mathrm{can}}^{\lambda,\mu\nu}=0$) eliminates the antisymmetric part of the canonical energy-momentum tensor (equivalently obtained by expanding the Wigner-function energy-momentum tensor to first order in gradients \cite{Chen15,LL87,dGM84}). In the background rest frame, the momentum-density perturbation $\delta T^{0i}$ comprises the fluid momentum $w\delta u^i$ (with enthalpy density $w=\epsilon+P=\frac{19\pi^2}{9}T^4$ for three-flavor massless QCD) and half the curl of the helicity angular momentum density $(\mathbf j_\Omega,\Phi)$:
\begin{equation}
 \delta T^{0i}=w\delta u^i+\tfrac12\bigl[\nabla\times(\mathbf j_\Omega,\Phi)\bigr]^i,\qquad
 \int\dd^3x\,\mathbf x\times\tfrac12\nabla\times(\mathbf j_\Omega,\Phi)=\int\dd^3x\,(\mathbf j_\Omega,\Phi).
 \label{eq:momentumdensity}
\end{equation}
The stress tensor perturbation $\delta T^{ij}$ similarly contains the free-streaming flux of helicity angular momentum, whose divergence cancels the streaming contribution $\tfrac12\nabla\times(\mathbf j_\Omega,-\mathbf n\cdot\nabla\Phi)$ to $\partial_t[\tfrac12\nabla\times(\mathbf j_\Omega,\Phi)]$. Combining $\partial_\mu\delta T^{\mu i}=0$ with \eqref{eq:kinetic-supp} at $h=0$ yields the linearized transverse Navier--Stokes equation with collisional spin torque,
\begin{equation}
 w\,\partial_t\delta\mathbf u=\eta\nabla^2\delta\mathbf u+\tfrac12\nabla\times\bigl(\mathbf j_\Omega,C(\Phi-\delta\Phi_{\mathrm{rot}})\bigr),
 \label{eq:torque-supp}
\end{equation}
which gives (8) of the Letter. This mechanical torque transfers the angular momentum damped from the quasiparticle helicity distributions to the fluid flow and forms the Onsager-reciprocal partner \cite{dGM84,H21} of the vortical source $+C\delta\Phi_{\mathrm{rot}}$ in \eqref{eq:kinetic-supp}.

Including the magnetization current~\eqref{eq:magnetization}, the canonical spin density reads
\begin{equation}
 \mathbf s=(\mathbf j,\Phi)+\mathbf s_{\mathrm{mag}},
 \label{eq:spinreadout-rot}
\end{equation}
where $\mathbf s_{\mathrm{mag}}=2\chi\delta\bm\Omega$ in local equilibrium, and off-equilibrium spin-averaged perturbations contribute a magnetization current bounded by $\sqrt3k/(2\pi T)$ times the maximum spin allowed by their free energy. Eliminating the spin-averaged shear distribution in favor of $\eta$ in \eqref{eq:torque-supp} leaves $\mathbf s=(\mathbf j,\Phi)+2\chi\delta\bm\Omega$ at leading gradient order. Because $\Phi$ is traceless in helicity space, $(\delta\Phi_{\mathrm{flow}},\Phi)=0$, while the flow contribution $T(\delta\Phi_{\mathrm{flow}},\delta\Phi_{\mathrm{flow}})/2=w(\delta\mathbf u)^2/2$ reproduces the fluid kinetic energy density. The total quadratic free energy of the perturbation is therefore
\begin{equation*}
 \int\dd^3x\,\frac{T}{2}(\Phi_{\mathrm{full}},\Phi_{\mathrm{full}})
 =\int\dd^3x\,\left[\frac{w(\delta\mathbf u)^2}{2}+\frac{T}{2}(\Phi,\Phi)\right].
\end{equation*}
Under free evolution ($h=0$) with periodic boundary conditions or vanishing surface fluxes, free streaming conserves this functional, whereas integration by parts on the mechanical torque term gives
\begin{equation*}
 \frac12\int\dd^3x\,\delta\mathbf u\cdot\nabla\times\bigl(\mathbf j_\Omega,C(\Phi-\delta\Phi_{\mathrm{rot}})\bigr)
 =T\int\dd^3x\,(\delta\Phi_{\mathrm{rot}},C(\Phi-\delta\Phi_{\mathrm{rot}})).
\end{equation*}
Combining this relation with \eqref{eq:kinetic-supp} yields the exact free-energy dissipation identity
\begin{equation}
 \frac{\dd}{\dd t}\int\dd^3x\,\left[\frac{w(\delta\mathbf u)^2}{2}+\frac{T}{2}(\Phi,\Phi)\right]
 =-\int\dd^3x\,\left[\eta(\partial_i\delta u_j)^2+TQ[\Phi-\delta\Phi_{\mathrm{rot}},\Phi-\delta\Phi_{\mathrm{rot}}]\right]\le0,
 \label{eq:dissipation-supp}
\end{equation}
reproducing (9) of the Letter and combining viscous shear dissipation with collisional spin damping of the departure $\Phi-\delta\Phi_{\mathrm{rot}}$ from rotational equilibrium.

For a prescribed spatially uniform vorticity along a fixed Cartesian axis $i$ (no sum, with $s=s^i$ and $j=j^i$) varying as $e^{-\ii\omega t}$, solving \eqref{eq:kinetic-supp} at $h=0$ yields the frequency-dependent spin response
\begin{equation}
 s(\omega)=2\chi\Omega^i(\omega)
 +\bigl(j,C(C-\ii\omega)^{-1}\Phi_{\mathrm{rot}}(\omega)\bigr).
 \label{eq:vortical-supp}
\end{equation}
The first term represents the instantaneous magnetization response, and $(j,\Phi_{\mathrm{rot}})=\chi\Omega^i$ recovers the static limit $s=3\chi\Omega^i$. Expanding at low frequencies $|\omega|\ll\gamma_{\mathrm{tr}}$ gives (7) of the Letter:
\begin{equation*}
 \mathbf s(t)=3\chi\bm\Omega(t)-\chi\tau_\Omega\partial_t\bm\Omega(t)+\cdots,
 \qquad
 \tau_\Omega=\frac{(j,C^{-1}\Phi_{\mathrm{rot}})}{\chi\Omega^i}.
\end{equation*}
Here $\tau_\Omega$ is independent of $|\bm\Omega|$ and equals the integrated relaxation time $\int_0^\infty\dd t\,(j,e^{-Ct}\Phi_{\mathrm{rot}})/(\chi\Omega^i)$, evaluated numerically in Sec.~\ref{sec:rotnum-supp}.

To bound the dynamic departure from rotational equilibrium, we use the slow spectral weight $W_s(\Gamma)$ defined in \eqref{eq:slowfraction}. Subtracting the static limit gives
\begin{equation*}
 \frac{s(\omega)}{3\chi\Omega^i(\omega)}-1
 =\frac{\bigl(j,\ii\omega(C-\ii\omega)^{-1}\Phi_{\mathrm{rot}}(\omega)\bigr)}{3\chi\Omega^i(\omega)}.
\end{equation*}
Splitting the spectral integral at an arbitrary threshold $\Gamma>0$ as in \eqref{eq:diagonalbound}, we bound the low-rate contribution ($\lambda\le\Gamma$, where $|\ii\omega/(\lambda-\ii\omega)|\le1$) via the Cauchy--Schwarz inequality in terms of $\sqrt{W_s(\Gamma)}$, and the high-rate contribution ($\lambda>\Gamma$) via $|\ii\omega/(\lambda-\ii\omega)|\le|\omega|/\Gamma$. Using $(j,j)=T\chi$ and $(\Phi_{\mathrm{rot}},\Phi_{\mathrm{rot}})=41T|\bm\Omega|^2/36$ from \eqref{eq:jomega} yields
\begin{equation}
 \left|\frac{s(\omega)}{3\chi\Omega^i(\omega)}-1\right|\le\frac{\sqrt{41}}9\left[\sqrt{W_s(\Gamma)}+\frac{|\omega|}{\Gamma}\right],
 \label{eq:adiabatic-supp}
\end{equation}
with $\sqrt{41}/9\simeq0.71$, where the two terms bound the slow-mode weight and the high-mode phase lag, respectively. Together with \eqref{eq:weightbound}, choosing $|\omega|\ll\Gamma\ll\gamma_{\mathrm{tr}}$ demonstrates that the spin density closely tracks instantaneous rotational equilibrium.

\subsection{Coupled hydrodynamic modes and bound on off-equilibrium spin polarization}
\label{sec:hydro-supp}
For perturbations with wave vector $\mathbf k$ about a static unpolarized background ($\mathbf u=0$, $\bm\Omega=0$), invariance under $\mathrm{SO}(2)$ rotations about $\hat{\mathbf k}$ and reflections across planes containing $\hat{\mathbf k}$ decouples the linearized dynamics into three symmetry sectors:
\begin{enumerate}
\item the \emph{transverse sector} of helicity $\pm1$ about $\hat{\mathbf k}$, containing transverse spin $\mathbf s_\perp$, transverse velocity $\delta\mathbf u_\perp$, and vorticity $\delta\bm\Omega=\tfrac{\ii}{2}\mathbf k\times\delta\mathbf u_\perp$;
\item the \emph{reflection-odd longitudinal sector} of helicity $0$, containing longitudinal spin $\hat{\mathbf k}\cdot\mathbf s$ and the axial-charge densities $J_{5f}^0$;
\item the \emph{reflection-even longitudinal sector} of helicity $0$ (sound and heat diffusion), which is parity-even and spin-unpolarized, carrying no canonical spin.
\end{enumerate}
Via the free-streaming operator $\ii(\mathbf k\cdot\mathbf n)\Phi$, the $J=1$ spin profiles couple to $J=2$ axial profiles in the transverse sector, and to both $J=0$ axial charges and $J=2$ axial profiles in the reflection-odd longitudinal sector.

Along a transverse shear mode, the local mean spin density satisfies the equilibrium constitutive relation $\mathbf s=3\chi\delta\bm\Omega$ \cite{H21,Li21}, whereas along a longitudinal axial-charge diffusion mode, the diffusive axial current fixes $\mathbf s=\tfrac12\mathbf J_5=-\tfrac D2\nabla J_5^0$. Promoting spin to an independent hydrodynamic variable would instead require a slowly decaying collective mode ($\operatorname{Re}\lambda\ll\gamma_{\mathrm{tr}}$) whose \emph{off-equilibrium} spin polarization $\mathbf s-3\chi\delta\bm\Omega$ carries a finite fraction of the maximum spin compatible with its perturbation free energy.

The dissipation identity~\eqref{eq:dissipation-supp} and the microscopic lower bound~\eqref{eq:lowerbound} rule out any such mode. For a spatial Fourier mode of wave vector $\mathbf k$, the perturbation free energy is $T(\Phi_{\mathrm{full}},\Phi_{\mathrm{full}})/2=[T(\Phi,\Phi)+w\delta\mathbf u^*\cdot\delta\mathbf u]/2$, and the off-equilibrium spin at leading gradient order reads
\begin{equation}
 s-3\chi\delta\Omega^i=(j,\delta\Phi),\qquad \delta\Phi\equiv\Phi-\delta\Phi_{\mathrm{rot}}.
 \label{eq:residual-supp}
\end{equation}
Since $\delta\bm\Omega^*\cdot\delta\bm\Omega\le k^2\delta\mathbf u^*\cdot\delta\mathbf u/4$, the norm of $\delta\Phi$ relative to $\Phi_{\mathrm{full}}$ obeys
\begin{equation*}
 \sqrt{\frac{(\delta\Phi,\delta\Phi)}{(\Phi_{\mathrm{full}},\Phi_{\mathrm{full}})}}\le
 1+\frac{k}{2T}\sqrt{\frac{T(j_\Omega,j_\Omega)}{w}}
 \simeq1+0.12\frac{k}{T},
\end{equation*}
while for any eigenmode decaying as $e^{-\lambda t}$, \eqref{eq:dissipation-supp} implies
\begin{equation*}
 \operatorname{Re}\lambda\,(\Phi_{\mathrm{full}},\Phi_{\mathrm{full}})
 =Q[\delta\Phi,\delta\Phi]+\frac{\eta k^2}{T}\delta\mathbf u^*\cdot\delta\mathbf u
 \ge Q[\delta\Phi,\delta\Phi].
\end{equation*}
Splitting the spin source into contributions inside and outside the thermal interval $aT<p<bT$ as in Sec.~\ref{sec:slowweight-supp} [with the exterior susceptibility fraction $r_{ab}$ given by \eqref{eq:outsidesource}], any state satisfying $Q[\delta\Phi,\delta\Phi]\le\Gamma(\Phi_{\mathrm{full}},\Phi_{\mathrm{full}})$ obeys
\begin{equation}
 \frac{|s-3\chi\delta\Omega^i|}{\sqrt{(j,j)(\Phi_{\mathrm{full}},\Phi_{\mathrm{full}})}}
 \le\sqrt{r_{ab}}\left(1+0.12\frac{k}{T}\right)
 +\sqrt{\frac{\Gamma}{c_{ab}\gamma_{\mathrm{tr}}}},
 \label{eq:slowstate-supp}
\end{equation}
where the denominator sets the maximum spin compatible with the perturbation free energy. Thus, any slowly decaying mode ($\Gamma\ll\gamma_{\mathrm{tr}}$) carries a vanishingly small off-equilibrium spin fraction, even when its conserved hydrodynamic components are supported at thermal momenta. Similarly, for arbitrary initial data $\Phi_{\mathrm{full}}(0)$, integrating \eqref{eq:dissipation-supp} over $[0,t]$ yields $\int_0^t\dd t'\,Q[\delta\Phi,\delta\Phi]\le(\Phi_{\mathrm{full}}(0),\Phi_{\mathrm{full}}(0))/2$, bounding the time average of $|s-3\chi\delta\Omega^i|^2/[(j,j)(\Phi_{\mathrm{full}}(0),\Phi_{\mathrm{full}}(0))]$ by the square of the right-hand side of \eqref{eq:slowstate-supp} with $\Gamma=1/(2t)$. Reversible free streaming and Coriolis precession leave this bound intact.

We also estimate the nonlocal collision corrections omitted from (6) of the Letter, which retains only the local operator $C$ and the leading vortical source $+C\delta\Phi_{\mathrm{rot}}$. For a nonhydrodynamic perturbation at wave number $k$, nonlocal corrections scale as $k$ times the spatial range of the microscopic interaction: $\cO(k/T)$ for hard $2\leftrightarrow2$ scattering ($q\sim T$), $\cO(k/(gT))$ for soft HTL exchange ($q\sim gT$), and $\cO(k/(g^2T))$ for collinear LPM splitting over the coherent formation length $t_{\mathrm{form}}\sim 1/(g^2T)$. For $k\lesssim\gamma_{\mathrm{tr}}$, the largest ratio is bounded by $\gamma_{\mathrm{tr}}/(g^2T)=g^2\ln(1/g)\ll1$, confirming that these nonlocal terms are suppressed by powers of $g$ at kinetic scales and are even smaller in the hydrodynamic regime $k\ll\gamma_{\mathrm{tr}}$. For the hard small-angle collisions ($q>q_*\gg k$) used in the lower bound~\eqref{eq:lowerbound}, where the momentum transfer $q$ lies below the thermal scale $T$, shifting external momenta by $\mathbf k$ alters the collision rate by $\cO(g^4T^3k\int_{q_*}\dd q\,q^{-4})\sim g^{5/2}k\ll\gamma_{\mathrm{tr}}$ for $q_*=\sqrt g T$, leaving the transport-scale lower bound unaffected.

Projecting the coupled equations onto the $J=1$ polarization profiles and the equilibrium $J=0$ axial-charge profiles yields explicit moment equations whose small-$k$ solutions confirm the general bound~\eqref{eq:slowstate-supp}:
\begin{itemize}
\item \emph{Transverse shear sector.} Eliminating $\Phi$ from \eqref{eq:kinetic-supp} and \eqref{eq:torque-supp} for a transverse mode with decay rate $\lambda$ gives the dispersion relation
\begin{equation*}
 \lambda\left[1+\frac{k^2}{4wT}\bigl(j_\Omega,C(C-\lambda)^{-1}j_\Omega\bigr)\right]=\frac{\eta k^2}{w},
\end{equation*}
and spin-density amplitude $s=\chi\delta\Omega^i\,[2+(j,C(C-\lambda)^{-1}j_\Omega)/(T\chi)]=3\chi\delta\Omega^i[1+\tfrac13\lambda\tau_\Omega+\cO(\lambda^2)]$. As the shear mode decays, its spin density exceeds the instantaneous local-equilibrium value $3\chi\delta\Omega^i$ by a relative lag $\lambda\tau_\Omega/3\sim(\eta/w)\tau_\Omega k^2\sim(k/\gamma_{\mathrm{tr}})^2$. Conversely, releasing an initial off-equilibrium spin polarization $\chi h_0$ transfers only a fraction $3\chi k^2/(4w)\simeq0.009(k/T)^2$ of its amplitude into the hydrodynamic shear mode.

\item \emph{Longitudinal axial-charge sector.} For a flavor-symmetric collision operator, the singlet axial charge decouples from flavor-nonsinglet combinations, and $\mathbf J_5=2\mathbf s$ yields the axial-charge diffusion constant $D=\frac13\int_0^\infty\dd t\,(j,e^{-Ct}j)/(j,j)=\tau/3$. Along the diffusive mode ($\lambda\simeq Dk^2$), the spin density follows the axial-charge gradient via $\mathbf s=-\frac D2\nabla J_5^0$. Starting from an initial spin polarization $\Phi(0)=h_0j/T$ ($s(0)=\chi h_0$) and $J_5^0(0)=0$ at $k\ll\gamma_{\mathrm{tr}}$ and $\Gamma_5=0$, integrating the axial Ward identity $\partial_tJ_5^0+2\nabla\cdot\mathbf s=0$ across the initial kinetic decay gives
\begin{equation*}
 J_5^0\simeq-\frac{2\ii k h_0}{T}
 \int_0^\infty\!\dd t\,(j,e^{-Ct}j)
 =-2\ii k\chi h_0\tau.
\end{equation*}
On the subsequent diffusive time scale, the resulting diffusive axial current carries a spin density
\begin{equation*}
 \frac{s(t)}{\chi h_0}\simeq-Dk^2\tau\,e^{-Dk^2t}
 =-\frac{(k\tau)^2}{3}e^{-Dk^2t},
\end{equation*}
opposite in sign to the initial polarization. Eliminating the axial charge from the coupled singlet moment equations (Sec.~\ref{sec:rotnum-supp}) yields the secular equation for the longitudinal decay rate $\lambda$:
\begin{equation}
 \lambda=\frac{k^2}{3}\int_0^\infty\!\frac{\dd\mu_s(\Gamma)}{\Gamma-\lambda}.
 \label{eq:longsecular}
\end{equation}
At small $k$, its lowest root is the diffusive pole $\lambda\simeq Dk^2$. As $k$ increases toward the inverse mean free path, this diffusive root and the lowest nonhydrodynamic spin-relaxation root approach each other, coalesce at an exceptional point ($k\tau=\sqrt3/2$ in the single-relaxation-time Maxwell--Cattaneo (telegrapher) limit \cite{Romatschke16,Grozdanov19,H21}), and continue at larger $k$ as a complex-conjugate pair of damped propagating modes within the truncated moment system. For unequal quark masses, flavor mixing slightly shifts the diffusion eigenvalues from $\tau/3$; at the reference parameter set~\eqref{eq:finitebasisconfig}, however, the predominantly singlet diffusion constant $D\Gamma_{\mathrm{tr}}=0.287308$ matches $\tau\Gamma_{\mathrm{tr}}/3=0.287307$ to five significant digits.

\item \emph{Axial-charge relaxation.} Current quark masses $m_{0f}=\kappa_fgT$ induce helicity flips \cite{KRR20} via $N_m$ in soft-fermion exchange~\eqref{eq:conversionnorms} and the mass terms in collinear splitting~\eqref{eq:lpmtransfer}, damping the $\ell=0$ axial charge of a massive flavor at a parametric rate $\Gamma_5\sim\kappa_f^2g^4T\sim\kappa_f^2\gamma_{\mathrm{tr}}/\ln(1/g)$, while topological transitions also relax the flavor-singlet axial charge \cite{MooreTassler10}. We set $\Gamma_5=0$ in the baseline analysis to retain the slowest longitudinal modes; since the $J=0$ collision kernel is not evaluated explicitly here, Sec.~\ref{sec:rotnum-supp} treats the strange-quark axial damping rate $\Gamma_5$ as a free parameter and tracks its impact on the mode spectrum and spin residues.
\end{itemize}

\section{Rapid Rotation, Numerical Solutions, and Heavy-Ion Phenomenology}
\label{sec:rotframe-supp}

\subsection{Kinetic theory in a corotating background and inertial waves}
In a plasma undergoing global rotation $\bm\Omega$, the thermodynamic ratio $|\bm\Omega|/T$ controls the equilibrium vortical polarization~\eqref{eq:jomega}, whereas the dynamical ratio $|\bm\Omega|/\gamma_{\mathrm{tr}}$ compares the macroscopic rotation frequency with the microscopic collisional relaxation rate. The local collision operator $C$ can be evaluated at $\bm\Omega=0$ to leading order whenever the rotation period exceeds the microscopic duration of a collision ($|\bm\Omega|\ll g^2T$ for collinear LPM splitting, $|\bm\Omega|\ll gT$ for soft HTL exchange, and $|\bm\Omega|\ll T$ for hard scattering). Linear perturbations about a rigidly rotating background can then be treated to \emph{all orders} in $|\bm\Omega|/\gamma_{\mathrm{tr}}$ at leading order in $|\bm\Omega|/T\ll1$.

For a rigidly rotating background with inertial-frame velocity $\mathbf u=\bm\Omega\times\mathbf x$, we adopt spatial coordinates corotating at angular velocity $\bm\Omega$ and measure quasiparticle momenta $\mathbf p$ in the local corotating rest frame near the rotation axis ($|\bm\Omega|\rho\ll1$) \cite{Vilenkin80,LL87,AW14,KRR20}. A free quasiparticle then obeys $\dd\mathbf p/\dd t=-2\bm\Omega\times\mathbf p$, and its helicity-locked spin rotates alongside $\mathbf p$. Here $\Phi$ denotes the deviation of the distribution from the stationary rotational equilibrium $\Phi_{\mathrm{rot}}$, and $\delta\mathbf s$ is the spin-density perturbation relative to $3\chi\bm\Omega$. For a perturbation with velocity $\delta\mathbf u$, pressure $\delta P$, and vorticity $\delta\bm\Omega=\tfrac12\nabla\times\delta\mathbf u$ superimposed on the rotating background, the coupled kinetic and fluid equations read
\begin{equation}
\begin{gathered}
 \partial_t\Phi+\mathbf n\cdot\nabla\Phi-2(\bm\Omega\times\mathbf p)\cdot\nabla_{\mathbf p}\Phi=-C\Bigl(\Phi-\frac{\delta\Omega^ij_\Omega^i}T\Bigr),\\
 w\,(\partial_t\delta\mathbf u+2\bm\Omega\times\delta\mathbf u)=-\nabla\delta P+\eta\nabla^2\delta\mathbf u+\tfrac12\nabla\times\Bigl(\mathbf j_\Omega,C\Bigl(\Phi-\frac{\delta\Omega^ij_\Omega^i}T\Bigr)\Bigr),
\end{gathered}
 \label{eq:rotating-supp}
\end{equation}
where the momentum derivative in the first line is evaluated in the comoving helicity basis. In the fixed Pauli basis, the quark Coriolis operator takes the form $-2(\bm\Omega\times\mathbf p)\cdot\nabla_{\mathbf p}\Phi-\ii[\bm\Omega\cdot\bm\sigma,\Phi]$, and the gluon operator similarly rotates the momentum and physical polarization triad together. This simultaneous rotation of momentum and spin is anti-Hermitian under the thermal inner product, commutes with the isotropic collision operator $C$, and preserves the perturbation free energy. Because the fluid Coriolis force also does no work, $\delta\mathbf u\cdot(\bm\Omega\times\delta\mathbf u)=0$, three properties follow directly:
\begin{enumerate}
\item At $\mathbf k=0$, spin components parallel to $\bm\Omega$ are unaffected by the Coriolis force and relax along the static-medium curve in Fig.~2(a) of the Letter, whereas transverse components precess reversibly at frequency $2|\bm\Omega|$ relative to the corotating axes while damping at the static-medium rates $\lambda_n$.
\item The free-energy dissipation identity~\eqref{eq:dissipation-supp} and the upper bound~\eqref{eq:slowstate-supp} on off-equilibrium spin polarization hold unchanged (with $s$ replaced by $\delta s$) for arbitrary $|\bm\Omega|/\gamma_{\mathrm{tr}}$ throughout the regime $|\bm\Omega|\ll g^2T$.
\item For incompressible perturbations with $c_sk\gg2|\bm\Omega|$ propagating at angle $\theta$ relative to $\bm\Omega$, the Coriolis force in \eqref{eq:rotating-supp} converts the two overdamped transverse shear modes into circularly polarized inertial waves \cite{Greenspan68,LL87} with dispersion relation $\omega=\pm2|\bm\Omega|\cos\theta-\ii\eta k^2/w$, and renders axial-charge diffusion anisotropic. Diffusion along $\bm\Omega$ remains unchanged, whereas diffusion perpendicular to $\bm\Omega$ is suppressed and accompanied by a Hall-like axial current orthogonal to both $\bm\Omega$ and $\nabla J_5^0$. Specifically, the first angular moment of the Coriolis term yields $2\bm\Omega\times\mathbf J_5$, so the linear response to $\nabla J_5^0$ involves $(C+2\bm\Omega\times)^{-1}$, which acts as $(\Gamma-2\bm\Omega\times)/(\Gamma^2+4|\bm\Omega|^2)$ in the plane perpendicular to $\bm\Omega$ for a collision eigenvalue $\Gamma$. With $\hat{\bm\Omega}=\bm\Omega/|\bm\Omega|$, we parameterize the constitutive relation as
\begin{equation*}
 \mathbf J_5=-D_\parallel\hat{\bm\Omega}(\hat{\bm\Omega}\cdot\nabla J_5^0)-D_\perp\bigl[\nabla J_5^0-\hat{\bm\Omega}(\hat{\bm\Omega}\cdot\nabla J_5^0)\bigr]+D_{\mathrm H}\hat{\bm\Omega}\times\nabla J_5^0.
\end{equation*}
For a flavor-symmetric collision operator, spectral decomposition yields
\begin{equation}
 D_\parallel=\frac\tau3,\qquad
 D_\perp(|\bm\Omega|)=\frac13\int_0^\infty\!\dd\mu_s(\Gamma)\,\frac{\Gamma}{\Gamma^2+4|\bm\Omega|^2},\qquad
 D_{\mathrm H}(|\bm\Omega|)=\frac13\int_0^\infty\!\dd\mu_s(\Gamma)\,\frac{2|\bm\Omega|}{\Gamma^2+4|\bm\Omega|^2}.
 \label{eq:axialdiffusion-supp}
\end{equation}
\end{enumerate}

To determine the spin carried by an inertial wave of frequency $\omega=\pm2|\bm\Omega|\cos\theta-\ii\eta k^2/w$, we decompose its vorticity perturbation $\delta\bm\Omega$ into spherical-harmonic components with azimuthal quantum number $m=0,\pm1$ about $\bm\Omega$. Because component $m$ precesses at frequency $2m|\bm\Omega|$ under the Coriolis force, the leading-gradient quasiparticle response samples the shifted frequency $\omega+2m|\bm\Omega|$:
\begin{equation}
 \delta\mathbf s=\sum_{m=0,\pm1}\Bigl[2+\frac{\bigl(j,C(C-\ii\omega-2\ii m|\bm\Omega|)^{-1}j_\Omega\bigr)}{T\chi}\Bigr]\chi\,\delta\bm\Omega_m.
 \label{eq:inertialspin-supp}
\end{equation}
For wave propagation parallel to the rotation axis ($\theta=0$, $\mathbf k\parallel\bm\Omega$), the circularly polarized wave vorticity rotates at the Coriolis precession frequency ($\omega+2m|\bm\Omega|=-\ii\eta k^2/w\simeq0$), yielding $\delta\mathbf s=3\chi\delta\bm\Omega$ up to a small viscous lag. For oblique angles $0<\theta<\pi/2$ [Fig.~3(c) of the Letter], the relative off-equilibrium fraction $\{[(\delta s^i-3\chi\delta\Omega^i)^*(\delta s^i-3\chi\delta\Omega^i)]/[9\chi^2\delta\Omega^{j*}\delta\Omega^j]\}^{1/2}$ grows linearly as $|\omega+2m|\bm\Omega||\tau_\Omega/3\sim|\bm\Omega|\tau_\Omega$ at slow rotation and saturates at $1/3$ for $|\bm\Omega|\tau_\Omega\gg1$, where rapid precession averages out the quasiparticle polarization and leaves only the instantaneous magnetization contribution $2\chi\delta\bm\Omega$. This high-frequency lag reflects reversible Coriolis precession rather than a slow dissipative spin mode.

Causality further restricts $|\bm\Omega|/\gamma_{\mathrm{tr}}$ in the hydrodynamic regime. Rigid rotation requires the system to lie within the light cylinder, $|\bm\Omega|R<1$ for a plasma of transverse radius $R$ \cite{AW14,CG17}. In the collision-dominated hydrodynamic regime, the system size must exceed the mean free path, $R\gg\gamma_{\mathrm{tr}}^{-1}$, which immediately implies $|\bm\Omega|<R^{-1}\ll\gamma_{\mathrm{tr}}$. More generally, subluminal flow ($\mathbf u^2<1$) varying over a macroscopic gradient length $L\gg\gamma_{\mathrm{tr}}^{-1}$ has local vorticity $|\bm\Omega|\lesssim L^{-1}\ll\gamma_{\mathrm{tr}}$. The ratio $|\bm\Omega|/\gamma_{\mathrm{tr}}$ is thus bounded by the Knudsen number $\mathrm{Kn}\sim(\gamma_{\mathrm{tr}}L)^{-1}$, meaning that rotation rapid enough to generate an order-unity spin lag ($|\bm\Omega|\gtrsim\gamma_{\mathrm{tr}}$) lies outside the hydrodynamic regime.

\subsection{Numerical solution in a rotating plasma and variational convergence}
\label{sec:rotnum-supp}
We project the coupled fluid--kinetic equations onto the $J=1$ polarization profiles and the three equilibrium $J=0$ axial-charge profiles using the cubic B-spline bases of Sec.~\ref{sec:finitebasis-supp} ($N_b=24$ and $40$) and the matrix elements $Q[b_A,b_B]$ of the leading-order $J=1$ collision operator evaluated at $\mathbf k=0$ and $\bm\Omega=0$. The rotational source $j_\Omega$ shares the quark entries of $j$, and its projection onto the retained modes omits only a fraction $2.4\times10^{-3}$ of $(j_\Omega,j_\Omega)$ at both basis sizes (stemming from soft gluons below the numerical cutoff $p=0.01T$, where the off-equilibrium perturbation $\Phi-\Phi_{\mathrm{rot}}$ vanishes). In terms of the normalized eigenmodes $\Phi_n$ of \eqref{eq:galerkin}, the mixed spectral weights $v_n$ and the finite-basis response to a prescribed background rotation in \eqref{eq:vortical-supp} read
\begin{equation*}
 v_n=\frac{(j,\Phi_n)(\Phi_n,j_\Omega)}{(j,j)},
 \qquad
 \frac{s(\omega)}{\Omega^i(\omega)}\simeq2\chi+\chi\sum_n v_n\frac{\lambda_n}{\lambda_n-\ii\omega},
\end{equation*}
where the two factors in the numerator of $v_n$ represent the overlaps of $\Phi_n$ with the spin observable and the vortical source, respectively.

For perturbations about the static background ($\bm\Omega=0$), expanding $\Phi=\sum_n a_n(t)\Phi_n$ with $a_n=(\Phi_n,\Phi)$ and projecting \eqref{eq:kinetic-supp} and \eqref{eq:torque-supp} onto the transverse channel at wave number $k$ yields
\begin{equation}
\begin{aligned}
 \dot a_n&=-\lambda_n\bigl[a_n-(\Phi_n,\delta\Phi_{\mathrm{rot}})\bigr],\\
 \partial_t\delta\Omega^i&=-\frac\eta wk^2\delta\Omega^i+\frac{k^2}{4w}\sum_n\lambda_n(j_\Omega,\Phi_n)\bigl[a_n-(\Phi_n,\delta\Phi_{\mathrm{rot}})\bigr].
\end{aligned}
 \label{eq:galerkin-rot}
\end{equation}
In the reflection-odd longitudinal sector, the amplitudes $a_n(t)$ couple to the three flavor axial-charge densities $J_{5f}^0(t)$:
\begin{equation}
\begin{aligned}
 \dot a_n&=-\lambda_na_n-\frac{2\ii k}{T^3}\sum_{f\in\{u,d,s\}}J_{5f}^0\,(\Phi_n,j_f),\\
 \partial_tJ_{5f}^0&=-2\ii k\sum_n(j_f,\Phi_n)a_n,
\end{aligned}
 \label{eq:galerkin-rot-long}
\end{equation}
where $j_f$ is the flavor-$f$ component of $j=\sum_fj_f$. In both sectors, the total spin density is $s=\sum_n(j,\Phi_n)a_n+2\chi\delta\Omega^i$. For the flow sector, we use the leading-order shear viscosity $\eta=1.375\times10^{13}\,T^3$ at $g=10^{-3}$, obtained variationally in the $J=2$ tensor channel with each radial profile expanded in ten functions, $x^2(1+x^2)^{-1}$ times the powers $0,\dots,9$ of $x/(1+x)$ with $x=p/T$. With four, six, and eight functions the estimate is $1.242$, $1.343$, and $1.368\times10^{13}\,T^3$; the increments decrease geometrically by a factor of about four, so the ten-function value lies about $0.2\%$ below the complete-basis limit, and the spread of four independent quadrature replicas gives a statistical error of $0.13\%$. The ten-function value exceeds the next-to-leading-logarithmic result of Ref.~\cite{AMY03b}, $1.369\times10^{13}\,T^3$, by $0.4\%$. It is evaluated at $\kappa=(0,0,1)$ and $q_*=\sqrt g\,T$; with four functions, changing $\kappa_s$ to $0.5$ or $0$, which bracket the value $0.3$ of \eqref{eq:finitebasisconfig}, lowers $\eta$ by $0.10\%$ or $0.16\%$, and halving or doubling $q_*$ changes it by at most $0.24\%$. Together with $w=19\pi^2T^4/9$, the ten-function value gives $\eta k^2/w=0.1130\,k^2/\Gamma_{\mathrm{tr}}$. Because $(k/T)^2=2.9\times10^{-26}\,(k/\Gamma_{\mathrm{tr}})^2$ at $g=10^{-3}$, the back-reaction of the collisional torque in \eqref{eq:galerkin-rot} is of relative order $10^{-26}$ at $k\sim\Gamma_{\mathrm{tr}}$ and is evaluated perturbatively to first order. Refining the radial basis tests the momentum dependence within this angular truncation, though omitted nonhydrodynamic $J=0$ profiles, $J=2$ axial profiles, and higher angular moments can still modify the dispersion relations at finite $k$.

\begin{table}[t]
 \centering
 \begingroup\renewcommand{\arraystretch}{1.2}
 \small
 \begin{tabular}{@{}p{0.58\textwidth}rr@{}}\toprule
 Quantity & $N_b=24$ & $N_b=40$\\\midrule
 Integrated spin relaxation time $\tau\Gamma_{\mathrm{tr}}$ & $0.8637$ & $0.8619$\\
 Vortical relaxation time $\tau_\Omega\Gamma_{\mathrm{tr}}$ & $1.0339$ & $1.0316$\\
 Negative mixed spectral weights $v_n$: count; sum & $42$; $-0.21$ & $58$; $-0.28$\\
 Half-equilibration time of $(j,\Phi)$ after switching on $\bm\Omega$, $t\Gamma_{\mathrm{tr}}$ & $0.658$ & $0.657$\\
 Predominantly singlet axial diffusion: $\tau\Gamma_{\mathrm{tr}}/3$; mode at $k=0.01\,\Gamma_{\mathrm{tr}}$ & $0.2879$; $0.2879$ & $0.2873$; $0.2873$\\
 Relative spin amplitude in the axial diffusion mode at $k=0.1\,\Gamma_{\mathrm{tr}}$; asymptotic $-(k\tau)^2/3$ & $-0.00251$; $-0.00249$ & $-0.00250$; $-0.00248$\\
 Relative spin amplitude in the axial diffusion mode at $k=0.3\,\Gamma_{\mathrm{tr}}$ & $-0.0245$ & $-0.0243$\\
 Non-singlet axial diffusion constants $D\Gamma_{\mathrm{tr}}$ & $0.2710$, $0.2712$ & $0.2705$, $0.2707$\\
 Coalescence in the retained singlet system, $k\tau$ & $0.749$ & $0.749$\\
 Shear-mode spin deviation $s/(3\chi\delta\Omega^i)-1$ at $k=0.1\,\Gamma_{\mathrm{tr}}$ & $3.9\times10^{-4}$ & $3.9\times10^{-4}$\\
 \quad at $k=0.3\,\Gamma_{\mathrm{tr}}$ & $3.5\times10^{-3}$ & $3.5\times10^{-3}$\\
 \quad at $k=\Gamma_{\mathrm{tr}}$ & $0.045$ & $0.045$\\
 Off-equilibrium spin transferred to the shear mode, per $(k/T)^2$ & $0.0090$ & $0.0090$\\\bottomrule
 \end{tabular}
 \endgroup
 \caption{Numerical results for the retained angular moments in \eqref{eq:galerkin-rot} and \eqref{eq:galerkin-rot-long} for the parameter set~\eqref{eq:finitebasisconfig}, using $\eta k^2/w=0.1130\,k^2/\Gamma_{\mathrm{tr}}$. The coalescence entries use the flavor-singlet equations leading to \eqref{eq:longsecular}; the axial-diffusion entries retain all three flavors. Relative spin amplitudes are normalized by the initial longitudinal polarization.}
 \label{tab:rotation}
\end{table}

\begin{figure}[t]
 \centering
 \includegraphics[width=160mm]{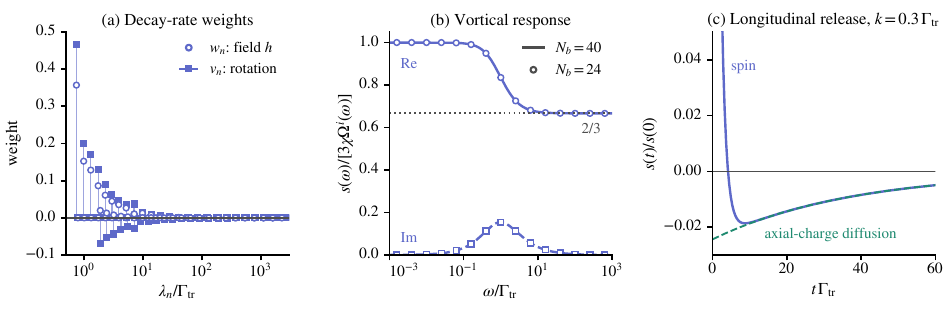}
 \caption{(a) Spectral weights of the decay eigenvalues $\lambda_n$ at $N_b=40$: spin weights $w_n$ for a conjugate field $h$ (open circles) and mixed weights $v_n$ for fluid rotation $\bm\Omega$ (filled squares). (b) Real (solid) and imaginary (dashed) parts of $s(\omega)/[3\chi\Omega^i(\omega)]$ from \eqref{eq:vortical-supp} at $N_b=40$, with $N_b=24$ shown as open markers; horizontal dotted line: high-frequency magnetization limit $2/3$. (c) Longitudinal spin density after releasing an initial polarization at $k=0.3\,\Gamma_{\mathrm{tr}}$, normalized to its initial value (off scale, solid purple), compared with the asymptotic axial-charge diffusion contribution (dashed green).}
 \label{fig:rotationsupp}
\end{figure}

The resulting spectral and hydrodynamic quantities are summarized in Table~\ref{tab:rotation} and Fig.~\ref{fig:rotationsupp}. As shown in Fig.~\ref{fig:rotationsupp}(b), the dynamic vortical response $s(\omega)/[3\chi\Omega^i(\omega)]$ approaches the rotational-equilibrium limit $1$ as $\omega\to0$ with low-frequency slope $\tau_\Omega/3$ ($\tau_\Omega\simeq1.032/\Gamma_{\mathrm{tr}}\simeq1.197\tau$), and asymptotes to the instantaneous magnetization fraction $2/3$ for $\omega\gg\Gamma_{\mathrm{tr}}$. Although the mixed spectral weights $v_n$ [Fig.~\ref{fig:rotationsupp}(a)] can take negative values at higher decay rates because $j\ne j_\Omega$ in the gluon sector, the cumulative sums of $v_n$ and $\lambda_nv_n$ ordered by increasing $\lambda_n$ remain positive in both bases. Summation by parts then guarantees that $(j,e^{-Ct}j_\Omega)/(j,j)$ is positive and monotonically decreasing for all $t\ge0$. When an initial longitudinal spin polarization is released at $k=0.3\,\Gamma_{\mathrm{tr}}$ [Fig.~\ref{fig:rotationsupp}(c)], it damps on the transport time scale $\Gamma_{\mathrm{tr}}^{-1}$, leaving an axial-charge diffusion tail of relative amplitude $-0.0243$, opposite in sign to the initial polarization and close to the small-$k$ estimate $-(k\tau)^2/3=-0.0223$. Within the retained singlet moment system, increasing $k$ brings the axial diffusion rate and the lowest nonhydrodynamic spin-relaxation rate $\lambda_0$ together until they coalesce at the exceptional point $k\tau=0.749$ (where both equal $0.415\,\Gamma_{\mathrm{tr}}$), beyond which they split into a complex-conjugate pair of damped propagating modes. Between $N_b=24$ and $40$, the frequency-dependent vortical response, the time-domain relaxation curve in Fig.~3(a) of the Letter, and the longitudinal relaxation curve differ by at most $3.3\times10^{-4}$, $2.5\times10^{-4}$, and $8.8\times10^{-4}$, respectively.

\begin{figure}[t]
 \centering
 \includegraphics[width=160mm]{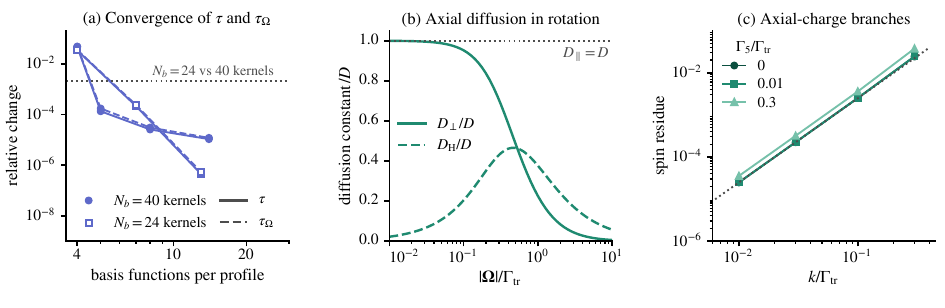}
 \caption{(a) Relative change of $\tau$ (solid) and $\tau_\Omega$ (dashed) from the finest basis across nested B-spline subspaces, evaluated on the quadrature meshes of the $N_b=40$ (filled circles) and $N_b=24$ (open squares) calculations; horizontal dotted line: difference between the two quadrature meshes. (b) Anisotropic diffusion coefficients $D_\perp(|\bm\Omega|)$ (solid) and $D_{\mathrm H}(|\bm\Omega|)$ (dashed) from \eqref{eq:axialdiffusion-supp}, normalized by $D=\tau/3$; horizontal dotted line: $D_\parallel/D=1$. (c) Magnitude of the summed spin residues of the three axial-charge branches, normalized by the initial longitudinal polarization, versus $k/\Gamma_{\mathrm{tr}}$ for strange-quark axial relaxation rates $\Gamma_5=0$, $0.01$, and $0.3\,\Gamma_{\mathrm{tr}}$ (shades of green), compared with the $\Gamma_5=0$ small-$k$ asymptote $(k\tau)^2/3$ (gray dotted).}
 \label{fig:continuumsupp}
\end{figure}

To separate basis truncation errors from multidimensional quadrature errors in the collision integrals, we exploit the knot-insertion property of B-splines \cite{deBoor78}: cubic B-splines defined on nested subsets of a uniform logarithmic grid form strictly nested variational subspaces. At fixed quadrature meshes, the variational principle guarantees that $(j,C^{-1}j)$ and each quadratic form on the right-hand side of $4(j,C^{-1}j_\Omega)=(j+j_\Omega,C^{-1}(j+j_\Omega))-(j-j_\Omega,C^{-1}(j-j_\Omega))$ increase monotonically with basis dimension. The resulting vortical relaxation time $\tau_\Omega$ likewise increases across all sequences in Table~\ref{tab:continuum}, even though monotonicity of the difference is not guaranteed a priori. As shown in Table~\ref{tab:continuum} and Fig.~\ref{fig:continuumsupp}(a), refining from $14$ to $40$ basis functions per profile at fixed quadrature meshes changes $\tau$ and $\tau_\Omega$ by at most $1.2\times10^{-5}$, and refining from $13$ to $24$ functions on the $N_b=24$ mesh changes them by only $5\times10^{-7}$. An asymptotic estimate of the tail $p>24T$, where the angular damping rate scales as $T^2/p^2$, yields a relative correction below $10^{-7}$ to $(j,C^{-1}j)$. These basis-truncation variations are far smaller than the $2.1\times10^{-3}$ difference between the $N_b=24$ and $40$ columns in Table~\ref{tab:rotation}, which reflects the simultaneous refinement of the angular quadratures and LPM meshes.

\begin{table}[t]
 \centering
 \begingroup\renewcommand{\arraystretch}{1.2}
 \small
 \begin{tabular}{@{}lrrrr@{}}\toprule
 Integration meshes; basis functions per profile & $\tau\Gamma_{\mathrm{tr}}$ & $\tau_\Omega\Gamma_{\mathrm{tr}}$ & $\lambda_0/\Gamma_{\mathrm{tr}}$ & Spin/current rate\\\midrule
 $N_b=40$ mesh; $5$ functions & $0.861807$ & $1.031437$ & $0.7978$ & $0.927871$\\
 $N_b=40$ mesh; $8$ functions & $0.861899$ & $1.031582$ & $0.7583$ & $0.927786$\\
 $N_b=40$ mesh; $14$ functions & $0.861912$ & $1.031601$ & $0.7506$ & $0.927776$\\
 $N_b=40$ mesh; $40$ functions & $0.861921$ & $1.031614$ & $0.7467$ & $0.927773$\\
 $N_b=24$ mesh; $7$ functions & $0.863557$ & $1.033687$ & $0.7776$ & $0.927749$\\
 $N_b=24$ mesh; $13$ functions & $0.863739$ & $1.033929$ & $0.7467$ & $0.927691$\\
 $N_b=24$ mesh; $24$ functions & $0.863739$ & $1.033930$ & $0.7455$ & $0.927691$\\
 $N_b=40$ mesh; $2$-function subspace & $0.603085$ & $0.719935$ & & \\
 $N_b=24$ mesh; $2$-function subspace & $0.604083$ & $0.721215$ & & \\\bottomrule
 \end{tabular}
 \endgroup
 \caption{Variational convergence of $\tau$, $\tau_\Omega$, and $\lambda_0$ across nested B-spline subspaces at fixed collision-integral quadrature meshes. Across each sequence, $\tau$ and $\tau_\Omega$ increase while $\lambda_0$ decreases with basis refinement.}
 \label{tab:continuum}
\end{table}

To compute the coupled mode spectrum and spin response as a function of $|\bm\Omega|$ [Fig.~3(b) of the Letter and Fig.~\ref{fig:continuumsupp}(b)], we retain all three Cartesian components of the $J=1$ polarization profiles, the three flavor axial charges, and the transverse incompressible flow in \eqref{eq:rotating-supp}. At $k=0.3\,\Gamma_{\mathrm{tr}}$ and sound speed $c_s=1/\sqrt3$, strict incompressibility ($c_sk\gg2|\bm\Omega|$) requires $|\bm\Omega|\ll0.0866\,\Gamma_{\mathrm{tr}}$; beyond this scale, the curves represent an extrapolation of the incompressible moment truncation. To evaluate the mode weights, we prepare and switch off a weak field conjugate to the off-equilibrium spin $\delta s^i-3\chi\delta\Omega^i$ along each Cartesian axis $i=1,2,3$ in turn, summing the three responses and normalizing by the total initial polarization. Over the entire range $|\bm\Omega|\le3\,\Gamma_{\mathrm{tr}}$ at $\mathbf k\perp\bm\Omega$, the summed magnitude of the slow-mode amplitudes ($\operatorname{Re}\lambda_n<0.1\,\Gamma_{\mathrm{tr}}$) remains below $0.00812$, while all nonhydrodynamic spin-relaxation rates exceed $0.7314\,\Gamma_{\mathrm{tr}}$ [Fig.~3(b) of the Letter]. At $\bm\Omega=0$, the longitudinal diffusive residue $-0.02435$ contributes $-0.02435/3$ to the three-component average, accounting for the factor-of-three reduction relative to the purely longitudinal response. The corresponding flavor-singlet projections of the anisotropic diffusion coefficients $D_\perp(|\bm\Omega|)$ and $D_{\mathrm H}(|\bm\Omega|)$ are shown in Fig.~\ref{fig:continuumsupp}(b).

We also examine the sensitivity of our results to a finite strange-quark axial-charge relaxation rate $\Gamma_5$ and to variations in the individual collision channels. Including $\Gamma_5$ in \eqref{eq:galerkin-rot-long} introduces a damping gap in the strange-quark-dominated axial branch while leaving the two light-flavor branches diffusive. Increasing $\Gamma_5$ from $0$ to $0.3\,\Gamma_{\mathrm{tr}}$ at $k=0.3\,\Gamma_{\mathrm{tr}}$ shifts the summed longitudinal spin residue of the three axial-charge branches only modestly, from $-0.02435$ to $-0.03760$, while preserving its $\propto k^2$ scaling at small $k$ [Fig.~\ref{fig:continuumsupp}(c)]. Likewise, individually rescaling the soft-fermion, collinear, or elastic contributions in the $N_b=24$ collision operator by a factor of $1/2$ or $2$ (normalized to the corresponding strange-current rate $\Gamma_{\mathrm{tr}}$) alters $\tau\Gamma_{\mathrm{tr}}$ by at most $5\%$ ($0.845$--$0.898$) and $\tau_\Omega/\tau$ by at most $6\%$ ($1.158$--$1.263$), comparable to the mild variation across the quark-mass points in Table~\ref{tab:masspoints} ($\tau\Gamma_{\mathrm{tr}}=0.829$--$0.865$ and $\tau_\Omega/\tau=1.188$--$1.197$).

\subsection{Quantitative estimates for relativistic heavy-ion collisions}
\label{sec:hic-supp}
To connect our microscopic kinetic results with the quark--gluon plasma (QGP) produced in relativistic heavy-ion collisions, Table~\ref{tab:hic} compiles the characteristic rotation scales extracted from experimental measurements and hydrodynamic simulations. Global spin polarization measurements of $\Lambda$ and $\bar\Lambda$ hyperons by the STAR and ALICE Collaborations probe the average thermal vorticity at hadronization via $|\bm\Omega|/T\simeq P_\Lambda+P_{\bar\Lambda}$ \cite{Bec13,STAR17,STAR18,ALICE20}. Relativistic viscous hydrodynamic simulations of noncentral $\mathrm{Au}+\mathrm{Au}$ collisions at $\sqrt{s_{NN}}=200\,\mathrm{GeV}$ yield a freeze-out thermal vorticity with spatial average $|\bm\Omega|/T\simeq5\times10^{-3}$ and local peaks up to $|\bm\Omega|/T\simeq0.05$ \cite{Bec15}, spanning the range $10^{-3}\lesssim|\bm\Omega|/T\lesssim5\times10^{-2}$.

Because the strongly coupled QGP does not admit a direct weak-coupling expansion in powers of $g$ \cite{GM05,S05,KSS05,HS13,BRS18}, we obtain a phenomenological estimate by expressing the light-quark spin relaxation time in units of the kinematic shear viscosity $\eta/w$. Both quantities share the parametric weak-coupling scale $[g^4T\ln(1/g)]^{-1}$ \cite{AMY03b,GMT18}, which cancels in their ratio at leading-logarithmic order. Extrapolating our calculated ratio $\tau w/\eta\simeq7.63$ (Sec.~\ref{sec:rotnum-supp}) and combining it with the phenomenological range $\eta T/w=0.08$--$0.2$ \cite{KSS05,HS13,BRS18} yields
\begin{equation*}
 \tau T=\left(\frac{\tau w}{\eta}\right)\left(\frac{\eta T}{w}\right)=0.61\text{--}1.53,
 \qquad \frac{1}{\tau T}=0.66\text{--}1.64.
\end{equation*}
Combining this result with the thermal-vorticity range above gives the dimensionless product of the spin relaxation time and rotation rate,
\begin{equation*}
 |\bm\Omega|\tau=\left(\frac{|\bm\Omega|}{T}\right)(\tau T)\simeq10^{-3}\text{--}0.08\ll1,
\end{equation*}
placing the QGP well within the slow-rotation regime where spin closely tracks its local-equilibrium constitutive relation. Independently of hyperon polarization data, causality ($|\bm\Omega|R<1$) \cite{AW14,CG17} bounds $|\bm\Omega|\tau<\tau/R\simeq0.06$--$0.4$ for a fireball of transverse radius $R\simeq5$--$7\,\mathrm{fm}$ at $T=150$--$300\,\mathrm{MeV}$. Using $\tau_\Omega/\tau=1.197$, the second-order vortical lag $-\chi\tau_\Omega\partial_t\bm\Omega$ in (7) of the Letter yields a relative correction $\tau_\Omega/(3t_\Omega)\simeq0.06$--$0.16$ for vorticity evolving over a characteristic expansion time $t_\Omega\simeq3\,\mathrm{fm}/c$ at $T=250\,\mathrm{MeV}$.

\begin{table}[t]
 \centering
 \begingroup\renewcommand{\arraystretch}{1.2}
 \small
 \begin{tabular}{@{}>{\raggedright\arraybackslash}p{0.43\textwidth}>{\raggedright\arraybackslash}p{0.34\textwidth}l@{}}\toprule
 Source & Value & $|\bm\Omega|/T$\\\midrule
 STAR, $\mathrm{Au}+\mathrm{Au}$ $7.7$--$39\,\mathrm{GeV}$, energy averaged & $P_\Lambda=1.08\%$, $P_{\bar\Lambda}=1.38\%$; $|\bm\Omega|\simeq9\times10^{21}\,\mathrm s^{-1}$ \cite{STAR17} & $0.025$\\
 STAR, $\mathrm{Au}+\mathrm{Au}$ $200\,\mathrm{GeV}$, $20$--$50\%$ centrality & $P_\Lambda=0.277\%$, $P_{\bar\Lambda}=0.240\%$ \cite{STAR18} & $0.005$\\
 ALICE, $\mathrm{Pb}+\mathrm{Pb}$ $2.76$, $5.02\,\mathrm{TeV}$, $15$--$50\%$ centrality & $P=(-0.01\pm0.05\pm0.03)\%$ \cite{ALICE20} & $\lesssim0.003$\\
 Viscous hydrodynamics, $\mathrm{Au}+\mathrm{Au}$ $200\,\mathrm{GeV}$, freeze-out at $130\,\mathrm{MeV}$ & thermal vorticity: mean $5\times10^{-3}$, local up to $0.05$; kinematic vorticity $\simeq0.03\,c/\mathrm{fm}$ \cite{Bec15} & $0.005$--$0.05$\\
 Spin relaxation from $\eta T/w=0.08$--$0.2$ and $\tau w/\eta=7.63$ & $1/(\tau T)=0.66$--$1.64$ ($\tau T=0.61$--$1.53$) & \\\bottomrule
 \end{tabular}
 \endgroup
 \caption{Characteristic rotation scales in relativistic heavy-ion collisions and the spin relaxation rate estimated from the specific shear viscosity $\eta T/w$. For the experimental data, $|\bm\Omega|/T\simeq P_\Lambda+P_{\bar\Lambda}$ without feed-down corrections; the STAR estimate of $|\bm\Omega|$ carries a factor-of-two systematic uncertainty from the chemical/kinetic freeze-out temperature.}
 \label{tab:hic}
\end{table}

\subsection{Rotational corrections to the collision operator}
\label{sec:rotkernel-supp}
While the estimates in Sec.~\ref{sec:hic-supp} evaluate the collision operator at $\bm\Omega=0$, fluid rotation also modifies the microscopic equilibrium distributions, self-energies, and scattering amplitudes entering the collision kernel. Here we estimate the resulting corrections to the spin relaxation time at fixed local temperature near the rotation axis for ultrarelativistic quarks with smoothly varying spectra. Because dynamical masses can change abruptly near rotation-induced phase transitions---as seen in Nambu--Jona-Lasinio models \cite{CG17}---our analysis applies within a single phase, away from critical points and narrow resonances.

The rotational energy shifts underlying \eqref{eq:jomega} are of order $|\bm\Omega|$ \cite{AW14,Lin26}, so expanding smooth thermal distributions and hard scattering amplitudes yields relative corrections of order $|\bm\Omega|/T$. In an otherwise isotropic medium, parity and time-reversal symmetries require the homogeneous longitudinal spin relaxation rate to be even in $\bm\Omega$, pushing analytic thermal corrections to quadratic order, $\mathcal O(|\bm\Omega|^2/T^2)$. For the heavy-ion vorticity range in Table~\ref{tab:hic}, $|\bm\Omega|/T\lesssim5\times10^{-2}$ and $(|\bm\Omega|/T)^2\lesssim2.5\times10^{-3}$, implying percent-level or smaller thermal modifications for order-unity coefficients. Soft propagators and coherent collinear splitting probe the lower scales $gT$ and $g^2T$, introducing corrections controlled by $|\bm\Omega|/(gT)$ and $|\bm\Omega|/(g^2T)$, which likewise remain small under the scale hierarchy of Sec.~\ref{sec:rotframe-supp}. Microscopic rotational corrections to $C$ are thus governed by the ratios of $|\bm\Omega|$ to the intrinsic energy scales of the collision processes, even when $|\bm\Omega|/\gamma_{\mathrm{tr}}$ is of order unity or larger.

The sensitivity of $\tau$ to such modifications can be bounded directly. Let $C(\bm\Omega)$ denote the dissipative collision operator in the rotating background, defined with respect to the same thermal inner product as $C$ after matching equilibrium weights and projecting out conserved modes. Assuming $C(\bm\Omega)$ satisfies the relative quadratic-form bound
\begin{equation}
 \bigl|(\Phi,[C(\bm\Omega)-C]\Phi)\bigr|
 \le\epsilon\,(\Phi,C\Phi),\qquad 0\le\epsilon<1,
 \label{eq:rotkernel-formbound}
\end{equation}
for all perturbations in the common form domain, the lower bound~\eqref{eq:lowerbound} remains valid with its prefactor rescaled by $1-\epsilon$. For homogeneous spin polarization parallel to $\bm\Omega$, the Coriolis force vanishes; holding the leading-order spin source $j$ fixed then isolates the effect of $C(\bm\Omega)-C$ on the integrated relaxation time. Because operator inversion reverses the ordering of positive-definite operators, \eqref{eq:rotkernel-formbound} implies
\begin{equation}
 \frac{\tau}{1+\epsilon}
 \le\frac{(j,C(\bm\Omega)^{-1}j)}{(j,j)}
 \le\frac{\tau}{1-\epsilon}.
 \label{eq:rotkernel-taubound}
\end{equation}
A relative perturbation of $\epsilon=0.3$ to the quadratic form thus confines the relaxation time to $[0.77\tau,\,1.43\tau]$, whereas a tenfold enhancement of $\tau$ would require $\epsilon\ge0.9$.

Finally, because small-angle momentum deflections damp both the helicity-locked spin and the dipole ($m=1$) distribution of the vector current through the same angular weight $1-\cos\theta$, smooth modifications to the scattering cross sections shift both relaxation rates in tandem. This is illustrated by the channel-sensitivity tests in Sec.~\ref{sec:rotnum-supp}: rescaling the elastic, soft-fermion, or collinear contributions individually by factors of $1/2$ or $2$ changes the ratio $\tau\Gamma_{\mathrm{tr}}$ to $0.845$--$0.898$, within $5\%$ of the baseline value. Together with \eqref{eq:rotkernel-taubound}, this confirms that light-quark spin relaxation remains at the transport scale under smooth rotational modifications to the microscopic scattering processes.

\bibliographystyle{apsrev4-2}
\bibliography{references}